\documentclass[12pt,authoryear]{elsarticle}
\usepackage[utf8]{inputenc}
\usepackage{placeins}
\usepackage{graphicx}
\usepackage[a4paper,width=160mm,top=25mm,bottom=25mm,bindingoffset=6mm]{geometry}
\usepackage[compact]{titlesec}
\usepackage{hyperref}
\usepackage{tikz}
\usetikzlibrary{calc}
\usepackage{eso-pic}
\usepackage{setspace}
\usepackage{lipsum}
\usepackage{url}
\usepackage[english]{babel}
\usepackage{fancyhdr}
\usepackage{array}
\usepackage{tabu}
\usepackage{multirow}
\usepackage{gensymb}
\usepackage{rotating}
 \usepackage{caption}
\usepackage{hanging}
\usepackage{todonotes}
\usepackage{changepage}
\usepackage{subcaption}
\usepackage{lscape}
\usepackage{chngcntr}
\usepackage{float}
\floatstyle{plaintop}
\restylefloat{table}
\usepackage{xcolor}
\usepackage{amsmath}
\usepackage{mathtools}
\usepackage{eufrak}
\usepackage[bbgreekl]{mathbbol}
\usepackage[makeroom]{cancel}
\usepackage{mathrsfs}
\DeclareSymbolFontAlphabet{\mathbbl}{bbold}
\usepackage{multirow,tabularx}
\usepackage{comment}
\hypersetup{
    colorlinks,
    linkcolor={blue},
    citecolor={blue},
    urlcolor={blue}
}
\usepackage{longtable}
\usepackage{ragged2e}
\usepackage{tikz, blindtext}
\usepackage{fix-cm}
\usepackage{cleveref}

\usepackage{enumitem}
\usepackage{multicol}
\usepackage{longtable}
\usepackage{amsfonts}
\usepackage[font=footnotesize]{caption}

\usepackage[english]{babel}
\usepackage[labelformat=simple]{subcaption}

\usepackage{amsthm}
\newtheorem*{remark}{Remark}

\newif\ifthesis
\thesisfalse

\begin{document}
\begin{frontmatter}
    
\title{Finite strain homogenization of periodic rod networks with application to semi-flexible biopolymers}

\author[a]{Vinayak}

\author[b]{Prashant K. Purohit}

\author[a]{Ajeet Kumar}


\affiliation[a]{organization={Department of Applied Mechanics},
addressline={Indian Institute of Technology}, 
state={Delhi},
country={India}}

\affiliation[b]{organization={Department of Mechanical Engineering and Applied Mechanics},
addressline={University of Pennsylvania}, 
city={Philadelphia},
postcode={19104}, 
state={Pennsylvania},
country={United States of America}}

\cortext[corauthor]{Corresponding author}

\begin{abstract}
In this work, we adopt a finite strain computational homogenization approach to characterize the response of semi-flexible biopolymer networks modeled as idealized 8- and 14-chain periodic networks. We use the geometrically exact special Cosserat rod theory to model the microscale fibers forming these 8- and 14-chain networks. This allows us to capture arbitrarily large microscale deformations. Both macroscopic strain- and stress-driven homogenization are performed to study the macroscopic uniaxial tension, compression and simple shear responses. Several phenomena unique to biopolymer networks are recovered such as strain-stiffening and volume shrinkage under uniaxial tension, softening under compression and reverse Poynting effect under simple shear. We find that nonlinearity and non-affine deformation at microscale, especially bending and buckling of microscale fibers, plays an important role in these phenomena. We obtain the postbuckled solutions of the homogenization problem using a nonlinear, imperfection-free path following approach and also check for their stability. We further compare our homogenization results with experimental data for uniaxial tension and compression of biofilament networks and find good agreement. When the fibers are replaced by helical rods in the 8-chain unit cell, we are also able to capture the enlarged stretching behaviour as shown in recently fabricated compliant metastructures.
\end{abstract}




\end{frontmatter}

\section{Introduction}
Natural and biological materials exhibit huge diversity in their mechanical behaviour - from hard sea shells to soft biological tissues making up animal organs. This diversity is enabled by their hierarchical nature and interactions therein \citep{piechocka2010structural}. This has also inspired a whole new class of bio-inspired artificial materials called the architected or metamaterials \citep{meza2015resilient,jiao2023mechanicalbeyond,jiao2023mechanical}. Biomaterials such as collagen or fibrin can be viewed as a random cross-linked network of biopolymer fibers providing them the necessary structural integrity \citep{picu2011mechanics,munster2013strain}. Individual fibers themselves have a multi-level architecture which imparts unique mechanical behaviour to them~\citep{collet2005elasticity,olive2024deformation}. When the bending stiffness of these fibers cannot be neglected compared to their axial stiffness, they are usually termed as being semi-flexible \citep{picu2011mechanics}. Characterizing the mechanical response of such materials is important not just for understanding existing biophysical phenomena but also to develop future bio-inspired/compatible materials \citep{li2022spider}. However, this endeavour is challenging not just due to the presence of multiple scales, but also because of the multiphysics interactions therein - such as chemical, thermal \citep{su2012semiflexible} or fluidic \citep{purohit2025fluid}. In this work, we restrict ourselves to the purely mechanical multiscale response of semi-flexible biopolymeric materials.\\\\
Polymeric-network materials exhibit highly nonlinear mechanical response \citep{storm2005nonlinear} and their constitutive modeling presents several challenges \citep{picu2021constitutive}. Based on experimental observations, continuum phenomenological material models have been proposed that mimic their behaviour. These can be classified into affine and non-affine models. Among non-affine models are 3-chain \citep{wang1952statistical}, 8-chain \citep{arruda1993three} and microsphere \citep{miehe2004micro} models. In these models, the fibers (or polymer chains) are \textit{a priori} assumed to be oriented in certain directions. The 8-chain model, for example, consists of eight polymer chains oriented along the body diagonals of a unit cell whose edges lie along the principal stretch directions. A brief review of these models is available in \citep{song2022hyperelastic}. The individual chains are modeled using the freely-jointed chain (Langevin function) or the worm-like chain model. Inspired by these models, \citep{brown2009multiscale} proposed an 8-chain model for fibrin networks wherein the fiber is considered as a straight rod that only undergoes stretching. They showed that by choosing an appropriate force-stretch law for individual fibers, one can recover the stretching response of collagen and fibrin networks \citep{purohit2011protein}. \citet{tutwiler2020rupture} extended this to a 14-chain model (which is a special case of the microsphere model and is inspired by \cite{wu1993improved}) in order to capture the rupture behaviour of blood clots. Other phenomenological models, such as the foam-like model \citep{gibson2003cellular} have also been used to capture the compressive response of collagen/fibrin networks \citep{kim2016foam}. \cite{islam2018effect,chen2023nonaffine,shivers2019normal,feng2016nonlinear} have shown that the microstructure has a crucial role to play in the overall response of the network. For example, the nature of cross-link joints (pin-jointed or welded), volume fraction of fibers, and coordination number are some of the important considerations \citep{wagner2006cytoskeletal}. Even though phenomenological models have a robust experimental basis, they do not \textit{a priori} provide insights into the interplay between macro- and micro- scales. Therefore, computational models have been proposed that can be broadly divided into particle-based and rod-network based models. The particle-based approaches use molecular dynamics to study the fiber network deformation behaviour \citep{rodney2005discrete,vahabi2016elasticity,filla2023multiscale,kliuchnikov2025strength}. Among rod-network based models, reduced theories such as Euler-Bernoulli beam \citep{zakharov2024clots} or Timoshenko beam \citep{dey2024evaluation} theories have been used to model the fiber network. These theories are limited to small deformations. To overcome this limitation, the geometrically exact (GE) rod theory \citep{simo1985finite} has been used recently. For example, \citet{prachaseree2025towards} studied the structure-function relationship in a two-dimensional random fiber network modeled using GE rods. Along similar lines, considering a three-dimensional network, \citet{lohr2025modeling} used GE isogeometric beams to show the effect of fiber undulations (intrinsic curvature). They also studied the contribution of fiber deformation modes (such as stretching, bending, and torsion) on the overall response under biaxial tension and simple shear. Computational models have the advantage of considering the full-scale fiber network, allowing them to capture the microscale behaviour accurately. However, this turns out to be computationally expensive, especially when large, arbitrary three-dimensional deformation of fibers needs to be considered. In this work, we use homogenization to obtain the effective behaviour of biopolymer networks using the 8- and 14-chain microstructures - thus avoiding the need to do full-scale simulation. We use the geometrically exact special Cosserat rod model for the microscale fibers, enabling us to capture large non-linear deformations at both the macro and micro scales.\\\\
Experiments have shown that fibers in biopolymer networks also have a tendency to buckle. For example, the strain-softening response of fiber networks under compression \citep{lakes1993microbuckling,kim2014structural,notbohm2015microbuckling} and shear \citep{zakharov2024clots} has been observed to be a consequence of buckling of fibers. Buckling has long been studied as a bifurcation problem wherein the stability of the solutions (to the elasticity problem, for example) is checked and new buckled (or bifurcated) solution paths are obtained that exhibit characteristics considerably different from the primary path \citep{elliott2006stability}. For rod structures, stability has been rigorously studied in \cite{pecknold1985snap,kumar2010generalized} and more recently for architected metamaterials in \cite{combescure2016post,zhang2021computational,azulay2023instability}. Linear buckling techniques and initial imperfection-based post-buckling analysis tools have been used in the rod network homogenization models of \cite{jamshidian2020multiscale,gartner2021nonlinear, herrnbock2022homogenization}. Herein, we study the stability of 8- and 14-chain solutions without introducing any imperfections in the system and obtain the post-buckled solutions using a nonlinear path following approach. The novel contributions of this work are the following:
 \begin{enumerate}
     \item Using a homogenization-based study on 8- and 14-chain geometrically exact rod networks, we show that several phenomena unique to some biomaterials can be recovered - such as strain stiffening and dramatic volume shrinkage under uniaxial tension \citep{purohit2011protein,ehret2017inverse}, sudden softening under compression \citep{kim2014structural} and reverse Poynting effect under simple shear \citep{horgan2017poynting,ramanujam2024mechanics}. Fiber buckling plays a critical role in all of these phenomena.
     \item We show that the homogenization model captures the uniaxial tension experiment of \cite{roeder2002tensile} on collagen and the 8-chain analytical model proposed in \cite{brown2009multiscale} for fibrin. We also recover the compression response of fibrin networks in \cite{liang2017phase}. Furthermore, by considering helix-like individual fibers in the 8-chain model, we show how the force-stretch relationship of individual helical fibers manifests in the macroscopic stress-strain response of a metastructure constructed from them.
 \end{enumerate}

The outline of the paper is as follows. In section \ref{geometrically exact rod theory}, a brief overview of the intrinsically curved geometrically exact special Cosserat rod theory is given. In Section \ref{sec:3d_homogenization_theory}, the homogenization model in both strain and stress-driven settings is presented along with various constraints that are needed to formulate the homogenization problem. The expressions for the macroscopic stress and elasticity tensor are also obtained. In Section \ref{sec:numerical_examples}, several results of numerical simulation performed on 8- and 14-chain RVEs are discussed. Section \ref{sec:conclusions} concludes our paper.\\\\
$\textbf{Notation}$:
The following notations are used unless specified otherwise. Vectors are denoted by lowercase bold letters whereas second-order tensors are denoted by uppercase bold letters. The symbol $(\cdot)^{\prime}$ represents derivative with respect to the rod’s undeformed arc-length. Repeated Latin indices imply summation from 1 to 3 whereas repeated Greek indices imply summation from 1 to 2.\\\\

\section{Brief description of the geometrically exact special Cosserat rod theory}\label{geometrically exact rod theory}
\begin{figure}[h!]
    \centering
    \includegraphics[width=.8\textwidth]{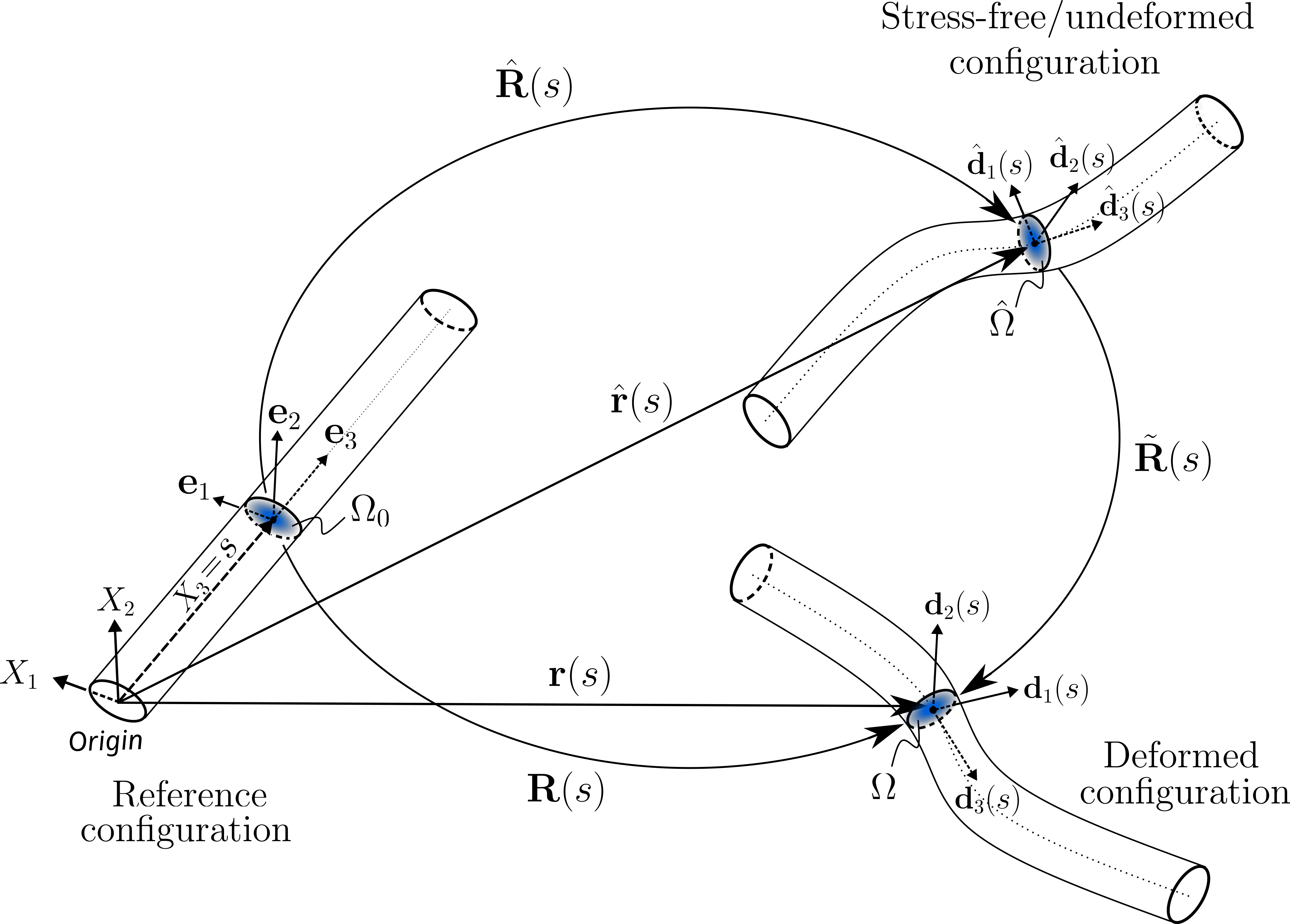}
    \caption{Kinematics of a special Cosserat rod}
    \label{fig:ge_rod_kinematics}
\end{figure}
In this section, we briefly review the theory of geometrically exact special Cosserat rods, adopting the notation and conventions presented by \cite{vinayak2026homogenizationrodlikemetamaterialsspecial}. A special Cosserat rod is described by a centerline curve together with cross-section attached to every point on this curve. The orientation of the cross-section is given by a triad of orthonormal directors as shown in Figure \ref{fig:ge_rod_kinematics}. The reference centerline of the rod is represented by the straight line $s\textbf{e}_3$ where $s\in[0,l]$ and the orthonormal director triads coincide with the global basis in the reference state. In the stress-free configuration, the centreline is given by $\hat{\textbf{r}}(s)$ whereas the director triads are given by $(\hat{\textbf{d}}_{1}(s),\hat{\textbf{d}}_{2}(s),\hat{\textbf{d}}_{3}(s))$. 
The associated rotation tensor in the stress-free configuration is $\hat{\textbf{R}}(s)$ such that
\begin{align}\label{eq:stress_free_directors2}
    \hat{\textbf{d}}_i(s) = \hat{\textbf{R}}(s)\textbf{e}_i.
\end{align}
The deformed configuration is given by the centreline curve $\textbf{r}(s)$ and the director triads $({\textbf{d}}_{1}(s),{\textbf{d}}_{2}(s),{\textbf{d}}_{3}(s))$ such that
\begin{align}\label{eq:spatial_directors}
    \textbf{d}_i(s) = \textbf{R}(s)\textbf{e}_i.
\end{align}
In this work, we parametrize the rotation tensor using a vector $\boldsymbol{\theta}\in\mathbbl{R}^3$ whose direction coincides with the axis of rotation and whose magnitude equals the angle of rotation. Thus, the variables $\textbf{r}(s)$ and $\boldsymbol{\theta}(s)$ are the kinematic variables of this theory. The strain measures in this theory are
\begin{align}
    &\textbf{v}_0 = \textbf{R}^T\textbf{r}^{\prime} = \text{v}_i\textbf{e}_i,\label{eq:v0}\\
    &\textbf{k}_0 = \text{axial}(\textbf{K}_0)=\kappa_i\textbf{e}_i~ \text{where} ~\textbf{K}_0 = \textbf{R}^T\textbf{R}^{\prime}\label{eq:k0}.
\end{align}
Here ($\text{v}_1$, $\text{v}_2$) are the shear strains, $\text{v}_3$ is the axial stretch, ($\kappa_1$, $\kappa_2$) are the bending curvatures and $\kappa_3$ is the twist. The $axial$ function above returns the axial vector of a skew-symmetric tensor. The spatial counterparts of the above strain measures are
\begin{align}
    &\textbf{v} = \textbf{R} \textbf{v}_0 = \textbf{r}^{\prime} = \text{v}_i\textbf{d}_i,\label{eq:v_spatial}\\
    &\textbf{k} = \text{axial}(\textbf{K}) = \kappa_i\textbf{d}_i\quad \text{where} ~\textbf{K} = \textbf{R}\textbf{K}_0\textbf{R}^T =\textbf{R}^{\prime}\textbf{R}^T.\label{eq:k_spatial}
\end{align}
The internal contact force $\textbf{n}(s)$ and internal contact moment $\textbf{m}(s)$ acting on a cross-section of the rod in the current configuration are given by
\begin{align}\label{component}
\textbf{n}= \text{n}_i \textbf{d}_{i}, \quad\quad \textbf{m}=\text{m}_i\textbf{d}_{i}
\end{align}
while its rotational pull-backs are given by
\begin{align}\label{component1}
\textbf{n}_0= \text{n}_i \textbf{e}_{i},\quad\quad \textbf{m}_0= \text{m}_i \textbf{e}_{i},
\end{align}
respectively. Together, they are also called the rod's stress resultants. The linear and angular momentum balance equations of the rod assuming statics are given by
\begin{align}\label{global_eq}
\textbf{n}^{\prime} + \hat{\textbf{n}}&=\boldsymbol{0},\nonumber\\
\textbf{m}^{\prime}+\text{\textbf{v}}\times\textbf{n} +\hat{\textbf{m}}&=\boldsymbol{0}.
\end{align}
Here $\hat{\textbf{n}}$ and $\hat{\textbf{m}}$ are the distributed force and couple, respectively, that act on the rod. One further assumes the existence of a scalar-valued function $\phi(\textbf{v}_0,\textbf{k}_0)$ denoting stored energy per unit undeformed length such that
\begin{align}
    \textbf{n}_0 = \frac{\partial \phi}{\partial \textbf{v}_0},~~~~~~~~\textbf{m}_0 = \frac{\partial \phi}{\partial \textbf{k}_0},~~~~~~~~~\mathbbl{C}_0^{rod}=\frac{\partial^2\phi}{\partial[\textbf{v}_0,\textbf{k}_0]\partial[\textbf{v}_0,\textbf{k}_0]}.
\end{align}
Here $\mathbb{C}_0^{{rod}}$ is the rod's elasticity tensor whose diagonal components are the shearing, stretching, bending and twisting stiffnesses whereas the off-diagonal components are the various coupling stiffnesses. It assumes a diagonal form for isotropic rods. Often, the following quadratic energy model is assumed with $\mathbb{C}_0^{{rod}}$ being a constant tensor: 
\begin{align}\label{eq:rod_energy_function_quadratic}
    \phi(\textbf{v}_0,\textbf{k}_0) = \frac{1}{2}\begin{bmatrix}
        \textbf{v}_0 - \hat{\textbf{v}}_0 \\
        \textbf{k}_0 - \hat{\textbf{k}}_0
    \end{bmatrix}^T\mathbbl{C}_0^{rod}
    \begin{bmatrix}
        \textbf{v}_0 - \hat{\textbf{v}}_0 \\
        \textbf{k}_0 - \hat{\textbf{k}}_0
    \end{bmatrix}.
\end{align}
Here $\hat{\textbf{v}}_0$ and $\hat{\textbf{k}}_0$ are the strain measures in the rod's stress-free/natural configuration. The quadratic model works for small departures of strain from its natural-state value. An unshearable and inextensible Kirchhoff rod can be incorporated using an additional constraint:
\begin{align}\label{eq:kirchhoff_constraint}
    \textbf{v}_0 = \textbf{e}_3.
\end{align}
We also use the following notation for the position of different nodes of a rod in the RVE:
\begin{align}
    \textbf{r}^{\alpha}_{k} &= \textbf{r}^{\alpha}(s^{\alpha}_k)\quad \text{for the $k^{th}$ internal node located at $s_k\in(0,L^{\alpha})$}\text{ in the $\alpha^{th}$ rod}.
\end{align}
For the rod's two end-points, we use the following special notation:
\begin{align}
    \textbf{r}^{\alpha}_{\beta} =     \begin{cases}
        \textbf{r}^{\alpha}_{+} & \text{for } \textbf{r}^{\alpha}(L^{\alpha})\\
        \textbf{r}^{\alpha}_{-} & \text{for } \textbf{r}^{\alpha}(0)\\
    \end{cases}.
\end{align}
\section{Three-dimensional finite strain homogenization problem}\label{sec:3d_homogenization_theory}
\begin{figure}[h!]
    \centering
    \includegraphics[width=0.8\linewidth]{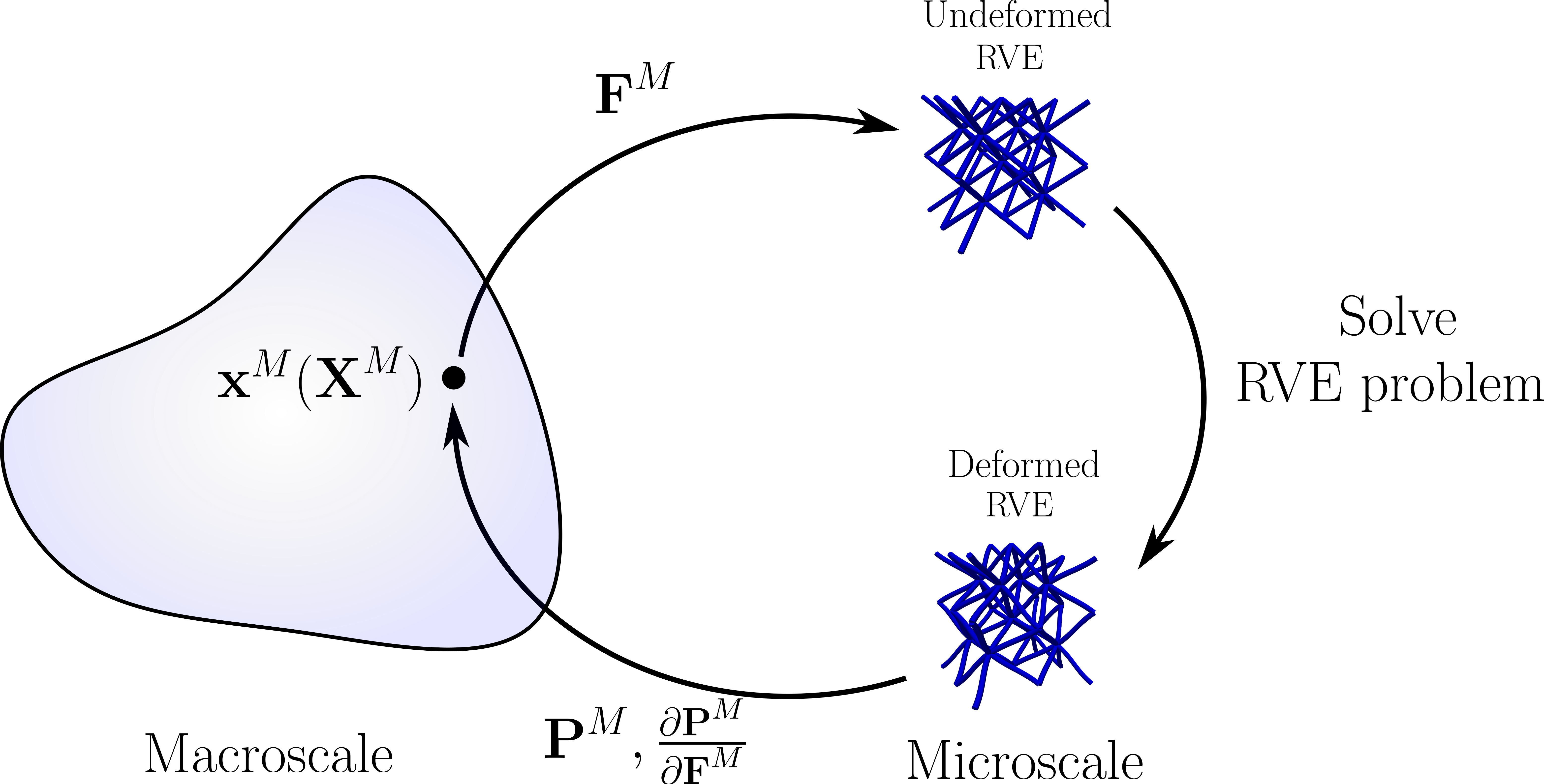}
    \caption{A schematic of a homogenization problem.}
    \label{fig:3d_homogenization_schematic}
\end{figure}
\begin{figure}[h!]
    \centering
    \includegraphics[width=0.8\linewidth]{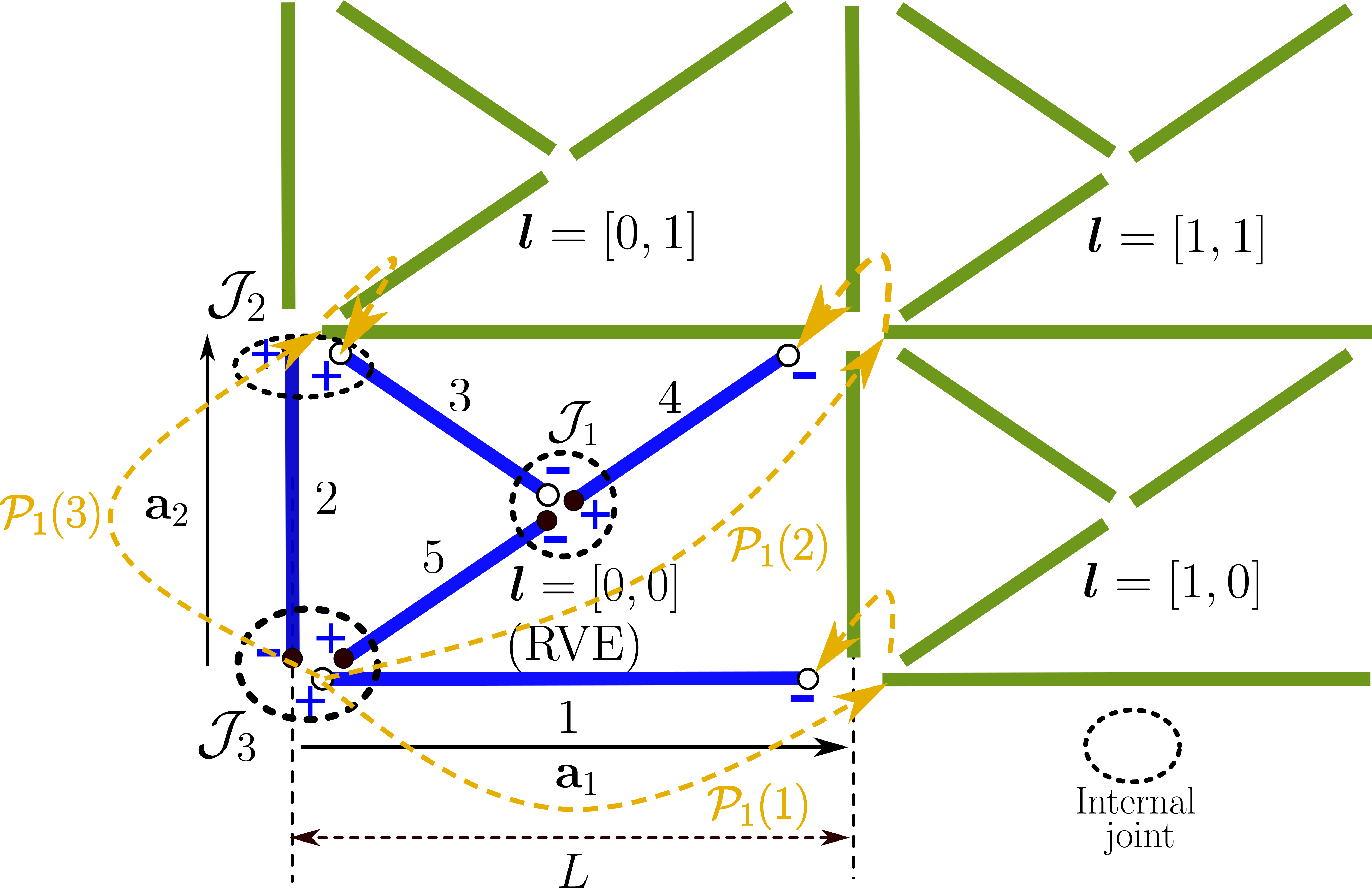}
    \caption{A schematic showing a generalized two dimensional unit cell or RVE (blue) and its neighbouring images (green). The internal joints of the RVE, $\mathcal{J}_i$, are represented in enclosed dashed ovals. The black nodes are the slave nodes whereas white ones are the master nodes.}
    \label{fig:1d_3dh:lattice_rve}
\end{figure}

 In homogenization, a heterogenous body is replaced by a fictitious homogeneous body whose material properties are unknown. It is the determination of these homogenized material properties that forms the homogenization or RVE problem. Figure \ref{fig:3d_homogenization_schematic} shows the deformed configuration of such a three dimensional homogeneous medium given by the map $\textbf{x}^M(\textbf{X}^M)$ where $\textbf{X}^M$ is the position of a material point in the undeformed configuration. Here superscript $(\cdot)^M$ implies the quantity to be macroscopic. The macroscopic deformation gradient is then given by \begin{align}
    \textbf{F}^M(\textbf{X}^M) = \frac{\partial \textbf{x}^M}{\partial \textbf{X}^M}.
\end{align}
In order to obtain the material properties of the fictitious homogeneous domain, an RVE of the microstructure is identified. Information from the macroscale is passed to the microscale (macro-to-micro transition), an RVE problem is solved and macroscopic  conjugates are obtained (micro-to-macro transition). This information of the micro-macro transitions is non-trivial. It depends on the continuum model chosen to represent the fictitious homogenized macroscopic domain. In this work, we restrict ourselves to first order macroscopic model whose energy density is local, i.e.,
\begin{align}\label{eq:3dh1:energy_density_macro_model}
    W^M = W(\textbf{F}^M).
\end{align}
The microscale fibers are modeled as geometrically exact rods as described in Section \ref{geometrically exact rod theory}. Finally, with the chosen microscale and macroscale models, the homogenization problem is given by
\begin{align}\label{eq:3dh1:energy_density_macro}
W^M =\frac{1}{\hat{V}}\min_{\{\textbf{r}^{\alpha},\boldsymbol{\theta}^{\alpha}\}} \mathcal{E}^{RVE}
\end{align}
where
\begin{align}
\mathcal{E}^{RVE} = \sum_{\alpha=1}^{N_{rods}}\int_0^{L_{\alpha}}\Phi^{\alpha} ds^{\alpha}    
\end{align}
is the total energy of the RVE which in this case is the sum of energy of each rod in the RVE and $\hat{V}$ is the volume of the RVE in the stress-free configuration. The first Piola-Kirchhoff stress tensor is then given by
\begin{align}
    \textbf{P}^M = \frac{\partial W^M}{\partial \textbf{F}^M}.
\end{align}
One can also formulate the homogenization problem in a \textit{stress-driven} sense in which the first Piola-Kirchhoff stress $\textbf{P}^M$ is prescribed from the macroscale to the microscale. In this approach, the macroscopic deformation gradient becomes an unknown. This formulation is useful when one needs to model deformation scenarios such as uniaxial tension where the stress free lateral surface condition needs to hold. For now we focus on the strain-driven approach and will present the stress-driven case later. The energy minimization problem \eqref{eq:3dh1:energy_density_macro} has to be solved in the presence of joint constraints, periodic boundary conditions and global constraints that restrict rigid body modes. Figure \ref{fig:1d_3dh:lattice_rve} shows a schematic of a periodic two-dimensional rod network with internal joints and periodic boundary condition constraints.
\subsection{Internal joints in the RVE}
 A biopolymer network consists of cross-links that are usually modeled as joints between multiple fibers. The nature of these joints is an important consideration \citep{wagner2006cytoskeletal,picu2011mechanics}. They can be classified according to the relative rotation that is allowed between the cross-linked fibers. Let us assume, for now, that the joints between the rods are welded or rigid, i.e., no relative displacement or rotation is allowed. Such joints are appropriate for semi-flexible fibers as they allow transfer of moment across connected fibers. Here too, we follow the approach presented in \citet{vinayak2026homogenizationrodlikemetamaterialsspecial}. Consider that the RVE is a network of rods consisting of $N_J$ internal joints. By \textit{internal}, we imply that all the rods participating in a joint are internal to the RVE. The internal joints of a representative RVE are shown in Figure \ref{fig:1d_3dh:lattice_rve}. We use the master-slave approach to formulate the internal joint constraints \citep{jelenic1996non,ibrahimbegovic2000rigid}. For the $i^{th}$ joint in the RVE, we assume it to be formed by $J_i$ number of nodes each belonging to different rods that join there and denote this set of nodes by $\mathcal{J}_i$. It is defined as follows:
 \begin{align}
     \mathcal{J}_i = \{(\alpha,\beta):\textbf{r}_{\beta}^{\alpha}\text{ lies at the $i^{th}$ joint\}},~ \vert\mathcal{J}_i\vert = J_i \quad\forall~ i=1: N_J
 \end{align}
Here $\beta$ takes either `+' or `-' value. Each set $\mathcal{J}_i$ is an ordered set such that its first element $\mathcal{J}_i(1)=(\alpha_1,\beta_1)$ is the master node and all the other elements $\mathcal{J}_i(k)=(\alpha_k,\beta_k)$ where $k\in[2,J_i]$ are slave nodes. Once internal joints are defined, for an $i^{th}$ joint, we have the following translational constraint relating the positions of master and slave nodes:
 \begin{align}\label{eq:internal_jt_trans_constraint}
     \boldsymbol{\mathcal{J}}^{trans}_{i,k} \equiv \textbf{r}_{\beta_1}^{{\alpha}_1} - \textbf{r}_{\beta_k}^{\alpha_k} = \textbf{0} \quad \forall ~ k\in [2,J_i] ~\text{where}~(\alpha_k,\beta_k) = \mathcal{J}_i(k).
 \end{align}
Likewise, the relative angle between the directors of master and slave nodes is constrained by the following equation:
\begin{align}\label{eq:internal_joint_constraint_directors}
    \textbf{d}^{\alpha_1}_{\beta_1,p}\cdot\textbf{d}^{\alpha_k}_{\beta_k,q} = \hat{\textbf{d}}^{\alpha_1}_{\beta_1,p}\cdot\hat{\textbf{d}}^{\alpha_k}_{\beta_k,q}  \quad \forall~ k\in [2,J_i] ~ \text{and} ~ p,q\in \{1,2,3\}.
\end{align}
This results in the following rotational constraint for the joint:
\begin{equation}\label{eq:internal_jt_rotation_constraint2}
     \boldsymbol{\mathcal{J}}^{rot}_{i,k} \equiv axial\left(\log\left(\left((\textbf{R}^{\alpha_{k}}_{\beta_{k}})^T~\textbf{R}^{\alpha_1}_{\beta_1}\right)\left((\hat{\textbf{R}}^{\alpha_k}_{\beta_k})^T~\hat{\textbf{R}}^{\alpha_1}_{\beta_1}\right)^T\right)\right)=\textbf{0}\quad \forall ~ k\in [2,J_i].
\end{equation}
\begin{figure}[h!]
    \centering
        \includegraphics[width=.55\linewidth]{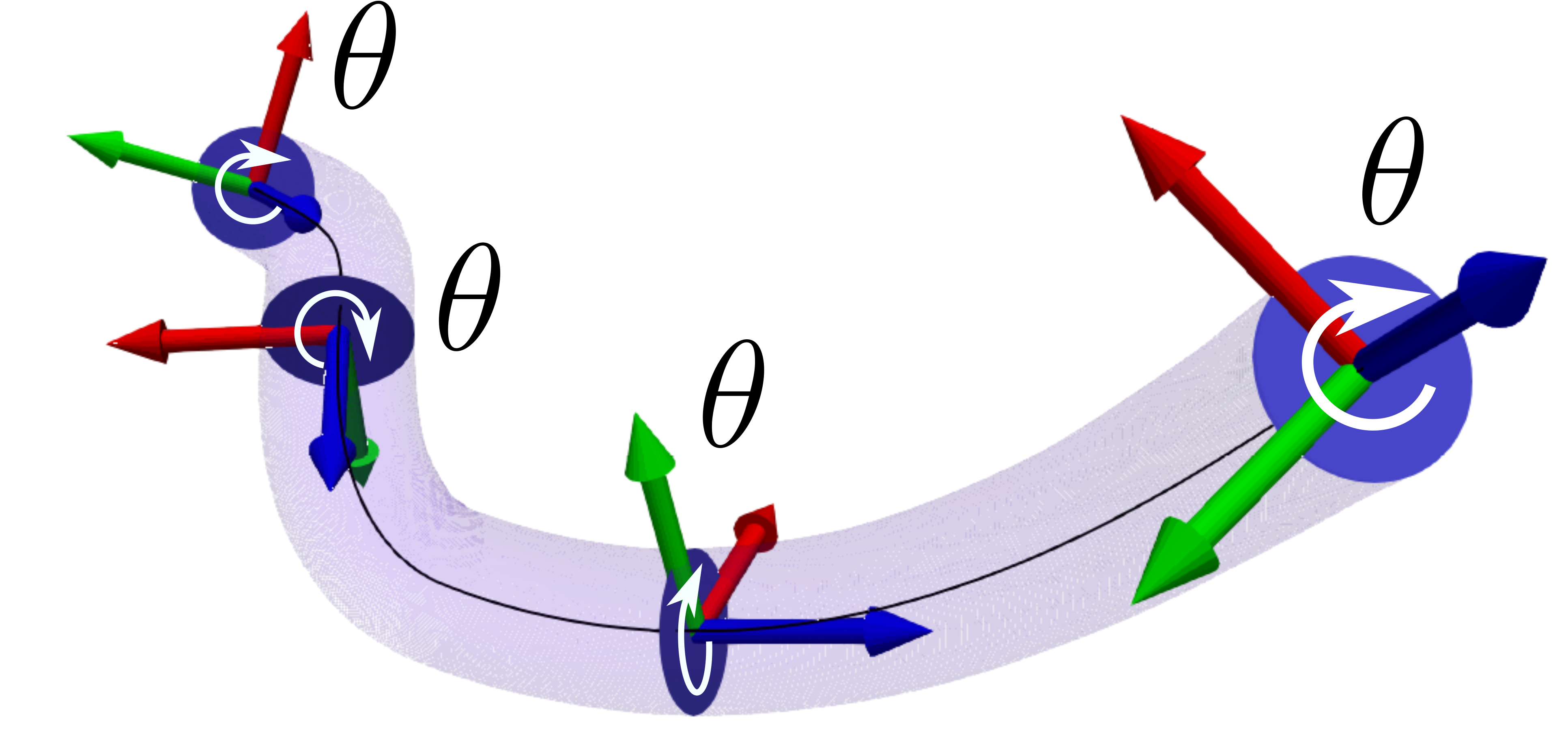}
    \caption{Free rotation of individual isotropic rods: every cross-section can be freely rotated by the same angle $\theta$ but about their own normal}
    \label{fig:ball-and-socket_joint}
\end{figure}
In case we have ball-and-socket joints, only the relative displacement between the nodes
at the joint is constrained, the relative rotation is free to take any value, i.e. \cref{eq:internal_jt_rotation_constraint2} is no longer required. However, for isotropic rods, one needs an additional constraint to fix the energetically free uniform rotation of every rod's cross-section about their local cross-section normal - we are not talking about rigid rotation here since the local axis of rotation is changing along the rod's arc-length. This is achieved by applying the following constraint on the cross-sectional director at any one of the nodes of a rod:
 \begin{align}
    \textbf{d}_{\beta,2}^{\alpha}\cdot \textbf{p} = 0
 \end{align}
where the vector $\textbf{p}$ could be a fixed global direction. For example, a possible choice is $\textbf{p} = \textbf{e}_3$. This constraint ensures that $\textbf{d}_2$ at that node lies along the line formed by the intersection of cross-sectional plane spanned by $(\textbf{d}_1-\textbf{d}_2)$ and the one spanned by $(\textbf{e}_1-\textbf{e}_2)$. We note that rigid joints inhibit floppy modes in a network whereas ball-and-socket joints lead to floppy modes unless an appropriate network topology is maintained \citep{picu2011mechanics}. 
\subsection{Periodic boundary conditions}
The periodic lattice of rods can be viewed as a multi-lattice crystal structure. Each point (basis node) of the rod can be identified to be associated with a simple lattice defined by lattice vectors $(\hat{\textbf{a}}_1,\hat{\textbf{a}}_2,\hat{\textbf{a}}_3)$\footnote{These lattice vectors need not be ``primitive", i.e., they can have more than one lattice site.} in the reference state. The lattice vectors form an RVE containing the rod network which when repeated in all three directions forms an infinite rod network. The undeformed centerline position of any rod in this network is given by
\begin{align}\label{eq:image_rod_position_stress_free}
    \hat{\textbf{r}}^{\boldsymbol{l},\alpha} = \hat{\textbf{H}}\boldsymbol{l} + \hat{\textbf{r}}^{\alpha}, \quad \boldsymbol{l} \in \mathbb{Z}^3\quad \text{and} \quad \alpha = 1,...,N_{rods}
\end{align}
where
\begin{align}
    \hat{\textbf{H}} = \begin{bmatrix}
        \hat{\textbf{a}}_1 & \hat{\textbf{a}}_2 & \hat{\textbf{a}}_3
    \end{bmatrix},
\end{align}
  $\hat{\textbf{r}}^\alpha$ is the centreline position of the $\alpha^{th}$ ``basis" rod belonging to the RVE and
 $\hat{\textbf{r}}^{\boldsymbol{l},\alpha}$ is the centreline position of the image of $\alpha^{th}$ rod
  in the $\boldsymbol{l}^{th}$ periodic box or image. The undeformed volume of the RVE is given by
\begin{align}
    \hat{V} = \text{det}(\hat{\textbf{H}}) = \hat{\textbf{a}}_1\cdot(\hat{\textbf{a}}_2 \times \hat{\textbf{a}}_3).
\end{align}
  Now, a macroscale deformation gradient $\textbf{F}^M$ is assumed to be imposed from the macroscale on the microscale. The position of microscale rods in the presence of $\textbf{F}^M$ is then given by the Cauchy-Born rule as follows: 
\begin{align}
    \textbf{r}^{\boldsymbol{l},\alpha}(s^{\alpha}) = \textbf{F}^M\hat{\textbf{r}}^{\boldsymbol{l},\alpha}(s^{\alpha}) + \textbf{w}^{\alpha}(s^{\alpha})
\end{align}
where the function $\textbf{w}^{\alpha}$ is the internal shift of the $\alpha^{th}$ rod's centreline in the RVE. The above map is \textit{quasi-uniform}\footnote{A quasiuniform deformation is the one in which internal shifts of the multilattice system are admitted \citep{tadmor2011modeling}, i.e., $\textbf{w}^{\alpha}\neq  \textbf{0}$ for all $\alpha \in [1,N_{rods}]$.} and thus appropriate to capture bending of microscale rods \citep{10.1115/1.2165699,picu2011mechanics}. Substituting \cref{eq:image_rod_position_stress_free} in the above equation, we get
\begin{align}\label{eq:periodic_node_pos_3dh}
    \textbf{r}^{\boldsymbol{l},\alpha} = \textbf{H}\boldsymbol{l} + \textbf{r}^{\alpha}
\end{align}
where
\begin{align}\label{eq:deformed_lattice_vectors_rod_pos}
    \textbf{H}= \begin{bmatrix}
        \textbf{a}_1 & \textbf{a}_2 & \textbf{a}_3
    \end{bmatrix} = \textbf{F}^M\hat{\textbf{H}}, \quad \quad \textbf{r}^{\alpha} = \textbf{F}^M\hat{\textbf{r}}^{\alpha} + \textbf{w}^{\alpha},
\end{align}
$(\textbf{a}_1,\textbf{a}_2,\textbf{a}_3 )$ are the deformed lattice vectors and the volume of the deformed RVE box is given by
\begin{align}
    V = \text{det}(\textbf{H}) = \textbf{a}_1\cdot(\textbf{a}_2\times \textbf{a}_3).
\end{align}
Next, assuming the periodicity of director vector field corresponding to each rod, we have
\begin{align}\label{eq:periodic_node_directors_3dh_stress_free}
    \hat{\textbf{d}}^{\boldsymbol{l},\alpha}_i = \hat{\textbf{d}}^{\alpha}_i \quad \quad \forall \quad \boldsymbol{l} \in \mathbb{Z}^3\quad \text{and} \quad \alpha = 1,...,N_{rods}
\end{align} 
for each $i=\{1,2,3\}$. In this work, the directors are assumed not depend explicitly on macroscale deformation gradient since we have assumed a standard Cauchy-continuum type macroscale model (see \cref{eq:3dh1:energy_density_macro_model}).\footnote{In higher order theories such as a Cosserat continuum, one would have to impose a macroscopic rotation on microscopic directors as well \citep{glaesener2019continuum,vinayak2026homogenizationrodlikemetamaterialsspecial}.} The micro-macro connection only exists between the centerline positions of the rods and the deformation gradient. Therefore, the microscale directors simply obey the following periodicity relationship:
\begin{align}\label{eq:periodic_node_directors_3dh_def}
    \textbf{d}^{\boldsymbol{l},\alpha}_i = \textbf{d}^{\alpha}_i \quad \quad \forall \quad \boldsymbol{l} \in \mathbb{Z}^3\quad \text{and} \quad \alpha = 1,...,N_{rods}.
\end{align}
Finally, \cref{eq:periodic_node_pos_3dh,eq:periodic_node_directors_3dh_def} establish the periodicity of microscale unknowns ($\textbf{r},\boldsymbol{\theta}$) across the infinite lattice. Using this periodicity, we now derive the periodic boundary conditions so that the infinite rod network problem is reduced to just an RVE. We follow the approach of \cite{vigliotti2014non} wherein the periodicity is assumed and the periodic boundary conditions are applied to the kinematic unknowns. This is in contrast with the Hill-Mandel approach in which periodic boundary conditions are derived from macro-micro averaging rules \citep{mandadapu2012homogenization,saeb2016aspects}. 

Without loss of generality, let us again consider the two-dimensional lattice  in figure \ref{fig:1d_3dh:lattice_rve}. The rod's end points that lie on the boundary of the RVE are called the boundary nodes and are further split into independent and dependent boundary nodes. An internal joint (as defined earlier) is also allowed to be on the boundary. In that case, only the master node of the joint is considered as the boundary node of the RVE. Let us denote the number of independent boundary nodes by $N_P$. Each independent boundary node can be related to multiple dependent boundary nodes through periodicity rule and joint connection.\footnote{We assume that the rod network is fully connected, i.e., there are no dangling rods without a joint connection.} For example, the RVE in Figure \ref{fig:1d_3dh:lattice_rve} consists of four corner boundary nodes. Out of those only one is independent and the rest are dependent. Let us now construct a set $\mathcal{P}_i$ consisting of the $i^{th}$ independent boundary node of the RVE and its dependent nodes as follows:
\begin{align}
    \mathcal{P}_i &= \{(\alpha,\beta): \textbf{r}^{\alpha}_{\beta} \text{ lies in the periodic set of $i^{th}$ independent boundary node}\}\nonumber\\
    &, |\mathcal{P}_i| = P_i \quad \forall i = 1,...,N_{P}
\end{align}
Here $\beta$ takes either `$+$' or `$-$' value. Each set $\mathcal{P}_i$ is an ordered set such that its first
element $\mathcal{P}_i(1) = (\alpha_{1}, \beta_{1})$ is the $i^{th}$ independent boundary node and all the other elements $\mathcal{P}_i(k) = (\alpha_{k}, \beta_{k})$
where $k \in [2, P_i]$ are the dependent boundary nodes of this independent boundary node. We can relate each independent node with its dependent nodes. First, using \eqref{eq:periodic_node_pos_3dh}, the neighbouring images of the independent nodes are related to the independent nodes through
\begin{align}\label{eq:periodic_node_pos_3dh_image}
    \textbf{r}^{\boldsymbol{l}_{i,k},~\alpha_{1}}_{\beta_{1}} = \textbf{r}^{\alpha_{1}}_{\beta_{1}} + \textbf{H}\boldsymbol{l}_{i,k}, \quad \quad (\boldsymbol{l}_{i,k})_j \in \{0,1\},\quad \boldsymbol{l}_{i,k}\neq \textbf{0} \quad \text{and} \quad (\alpha_1,\beta_1) = \mathcal{P}_i(1) ~\forall~i=1,...,N_P.
\end{align}
In the above equation, the quantity $\boldsymbol{l}_{i,k}$ is the id of the neighbouring image RVE on the boundary of which the $k^{th}$ dependent node in $\mathcal{P}_i$ lies. Since the RVE and its neighbouring images share their boundary, we have
\begin{align}\label{eq:independent_node_dependent_node_pos_connection}
   \textbf{r}^{\boldsymbol{l}_{i,k},~\alpha_{1}}_{\beta_{1}} = \textbf{r}^{\alpha_{k}}_{\beta_{k}}.
\end{align}
Using \cref{eq:periodic_node_pos_3dh_image,eq:independent_node_dependent_node_pos_connection}, the periodic boundary condition constraint is thus given by
\begin{align}\label{eq:periodic_jt_pos_constraint}
 \boldsymbol{\mathcal{P}}^{trans}_{i,k} \equiv \textbf{r}^{\alpha_{k}}_{\beta_{k}} - \textbf{r}^{\alpha_{1}}_{\beta_{1}} - \textbf{H}\boldsymbol{l}_{i,k} = \textbf{0} \quad \quad \forall \quad k\in[2,P_i] \quad \text{and} \quad (\alpha_1,\beta_1) = \mathcal{P}_i(1)~\forall~i=1,...,N_P.
\end{align}
Next we consider the periodic constraint for the director vectors. Using \cref{eq:periodic_node_directors_3dh_stress_free,eq:periodic_node_directors_3dh_def}, we have
\begin{align}\label{eq:director_image_rve}
    \hat{\textbf{d}}^{\boldsymbol{l}_{i,k},~\alpha_1}_{\beta_1,j} = \hat{\textbf{d}}^{\alpha_1}_{\beta_1,j},\quad \quad \text{and}\quad\quad \textbf{d}^{\boldsymbol{l}_{i,k},~\alpha_1}_{\beta_1,j} = \textbf{d}^{\alpha_1}_{\beta_1,j}, \quad \quad &(\boldsymbol{l}_{i,k})_j \in \{0,1\},\quad \boldsymbol{l}_{i,k}\neq \textbf{0} \quad \text{and} \quad \nonumber\\
    &(\alpha_1,\beta_1) = \mathcal{P}_i(1) ~\forall~i=1,...,N_P.
\end{align}
and with the rigid joint assumption, using \cref{eq:internal_joint_constraint_directors}, we can write
\begin{align}\label{eq:director_image_rve_joint}
    \textbf{d}^{\alpha_k}_{\beta_k,p}\cdot\textbf{d}^{\boldsymbol{l}_{i,k},\alpha_1}_{\beta_1,q} 
    =  \hat{\textbf{d}}^{\alpha_k}_{\beta_k,p}\cdot\hat{\textbf{d}}^{\boldsymbol{l}_{i,k},\alpha_1}_{\beta_1,q}
\end{align}
Thus, using \cref{eq:director_image_rve,eq:director_image_rve_joint}, the director triads of the independent and dependent boundary nodes obey the following constraint:
\begin{align}
    \textbf{d}^{\alpha_k}_{\beta_k,p}\cdot\textbf{d}^{\alpha_1}_{\beta_1,q} 
    =  \hat{\textbf{d}}^{\alpha_k}_{\beta_k,p}\cdot\hat{\textbf{d}}^{\alpha_1}_{\beta_1,q}\quad \quad \forall \quad k\in[2,P_i] \quad \text{and} \quad (\alpha_1,\beta_1) = \mathcal{P}_i(1)~\forall~i=1,...,N_P.
\end{align}
or
\begin{equation}\label{eq:periodic_jt_rotation_constraint2}
     \boldsymbol{\mathcal{P}}^{rot}_{i,k} \equiv axial\left(\log\left(\left((\textbf{R}^{\alpha_{k}}_{\beta_{k}})^T~\textbf{R}^{\alpha_1}_{\beta_1}\right)\left((\hat{\textbf{R}}^{\alpha_k}_{\beta_k})^T~\hat{\textbf{R}}^{\alpha_1}_{\beta_1}\right)^T\right)\right)=\textbf{0}.
\end{equation}
The constraint \cref{eq:periodic_jt_pos_constraint,eq:periodic_jt_rotation_constraint2} together are called the periodic boundary condition constraints. They ensure that the deformation of the RVE is locally periodic under quasiuniform deformation. Note that, in case of ball-and-socket joints, the constraint on periodicity of director vectors will not be present.
\subsection{Global constraints}
For solving the minimization problem \eqref{eq:3dh1:energy_density_macro}, the rigid body translation and rotation of the RVE needs to be restricted.
In case of strain-driven homogenization, one only needs to restrict rigid translation because
the rigid rotation is automatically prevented by periodic boundary condition constraints.
However, in case of stress-driven homogenization which we present later, both need to be constrained independently. \\\\
The rigid body translation can be constrained by fixing $\textbf{w}^{\alpha}$ at any one of the independent nodes or by fixing the geometric center of the RVE at the origin. For fixing the geometeric center, it is sufficient to only consider the nodes lying on the 
 boundary of the RVE described by a set $\mathcal{G}$. The rigid translation constraint is then given by 
\begin{align}\label{eq:constraint_mass_centre_continuum}
    \boldsymbol{\mathcal{G}}^{trans} \equiv \sum_{\substack{k=1\\(\alpha_k,\beta_k)=\mathcal{G}(k)}}^{N_G} \textbf{r}^{\alpha_{k}}_{\beta_{k}} = \textbf{0}.
 \end{align}
where $N_G = |\mathcal{G}|$. The rigid rotation of the RVE can be restricted by fixing the principle axes of faces of the box enclosing the RVE. This is done by setting the mixed moment of area of the faces to zero, i.e., 
\begin{align}\label{eq:mixed_moment_gc}
    \boldsymbol{\mathcal{G}}^{rot} \equiv \sum_{\substack{k=1\\(\alpha_k,\beta_k)=\mathcal{G}(k)}}^{N_G}\boldsymbol{\mathfrak{m}}^{\alpha_k}_{\beta_k}=\textbf{0}
\end{align}
where
\begin{align}
	\boldsymbol{\mathfrak{m}}^{\alpha_k}_{\beta_k} =  r^{\alpha_k}_{\beta_k,2}~r^{\alpha_k}_{\beta_k,3}\textbf{e}_1 + r^{\alpha_k}_{\beta_k,1}~r^{\alpha_k}_{\beta_k,3}\textbf{e}_2 + r^{\alpha_k}_{\beta_k,1}~r^{\alpha_k}_{\beta_k,2}\textbf{e}_3. 
\end{align}
 For symmetric RVE box faces (having normal along $\textbf{e}_i$ axis), one needs to modify the $i^{th}$ constraint in \eqref{eq:mixed_moment_gc}. For example, in case of symmetric box faces whose normal lie along $\textbf{e}_3$ direction,
\begin{align}
   \mathfrak{m}^{\alpha_k}_{\beta_k,3} =
	\bigg[\arctan\bigg(\frac{r^{\alpha_k}_{\beta_k,2}}{r^{\alpha_k}_{\beta_k,1}}\bigg)-\arctan\bigg(\frac{\hat{r}^{\alpha_k}_{\beta_k,2}}{\hat{r}^{\alpha_k}_{\beta_k,1}}\bigg)\bigg].
\end{align}
 Note that the positional coordinates of nodes in $\mathcal{G}$ are used to preserve the orientation of the RVE and not the directors. 
 \begin{remark}
The rigid rotation in the stress-driven case is also avoided by restricting $\textbf{F}^M$ to be $\textbf{U}^M$ where $\textbf{U}^M$ is the right stretch tensor. In this case, the stress drivers are taken to be the macroscopic Biot stress \citep{elliott2011reversible} or the second Piola-Kirchhoff stress tensor \citep{van2016formulation,kumar2025second}. However, the first Piola-Kirchhoff stress tensor (or engineering stress) is a natural description in experiments. 
\end{remark}
\subsection{RVE energy minimization problem}
In this section, we first derive the weak form for strain-driven homogenization of a rigidly connected RVE. With all the constraints defined in the previous sections, the constrained energy functional for the minimization problem in  \cref{eq:3dh1:energy_density_macro} is formulated as
\begin{align}
    \mathcal{E}^{cons,strain-driven} = \mathcal{E}^{RVE} + \mathcal{C}^G+ \mathcal{C}^P + \mathcal{C}^J 
\end{align}
where the constraint terms are as follows:
\begin{align}\label{eq:strain_driven_constraints}
    \mathcal{C}^G &=
        \boldsymbol{\lambda} \cdot
        \boldsymbol{\mathcal{G}}^{trans},\quad \quad
    \mathcal{C}^P = \sum_{i=1}^{N_{P}}\sum_{k=2}^{P_i}
    \begin{bmatrix}
        \textbf{p}^n_{i,k} \\
        \textbf{p}^m_{i,k}
    \end{bmatrix}\cdot
    \begin{bmatrix}
        \boldsymbol{\mathcal{P}}^{trans}_{i,k} \\
        \boldsymbol{\mathcal{P}}^{rot}_{i,k}
    \end{bmatrix}, \nonumber\\
    \mathcal{C}^J &=\sum_{i=1}^{N_{J}}\sum_{k=2}^{J_i} \begin{bmatrix}
        \textbf{J}^n_{i,k} \\
        \textbf{J}^m_{i,k}
    \end{bmatrix}\cdot
    \begin{bmatrix}
        \boldsymbol{\mathcal{J}}^{trans}_{i,k} \\
        \boldsymbol{\mathcal{J}}^{rot}_{i,k}
    \end{bmatrix}.
\end{align}
Here $(\boldsymbol{\lambda},\boldsymbol{\mu}), (\textbf{J}^n_{i,k},\textbf{J}^m_{i,k}), (\textbf{p}^n_{i,k},\textbf{p}^m_{i,k})$ are the Lagrange multipliers
enforcing the global, internal joint and periodic boundary condition constraints. Since this is a strain-driven case, we have global constraint only for rigid translation. The first variation of the above functional is obtained by varying the unknowns as follows:
\begin{align}\label{eq:perturb_config}
    \textbf{r}^{\epsilon}(s) &= \textbf{r}(s) + \epsilon \delta \textbf{r}(s),\quad \quad \textbf{R}_{\epsilon}(s) = e^{\epsilon\delta\boldsymbol{\theta}(s)}\textbf{R}(s)\nonumber\\
    \boldsymbol{\lambda}^{\epsilon} &= \boldsymbol{\lambda} + \epsilon \delta \boldsymbol{\lambda},\nonumber\\
    \textbf{p}^{n,\epsilon}_{i,k} &= \textbf{p}^n_{i,k} + \epsilon \delta \textbf{p}^n_{i,k}, \quad \quad {\textbf{p}}^{m,\epsilon}_{i,k} = \textbf{p}^m_{i,k} + \epsilon \delta \textbf{p}^m_{i,k}\nonumber\\
    \textbf{J}^{n,\epsilon}_{i,k} &= \textbf{J}^n_{i,k} + \epsilon \delta \textbf{J}^n_{i,k}, \quad \quad \textbf{J}^{m,\epsilon}_{i,k} = \textbf{J}^m_{i,k} + \epsilon \delta \textbf{J}^m_{i,k}
\end{align}
The variation of the constrained energy functional or the weak form is obtained as follows:
\begin{align}\label{eq:weak_form_continuum}
    G &= \frac{d}{d\epsilon}\mathcal{E}^{cons,strain-driven,\epsilon}\bigg |_{\epsilon=0}= \frac{d\mathcal{E}^{RVE,\epsilon}}{d\epsilon}\bigg |_{\epsilon=0} + \frac{d\mathcal{C}^{G,\epsilon}}{d\epsilon}\bigg |_{\epsilon=0} + \frac{d\mathcal{C}^{P,\epsilon}}{d\epsilon}\bigg |_{\epsilon=0} + \frac{d\mathcal{C}^{J,\epsilon}}{d\epsilon}\bigg |_{\epsilon=0} \nonumber\\
    &= \sum_{i=0}^{N_{rods}}\int_0^{L^{\alpha}}(\textbf{n}^{\alpha}\cdot\delta(\textbf{r}^{\alpha})^{\prime} + 
                (\textbf{n}^{\alpha}\times(\textbf{r}^{\alpha})^{\prime})\cdot\delta\boldsymbol{\theta}^{\alpha} +\textbf{m}^{\alpha}\cdot(\delta\boldsymbol{\theta}^{\alpha})^{\prime})ds^{\alpha} \nonumber\\
                &+\frac{d\mathcal{C}^{G,\epsilon}}{d\epsilon}\bigg |_{\epsilon=0} + \frac{d\mathcal{C}^{P,\epsilon}}{d\epsilon}\bigg |_{\epsilon=0} + \frac{d\mathcal{C}^{J,\epsilon}}{d\epsilon}\bigg |_{\epsilon=0}
\end{align}
For the linearization of the constraint terms, we refer the reader to \cite{vinayak2026homogenizationrodlikemetamaterialsspecial}. At equilibrium, the following must hold:
\begin{align}
    \frac{d\mathcal{E}^{cons,strain-driven}}{d\epsilon}\bigg|_{\epsilon=0} = 0.
\end{align}
The above equation can be discretized using standard finite elements or any other discretization scheme yielding the following:
\begin{align}\label{eq:strain_driven_residual}
    \mathcal{R}^{strain-driven}\left(\{\textbf{r}^{\alpha},\boldsymbol{\theta}^{\alpha}\},\boldsymbol{\lambda},\textbf{p}^n,\textbf{p}^m,\textbf{J}^n,\textbf{J}^m\right) = \textbf{0}.
\end{align}
We use the discrete helical element rod theory presented in \cite{vinayak2026homogenizationrodlikemetamaterialsspecial} for discretization of rods and solve the above equation using the standard Newton-Raphson scheme augmented with arc-length continuation for path following. After solving, the solution of the RVE problem in \eqref{eq:3dh1:energy_density_macro} is obtained and the rod's energy density can then be written as
 \begin{align}\label{eq:macro_energy}
     W^M(\textbf{F}^M) = \frac{1}{\hat{V}} \tilde{\mathcal{E}}^{RVE}\left(\tilde{\textbf{r}}^{\alpha}(\textbf{F}^M),\tilde{\boldsymbol{\theta}}^{\alpha}(\textbf{F}^M)\right)
 \end{align}
where $\{\tilde{\textbf{r}}^{\alpha}(\textbf{F}^M),\tilde{\boldsymbol{\theta}}^{\alpha}(\textbf{F}^M)\}_{\alpha=1}^{N_{rods}}$ denotes the configuration of the RVE at equilibrium. As the constraints get automatically satisfied at equilibrium, one can also write 
\begin{align}\label{eq:macro_energy_constrained}
    W^M(\textbf{F}^M) = \frac{1}{\hat{V}}\tilde{\mathcal{E}}^{cons,strain-driven}(\textbf{F}^M;\{\tilde{\textbf{r}}^{\alpha}(\textbf{F}^M),\tilde{\boldsymbol{\theta}}^{\alpha}(\textbf{F}^M)\}).
\end{align}
Henceforth, the superscript $\tilde{(\cdot)}$ is omitted for brevity. The macroscopic first Piola-Kirchhoff stress tensor is given by
\begin{align}
    \textbf{P}^M &= \frac{\partial W^M}{\partial \textbf{F}^M} = \frac{1}{\hat{V}}\frac{\partial \mathcal{E}^{cons,strain-driven}}{\partial \textbf{F}^M}
\end{align}
In evaluating the partial derivative in the above expression, the implicit derivatives of $\mathcal{E}^{cons,strain-driven}$ vanish at equilibrium and only the explicit derivative, arising out of the periodic boundary condition constraint term, remains. This results in the following expression:
\begin{align}\label{eq:macro_P_analytic}
                \textbf{P}^M= -\frac{1}{\hat{V}}\sum_{i=1}^{N_P}\sum_{k=2}^{P_i}\textbf{p}^n_{i,k}\otimes\hat{\textbf{H}}\boldsymbol{l}_{i,k}.
\end{align}
The fourth order elasticity tensor is then given by
\begin{align}
    \mathbbl{K}^M = \frac{\partial \textbf{P}^M}{\partial \textbf{F}^M} = -\frac{1}{\hat{V}}\sum_{i=1}^{N_P}\sum_{k=2}^{P_i}\frac{\partial \textbf{p}^n_{i,k}}{\partial \textbf{F}^M}\otimes\hat{\textbf{H}}\boldsymbol{l}_{i,k}.
\end{align}
In the above formula for macroscopic stiffness tensor, the quantity $\frac{\partial \textbf{p}^n_{i,k}}{\partial \textbf{F}^M}$ is an unknown. It is obtained by taking the derivative of the weak form in \cref{eq:weak_form_continuum} with respect to $\textbf{F}^M$ and solving the linearized system of equations so obtained.

Next, we consider the stress-driven homogenization. In this case, the first Piola Kirchhoff stress tensor $\textbf{P}^M$ is prescribed from the macroscale to the microscale. The constrained energy functional in \cref{eq:3dh1:energy_density_macro} is modified as follows:
\begin{align}
    \mathcal{E}^{cons,stress-driven} = \mathcal{E}^{RVE} + \mathcal{C}^G+ \mathcal{C}^P + \mathcal{C}^J - \textbf{P}^M:\textbf{F}^M
\end{align}
The above energy functional needs to be minimized considering the macroscopic deformation gradient $\textbf{F}^M$ as an unknown too. This forms the ``stress-driven" homogenization problem. Here, the global constraint is augmented with the rotational constraint too, i.e., 
\begin{align}\label{eq:global_constraint_with_rotation}
    \mathcal{C}^G = \boldsymbol{\lambda}\cdot\boldsymbol{\mathcal{G}}^{trans} + \boldsymbol{\mu}\cdot\boldsymbol{\mathcal{G}}^{rot}
\end{align}
where $\boldsymbol{\mu}$ is the additional Lagrange multiplier. The first variation of the stress-driven constrained energy functional can now be derived by applying the following perturbations:
\begin{align}
    \textbf{F}^{M,\epsilon} &= \textbf{F}^M + \epsilon\delta\textbf{F}^M, \nonumber\\
    \boldsymbol{\mu}^{\epsilon} &= \boldsymbol{\mu} + \epsilon \delta\boldsymbol{\mu}
\end{align}
in addition to those in \cref{eq:perturb_config}. This results in the additional equations - namely  \cref{eq:mixed_moment_gc} corresponding to arbitrary $\delta\boldsymbol{\mu}$ and another one corresponding to arbitrary $\delta\mathbf{F}^M$ given by
\begin{align}\label{eq:stress_driven_first_pk}
  \textbf{P}^M +   \frac{1}{\hat{V}}\sum_{i=1}^{N_P}\sum_{k=2}^{P_i}\textbf{p}^n_{i,k}\otimes\hat{\textbf{H}}\boldsymbol{l}_{i,k} = \textbf{0}.
\end{align}
It is exactly the same as the micro-to-macro transition (see \cref{eq:macro_P_analytic}) of first Piola-Kirchhoff stress tensor in the strain-driven approach.

\section{Numerical examples}\label{sec:numerical_examples}
\begin{figure}[h!]
    \centering
    \begin{subfigure}{0.49\textwidth}
        \includegraphics[width=0.95\linewidth]{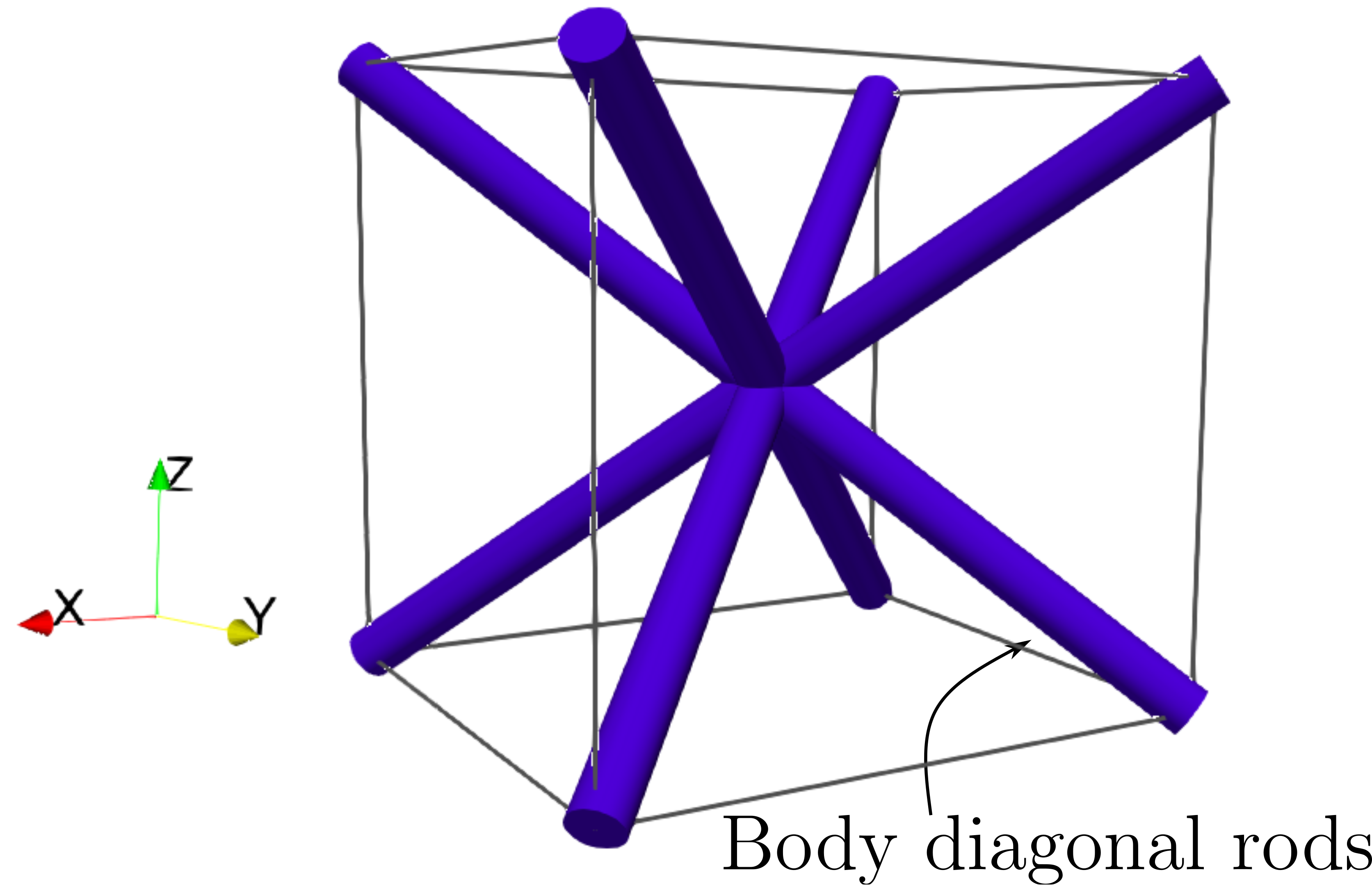}
            \caption{8-chain RVE}
    \end{subfigure}
        \begin{subfigure}{0.49\textwidth}
            \includegraphics[width=\linewidth]{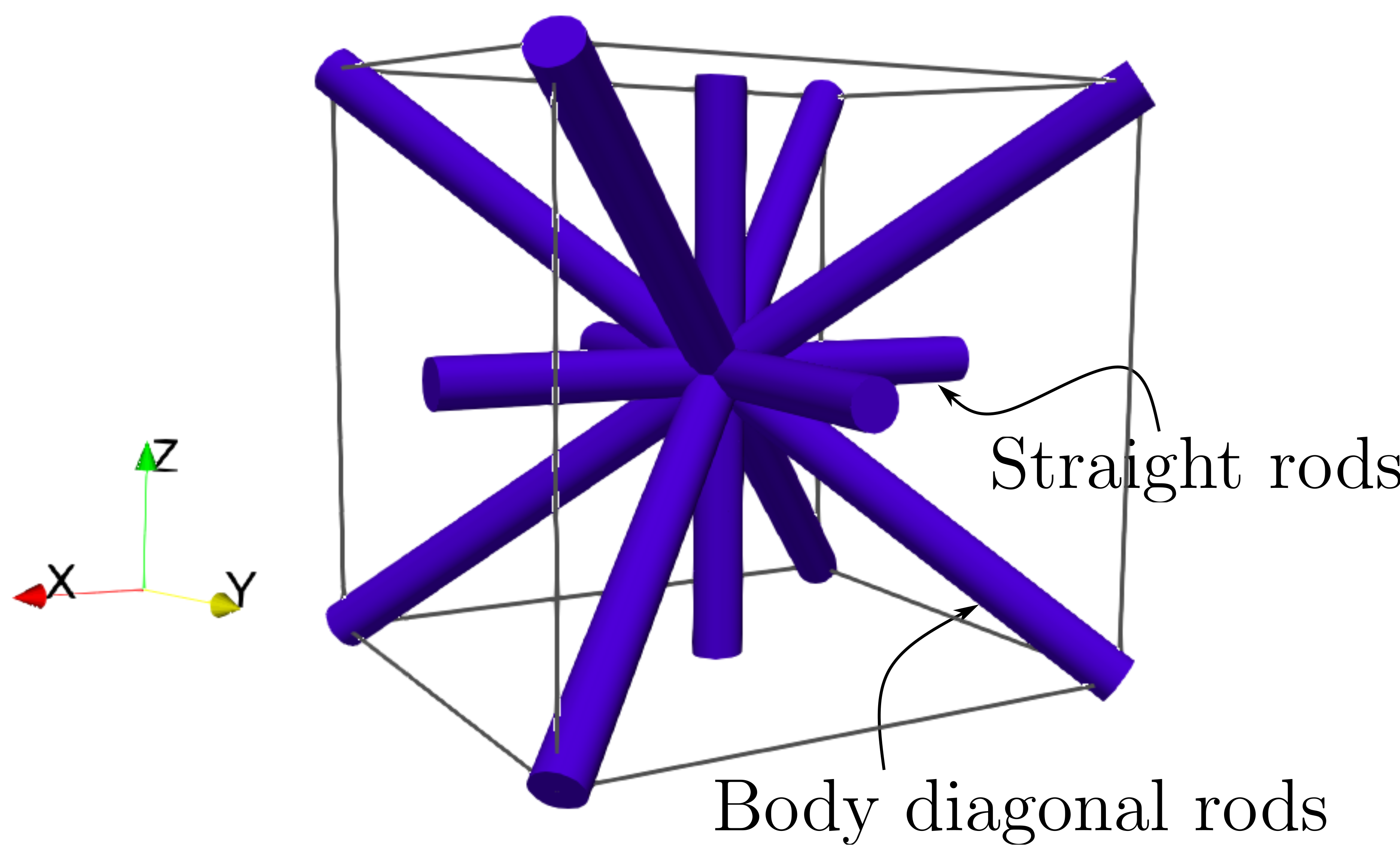}
            \caption{14-chain RVE}
    \end{subfigure}
    \caption{8- and 14-chain RVEs.}
    \label{fig:3dh:8chain_14chain_rve}
\end{figure}
In this section, using the stress/strain driven homogenization approach discussed above, we present extension, compression and shearing of 8- and 14-chain RVEs. The 8-chain model is nothing but a BCC lattice consisting of 8 rods connecting the center and the eight corner nodes of the cubic unit cell. The 14-chain model consists of 6 additional rods connecting the center and the 6 faces of the unit cell as shown in Figure \ref{fig:3dh:8chain_14chain_rve}. 
The RVEs are constructed with identical rods having a circular cross-section of radius $r$ and length $L^{\alpha} = l$ for inclined rods. Unless mentioned otherwise, the rods are modeled as special Cosserat rods. They obey the purely elastic material model with strain energy function given in equation \eqref{eq:rod_energy_function_quadratic} wherein $\mathbbl{C}^{rod}_0$ is taken to be a diagonal tensor with its components $\mathbbl{C}^{rod}_{11}=\mathbbl{C}^{rod}_{22}=kGA$, $\mathbbl{C}^{rod}_{33} = EA$, $\mathbbl{C}^{rod}_{44}=\mathbbl{C}^{rod}_{55} = EI$, $\mathbbl{C}^{rod}_{66}=GJ$. Here $E$ is the rod material's Young's modulus, $G$ is the shear modulus and $k$ is the shear correction factor. The symbols $A$, $I$ and $J$ denote cross-sectional area, second moment of area and polar moment of area, respectively. The RVE problem is non-dimensionalized as follows:
\begin{align}
    s &= \mathscr{L}\bar{s},\quad \textbf{r} = \mathscr{L}\bar{\textbf{r}}, \quad \boldsymbol{\theta} = \bar{\boldsymbol{\theta}},\quad \textbf{k}_0 = \frac{\bar{\textbf{k}}_0}{\mathscr{L}},\quad \textbf{n} = \frac{EI}{\mathscr{L}^2}\bar{\textbf{n}},\quad \textbf{m}=\frac{EI}{\mathscr{L}}\bar{\textbf{m}}.\end{align}
Here the quantities with overbar, i.e., $\bar{(\cdot)}$ are non-dimensionalized (or normalized) quantities and $\mathscr{L}$ is the non-dimensionalization length parameter. Using \cref{eq:3dh1:energy_density_macro}, this results in the following normalized energy density funtion:
\begin{align}
    W^M = \frac{EI}{\mathscr{L}^4}\bar{W}^M.
\end{align}
Unless specified otherwise, the length $\mathscr{L} = l = 1.0$, rod's radius $r = 0.05$ and the joints connecting the rods are modeled as rigid. 

As discussed earlier, two types of loading conditions are considered - strain-driven and stress-driven. The stability of the solution obtained from these two different problems is also studied in the numerical examples. To do so, we adopt the approach presented in \cite{kumar2010generalized,elliott2006stability}. We analyse the stress and strain-driven solutions using two different stability criteria, i.e., \textit{soft loading} and \textit{hard loading} stability criteria \citep{elliott2006stability}. In the soft loading criteria, perturbations with respect to both the configuration variables $\{\textbf{r}, \boldsymbol{\theta}\}$ and the macroscopic deformation gradient $\textbf{F}^M$ are considered whereas in the hard loading criteria, perturbations with respect to only the configuration variables are considered. Hence, it is said that soft loading stability criterion is a stricter condition on stability than the hard loading criterion. In the numerical examples, we emphasize the stability of the solutions obtained and show both the stable and unstable paths as per the two stability criteria. Note that the RVEs considered here possess inherent symmetries. Firstly, the 8-chain RVE has cubic symmetry and secondly, since we assume microscale rods of isotropic circular cross-section, the rods themselves possess symmetry. Therefore, the null space at a bifurcation point is bound to have multiple dimension which complicates tracking the bifurcated solution \citep{healey1988group,kumar2010generalized,combescure2016post}. 

The numerical examples presented in this section are solved using an in-house code developed on top of the deal.II library \citep{2024:africa.arndt.ea:deal}. The rods are discretized into $N^{el}$ discrete helical elements, and the corresponding value of $N^{el}$ is reported for each example below. The constraint set of 8- and 14-chain RVEs is given in \ref{appendix:14chain_rve_constraint_set}. The non-linear system of equations resulting from the constrained discrete energy functionals (either strain-driven or stress-driven) is solved using the Newton--Raphson method augmented with an arc-length continuation scheme. The stability analysis is performed using the SLEPc Krylov--Schur eigensolver with the shift-invert technique to compute the smallest set of eigenvalues.

\subsection{Extension}\label{sec:numerical_examples_extension}
First, the strain-driven extension of an RVE is considered. The prescribed macroscopic deformation gradient is of the form:
\begin{equation}
    \mathbf{F}^M = \begin{bmatrix}
    F^M_{11} & 0 & 0 \\
    0 & 1 & 0 \\
    0 & 0 & 1
    \end{bmatrix},\quad \quad F^M_{11}>1.
\end{equation}
\begin{figure}[h!]
\centering
    \includegraphics[width=0.49\textwidth]{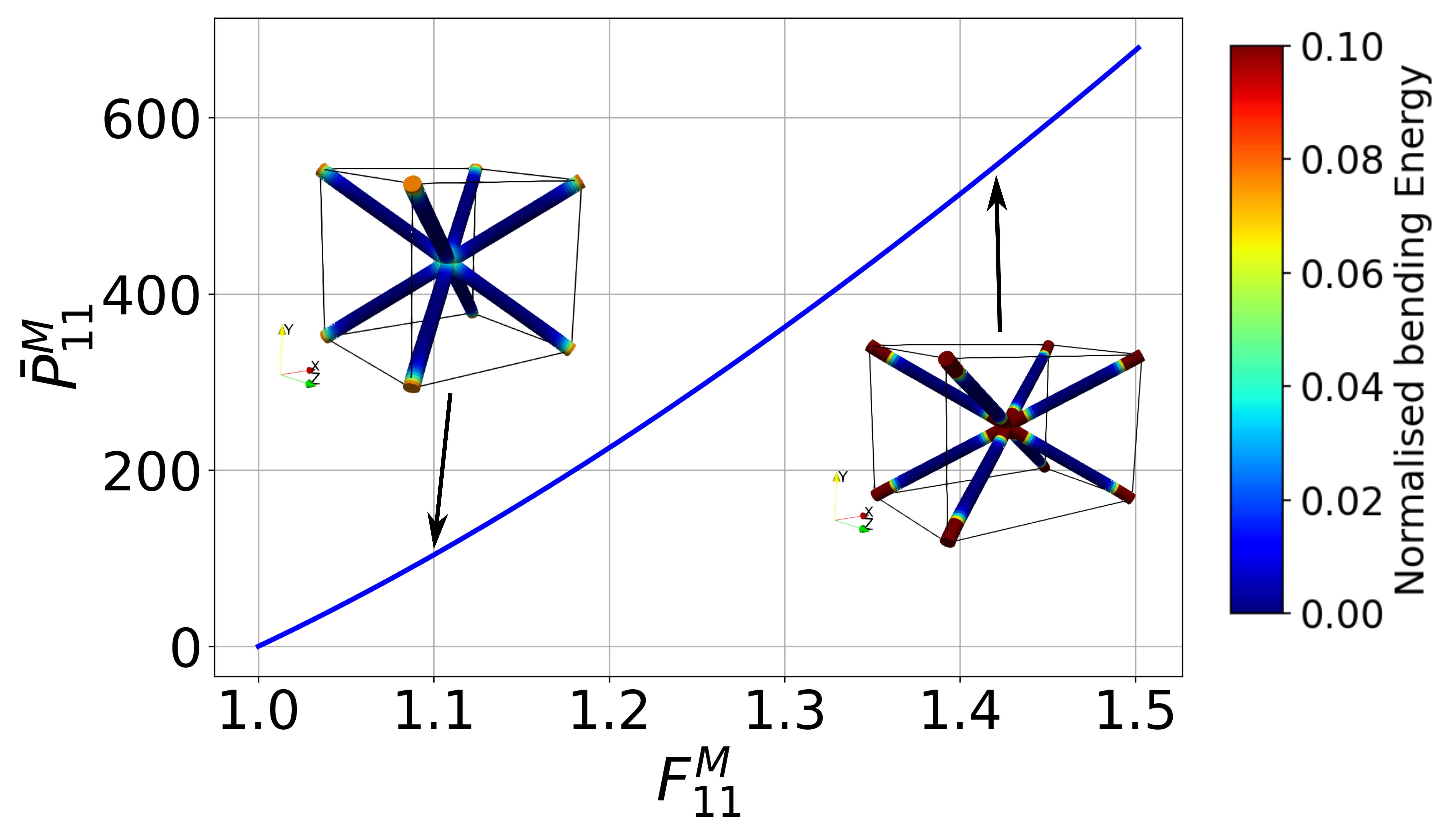}
    \includegraphics[width=0.49\textwidth]{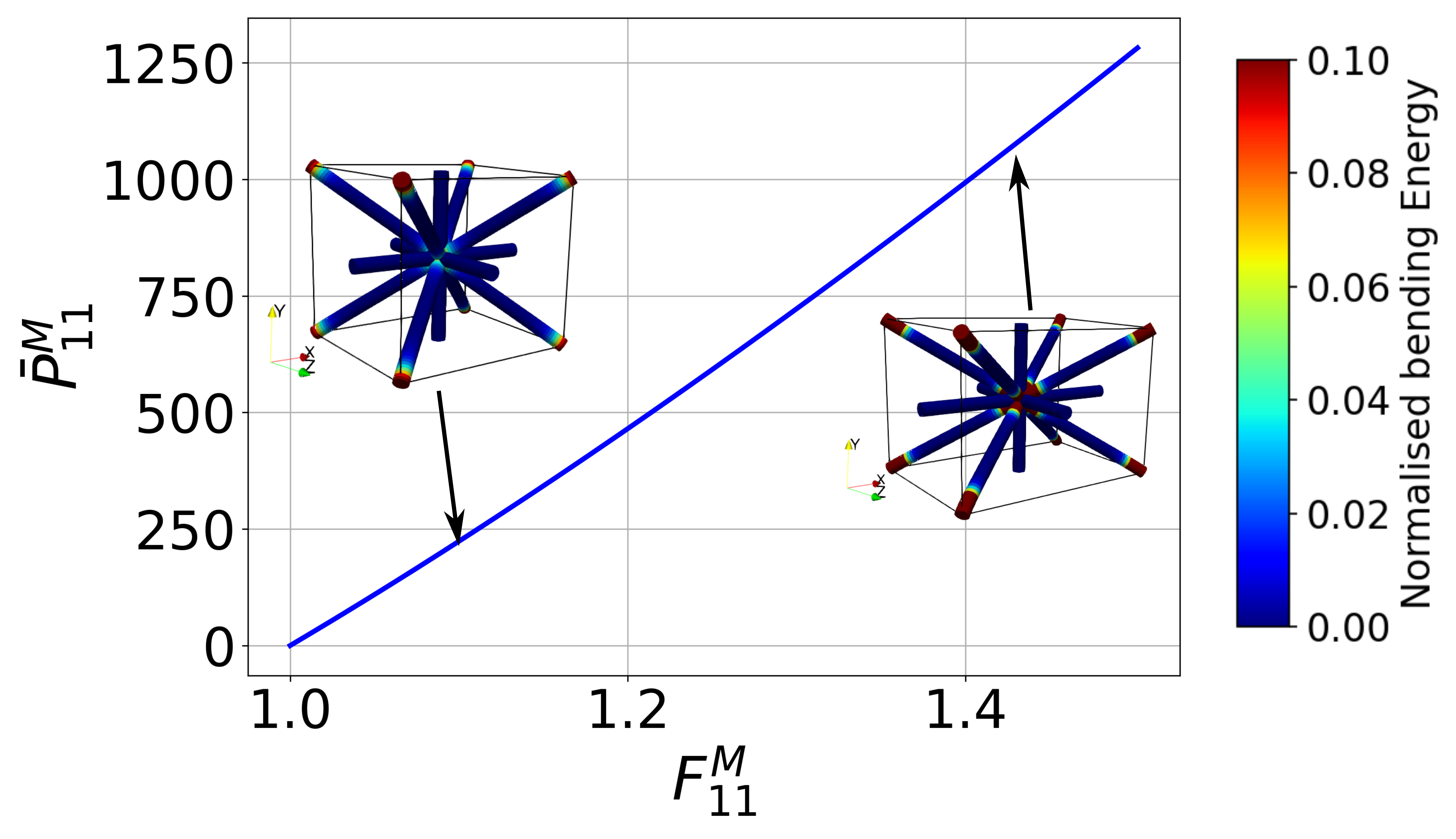}
    \caption{Strain-driven extension of 8 and 14-chain RVEs. $N^{el} = 20$}
    \label{fig:strain_driven_extension}
\end{figure}
Figure \ref{fig:strain_driven_extension} shows the variation of the macroscopic stress component $\bar{P}^M_{11}$ with the applied macroscopic deformation gradient component $F^M_{11}$. An approximately linear response (due to the constitutive law of the microscale rods in \cref{eq:rod_energy_function_quadratic}) in the stress-strain relationships of both the 8- and 14-chain RVEs is observed. The deformation mode of each of the rods is mostly stretching dominated (except for bending near the rod ends where the rigid joint constraint has to be maintained). In addition to the longitudinal tensile stress $P^M_{11}$, transverse stresses $P^M_{22}$ and $P^M_{33}$ are also generated. This is because as the RVE is stretched, the inclined rods have a tendency to bend in the transverse direction in order to align themselves along the longitudinal direction. However, this is restricted as the prescribed deformation gradient has to be maintained through the PBC. The configurations are stable all along the response in both hard and soft loading stability criteria. \\
\begin{figure}[h!]
    \centering
    \begin{subfigure}{0.6\textwidth}
    \centering
    \includegraphics[width=\textwidth]{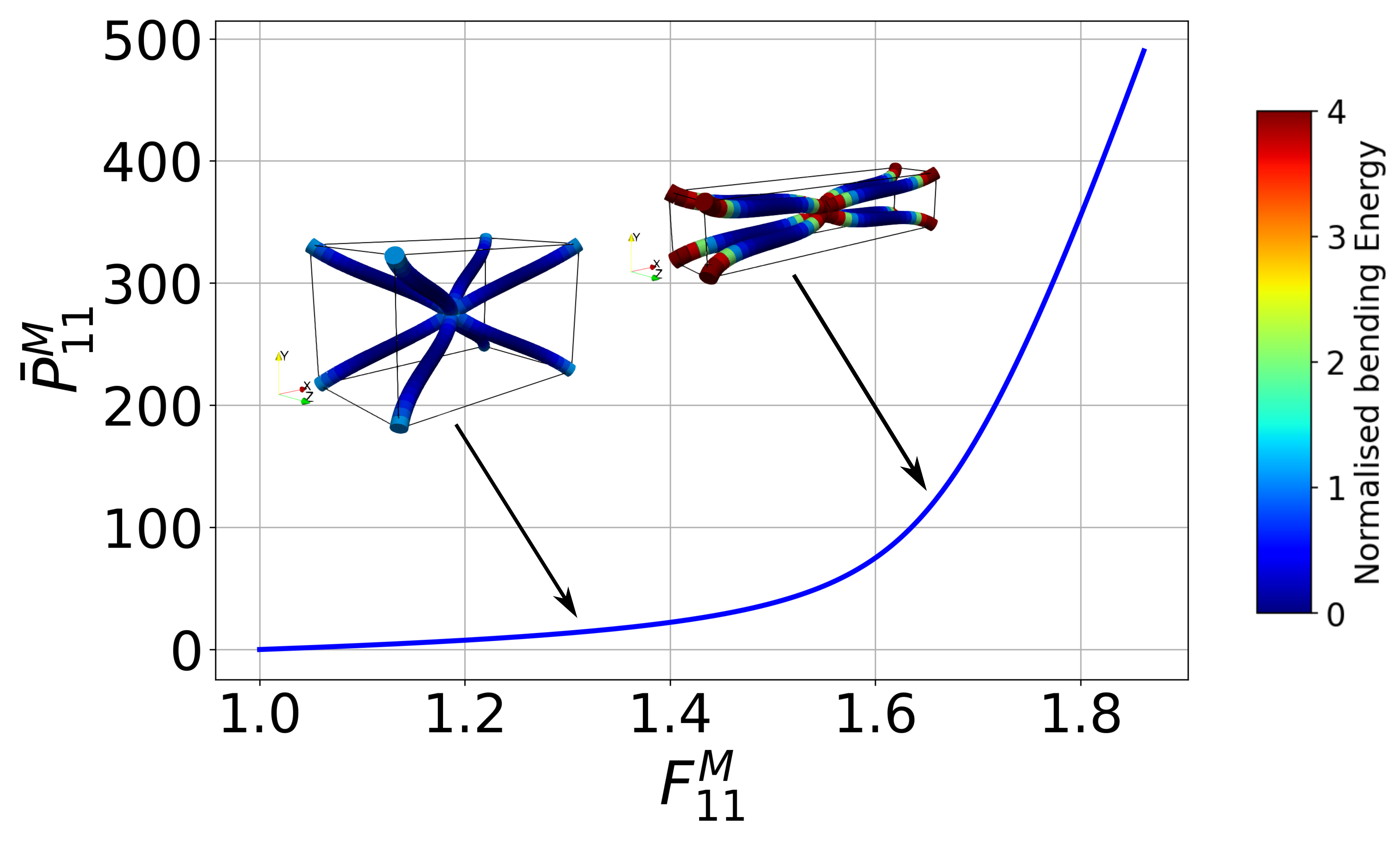}
    \caption{Longitudinal stress-strain response of 8-chain RVE under uniaxial tension.}
    \label{fig:stress_driven_tension_8chain_P11_F11}
    \end{subfigure}
        \begin{subfigure}{0.49\textwidth}
        \includegraphics[width=\textwidth]{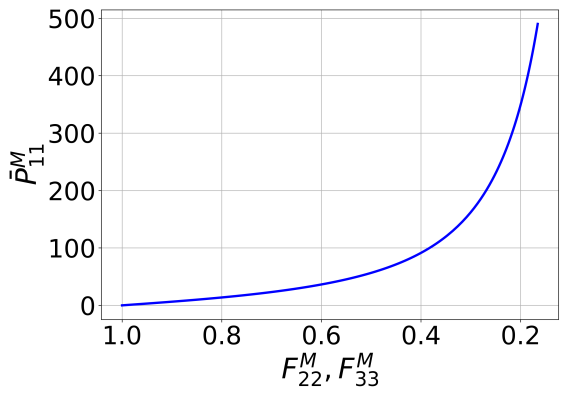}
    \caption{Stress-strain response in the transverse direction of the 8-chain RVE under uniaxial tension.}
    \label{fig:stress_driven_tension_8chain_P11_F22_33}
    \end{subfigure}
    \begin{subfigure}{0.49\textwidth}
    \includegraphics[width=\textwidth]{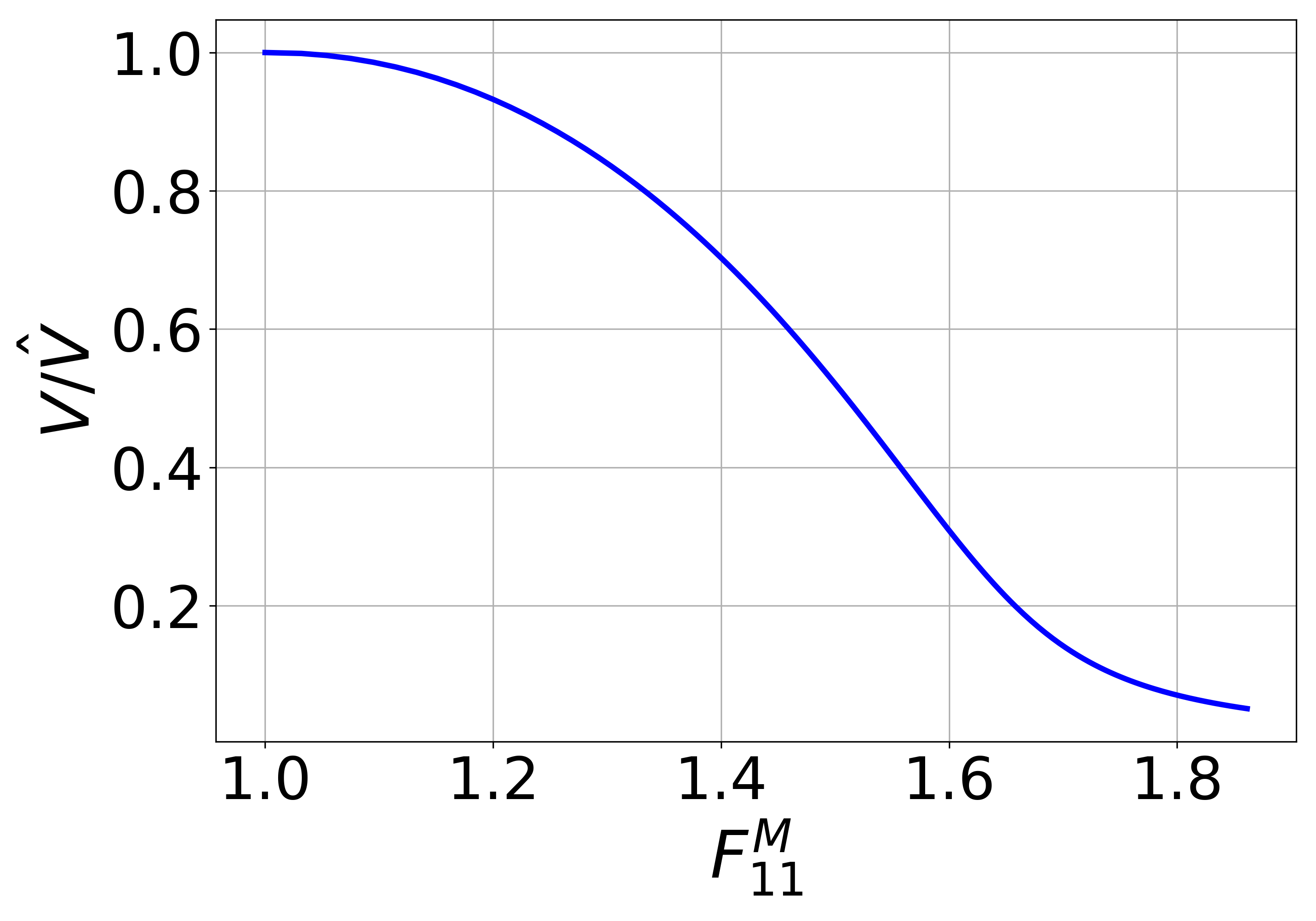}
    \caption{Volumetric stretch ($V/\hat{V}$) of 8-chain RVE under stress-driven uniaxial tension}
    \label{fig:stress_driven_tension_8chain_V_F}
    \end{subfigure}
    \caption{Stress-driven uniaxial tension of 8-chain RVE. $N^{el} = 20$}
    \label{fig:stress_driven_tension_8chain}
\end{figure}
\begin{figure}[h!]
    \centering
    \begin{subfigure}{\textwidth}
    \centering
        \includegraphics[width=0.6\textwidth]{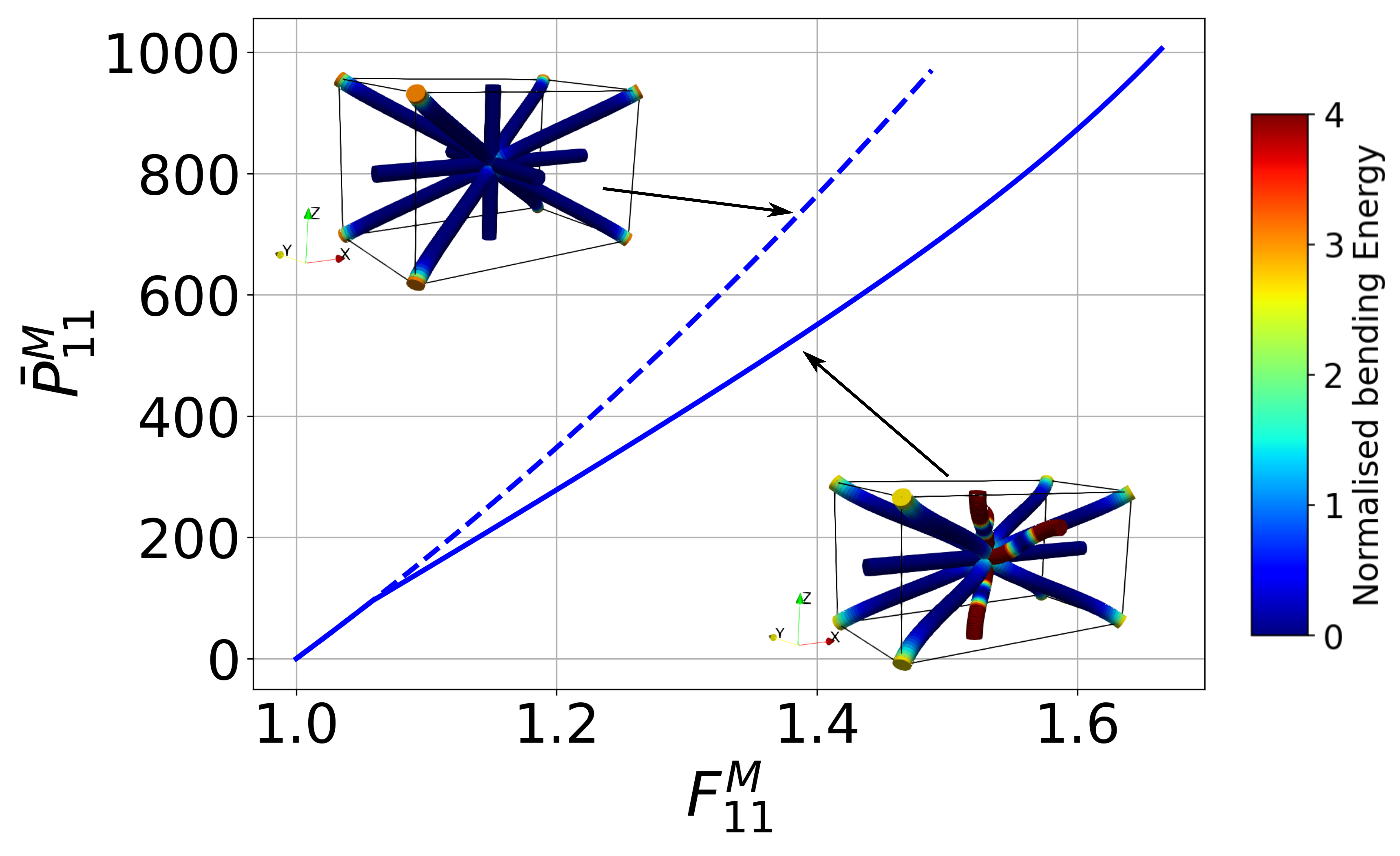}
        \caption{Stress-strain response of 14-chain RVE under uniaxial tension}
        \label{fig:stress_driven_tension_14chain_P11_F11}
    \end{subfigure}
    \begin{subfigure}{0.49\textwidth}
        \includegraphics[width=\textwidth]{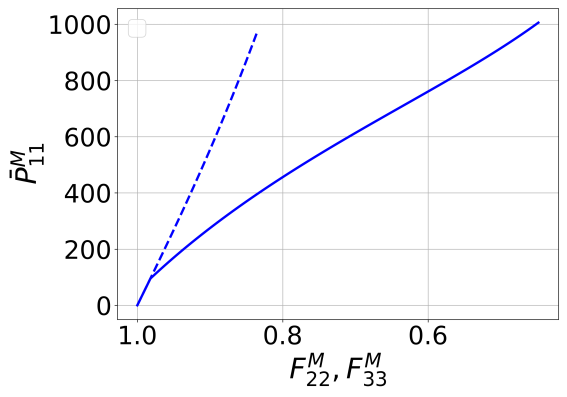}
        \caption{Stress-strain response of 14-chain RVE under uniaxial tension}
        \label{fig:stress_driven_tension_14chain_P11_F22F33}
    \end{subfigure}
    \begin{subfigure}{0.49\textwidth}
        \includegraphics[width=\textwidth]{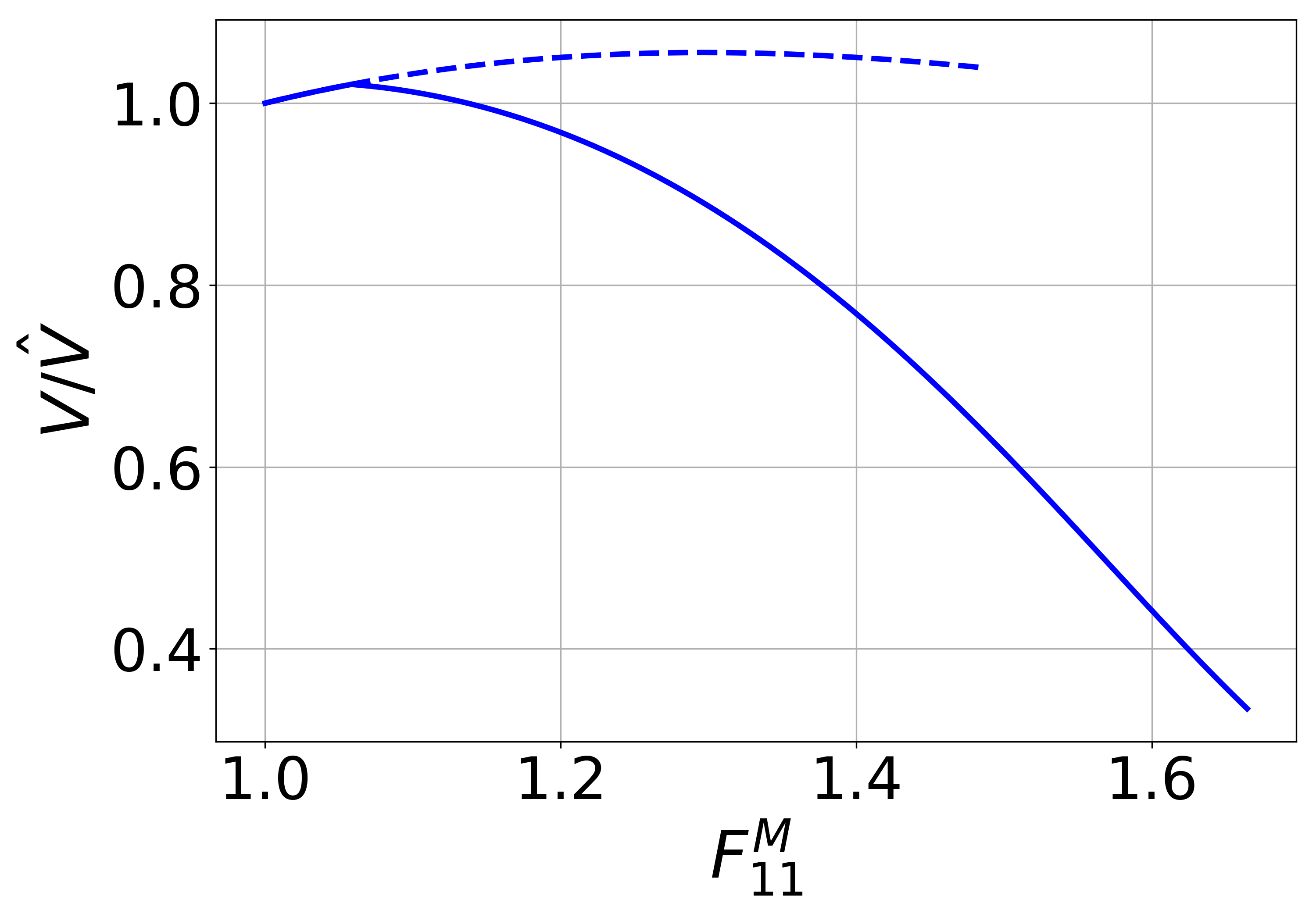}
        \caption{Volumetric stretch of 14-chain RVE under stress driven uniaxial tension}
        \label{fig:stress_driven_tension_14chain_V_F11}
    \end{subfigure}
    \begin{subfigure}{0.49\textwidth}
        \includegraphics[width=\textwidth]{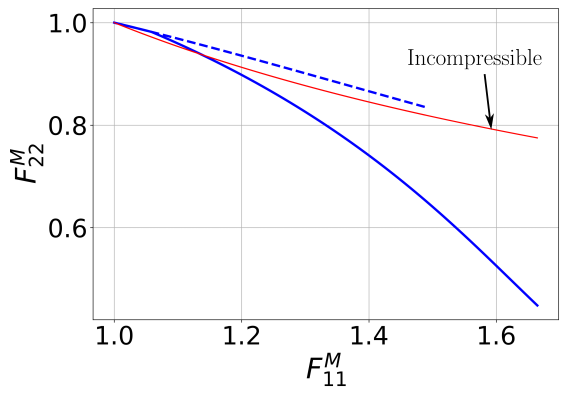}
        \caption{Lateral strain v/s longitudinal strain}
        \label{fig:stress_driven_tension_14chain_F22_F11}
    \end{subfigure}
    \caption{Stress-driven tension of 14-chain RVE. The dashed line shows the unstable solution. $N^{el} = 50$}
    \label{fig:stress_driven_tension_14chain}
\end{figure}
Uniaxial tension experiments \citep{brown2009multiscale,storm2005nonlinear,licup2015stress,garyfallogiannis2023fracture,kliuchnikov2025strength} usually consider traction-free lateral boundary conditions. To mimic such experiments, the stress-driven approach is considered, wherein the macroscopic first Piola-Kirchhoff stress is prescribed and is of the form:
\begin{equation}
    \mathbf{P}^M = \begin{bmatrix}
    P^M_{11} & 0 & 0 \\
    0 & 0 & 0 \\
    0 & 0 & 0
    \end{bmatrix},\quad\quad P^M_{11}>0 .
\end{equation}
Figures \ref{fig:stress_driven_tension_8chain} and \ref{fig:stress_driven_tension_14chain} show the macroscopic stress-strain 
response of 8- and 14-chain RVEs under stress-driven uniaxial tension, respectively. In the case of 8-chain RVE, an initially soft response gradually stiffens as the mode of deformation transitions from being bending-dominated to stretching-dominated. The stiffening transition happens around $F^M_{11}=\sqrt{3}$ when the inclined rods are almost aligned with the stretching direction. This contrasts with the constant, stiff response obtained in the case of strain-driven extension, as it is stretching-dominated throughout. Here, the RVE is allowed to contract laterally to maintain the traction-free boundary condition. This results in a highly non-linear Poisson's effect and an overall reduction in the volume of the RVE as shown in Figure \ref{fig:stress_driven_tension_8chain_V_F}. This phenomenon has important biophysical implications and is characteristic of biological tissues such as fibrin gels \citep{brown2009multiscale}. Phenomenological models have been proposed to capture it \citep{PhysRevE.110.014502,garyfallogiannis2024cracks}. The 8-chain RVE remains stable all along in both soft and hard loading criteria. Now consider the 14-chain RVE uniaxial tension response as shown in Figure \ref{fig:stress_driven_tension_14chain}. In this case, the initial response is stretching-dominated until the lateral rods, which are under compression due to the RVE's tendency to contract laterally, buckle. At the buckling point, the primary branch of the 14-chain RVE encounters four zero eigenvalues simultaneously. Due to the symmetry of the RVE and the loading direction, the various mode shapes correspond to the straight transverse rods (in $\textbf{e}_2$ and $\textbf{e}_3$ directions) buckling in two different planes (due to isotropic cross-section). Since a linear combination of the mode shapes is also a mode shape, we combine two modes such that the straight rods along both the transverse directions are buckled simultaneously. Branch switching with such a mode shape leads to sudden softening of the tensile response, wherein the deformation of lateral rods turn into being bending-dominated. This triggers the non-linear Poisson's effect and results in a dramatic volume shrinkage (see Figure \ref{fig:stress_driven_tension_14chain_V_F11}). The volumetric stretch ($V/\hat{V}$) is qualitatively different from the 8-chain case. Here, the volume initially increases and starts decreasing after the bifurcation point. In the case of the unbuckled solution path, the volume remains in the expansion regime (see figures \ref{fig:stress_driven_tension_14chain_V_F11} and \ref{fig:stress_driven_tension_14chain_F22_F11}).
\begin{figure}[h!]
    \centering
    \includegraphics[width=0.6\textwidth]{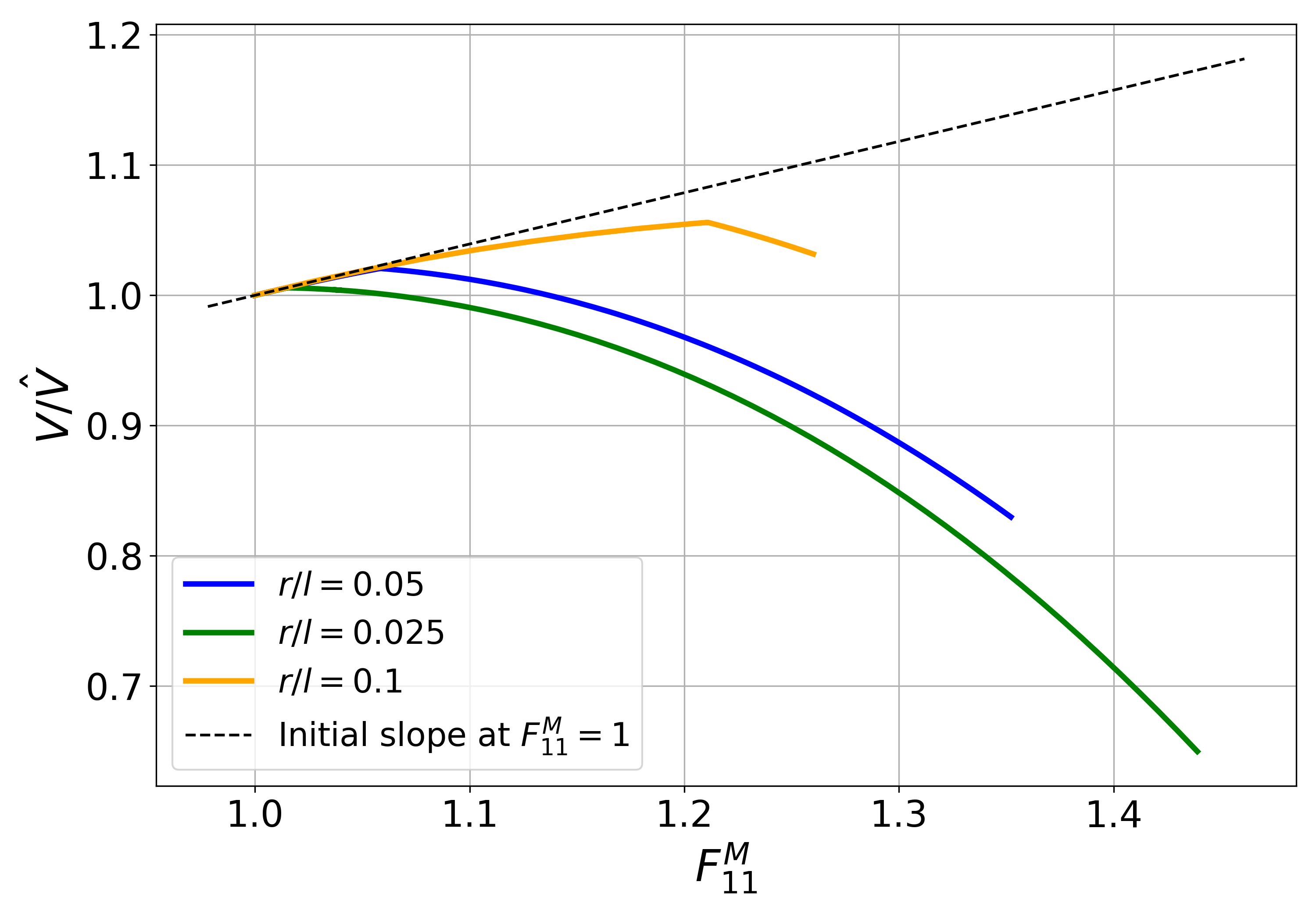}
    \caption{The variation of volumetric stretch in 14-chain RVE with the slenderness parameter $\frac{r}{l}$. The simulation is terminated when the fibers come in contact.}
    \label{fig:14chain_tension_volume_change_comparison}
\end{figure}
The increasing-decreasing trend of volumetric stretch can be tuned through the slenderness parameter $\frac{r}{l}$ \citep{PhysRevE.110.014502}. Figure \ref{fig:14chain_tension_volume_change_comparison} shows that the curve remains more above the incompressible limit ($V/\hat{V}=1$) as the $\frac{r}{l}$ parameter increases or the rods become thicker. Assuming the initial deformation to be stretching dominated, we determine the initial (at $F^M_{11}=1$) slope of volumetric stretch with respect to stretch in \ref{sec:fractional_volume_change_14chain} and show that it indeed turns out to be positive, leading to a definite increase in volume at small strains.

\begin{figure}[h!]
    \centering
    \includegraphics[width=\textwidth]{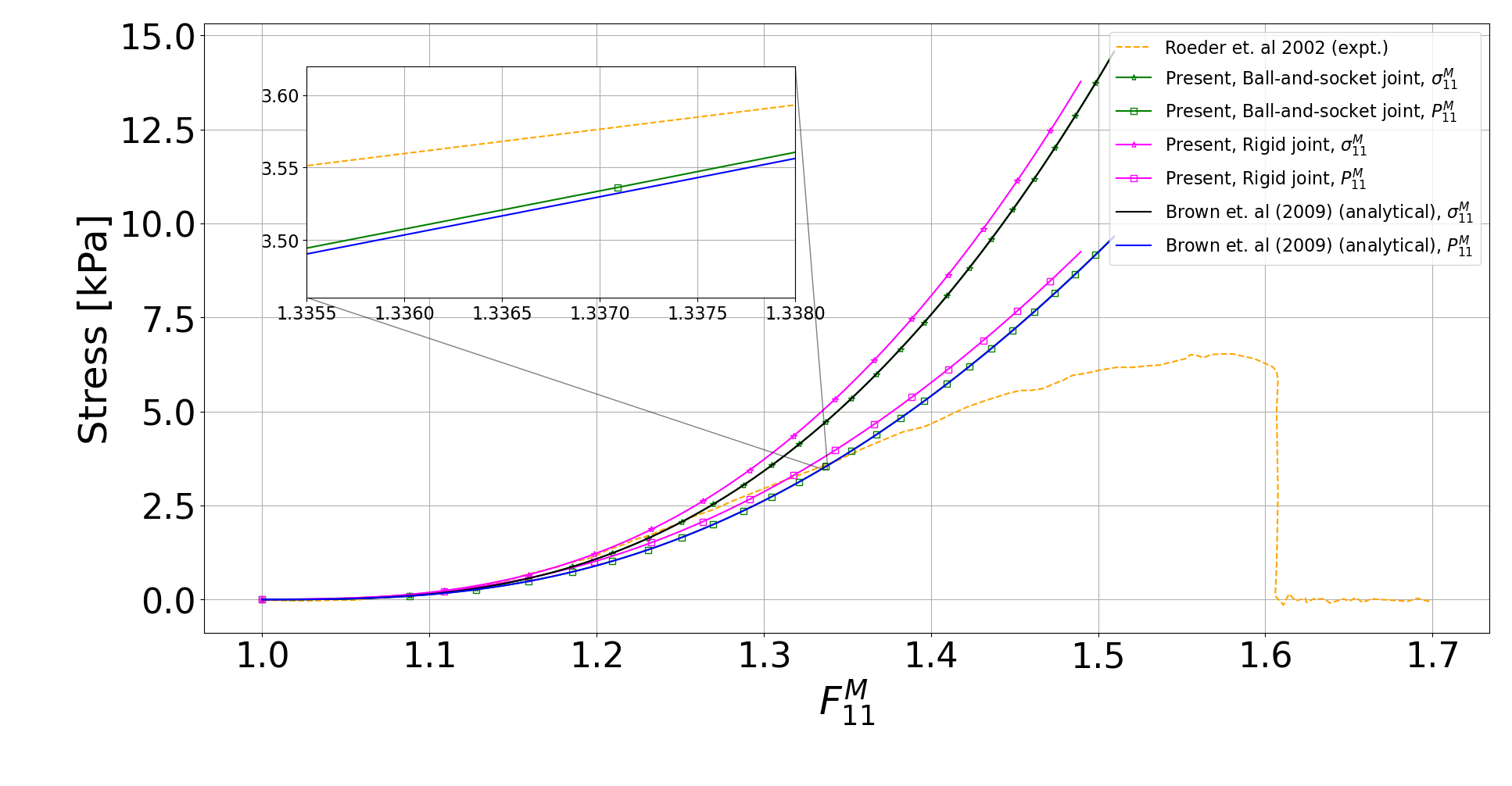}
    \caption{Comparison of homogenized (present model) response of 8-chain RVE with uniaxial tension experiment \citep{roeder2002tensile} of collagen network and 8-chain analytical model of \cite{brown2009multiscale}. The rod parameters (for collagen fibers) are: $E = 40 \text{MPa}$, $r=217.5$nm and $l=10 \mu m$. $N^{el} = 20$}
    \label{fig:8chain_uniaxial_tension_validation}
\end{figure}
Next, we quantitatively compare the macroscopic response obtained using the present homogenization-based model with the experimental results in \citet{roeder2002tensile} and the analytical model (based on 8-chain RVE) described in \cite{brown2009multiscale} for uniaxial tension on collagen gels. The uniaxial tension experiment is carried out with a traction-free stress state in the lateral directions. Further, collagen gels are considered to be incompressible. We incorporate these constraints in our setting as shown in \ref{sec:uniaxial_tension_appendix}. The deformation gradient is taken to be of the form
\begin{align}\label{eq:incompressible_def_gradient}
        \mathbf{F}^M = \begin{bmatrix}
    F^M_{11} & 0 & 0 \\
    0 & \frac{1}{\sqrt{F^{M}_{11}}} & 0 \\
    0 & 0 & \frac{1}{\sqrt{F^{M}_{11}}}
    \end{bmatrix},\quad\quad F^M_{11}>0
\end{align}
to satisfy incompressibility. We also consider the response with a ball-and-socket type joint, which mimics the stretching-only energy model for a single fiber in \cite{brown2009multiscale}. Figure \ref{fig:8chain_uniaxial_tension_validation} shows the variation in both Cauchy-stress component ($\sigma_{11}^M$) and the first Piola-Kirchhoff stress component ($P_{11}^M$). We note that a good agreement is observed with the experiment and the analytical model (briefly recalled in \ref{sec:uniaxial_tension_straight_appendix}). Both the homogenization and analytical models deviate from the experimental results once damage and/or fracture is said to initiate in the network. The figure shows that considering rigid joints has a negligible influence on the macroscopic response under uniaxial tension since the overall deformation is governed primarily by fiber rotation (rather than by fiber stretching or bending) due to incompressibility constraint. The network ruptures around the stretch value of about 1.6 (as shown in Figure \ref{fig:8chain_uniaxial_tension_validation}) which is close to the theoretical stretch (using the isotropic network) of about 1.5707 when all the randomly oriented fibers become aligned with the stretching direction \citep{brown2009multiscale}. Even at this macroscopic stretch value, the microscopic stretch in individual fibers are very small which justifies using linear force-stretch model for individual fibers to model collagen networks.
\begin{figure}[h!]
\centering
\begin{subfigure}{{0.4\textwidth}}
    \centering
    \includegraphics[width=\textwidth]{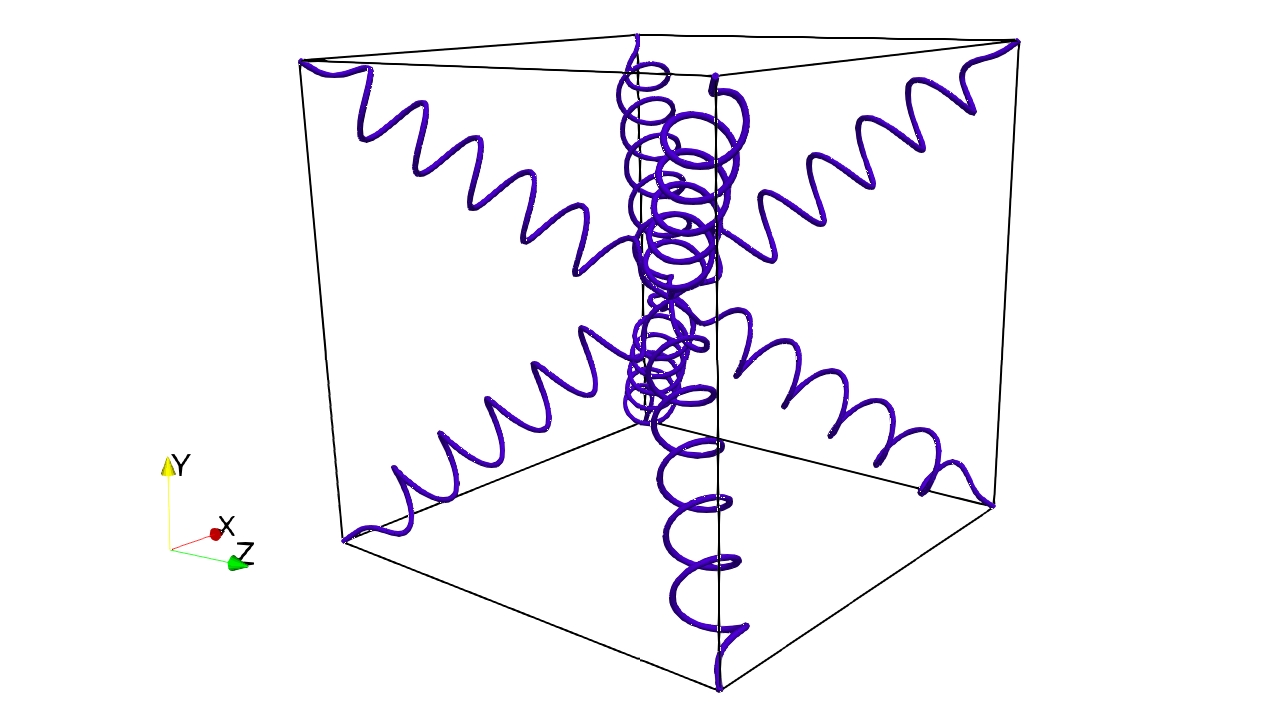}
    \caption{Helical 8-chain RVE}
\end{subfigure}
\begin{subfigure}{{0.59\textwidth}}
    \includegraphics[width=\textwidth]{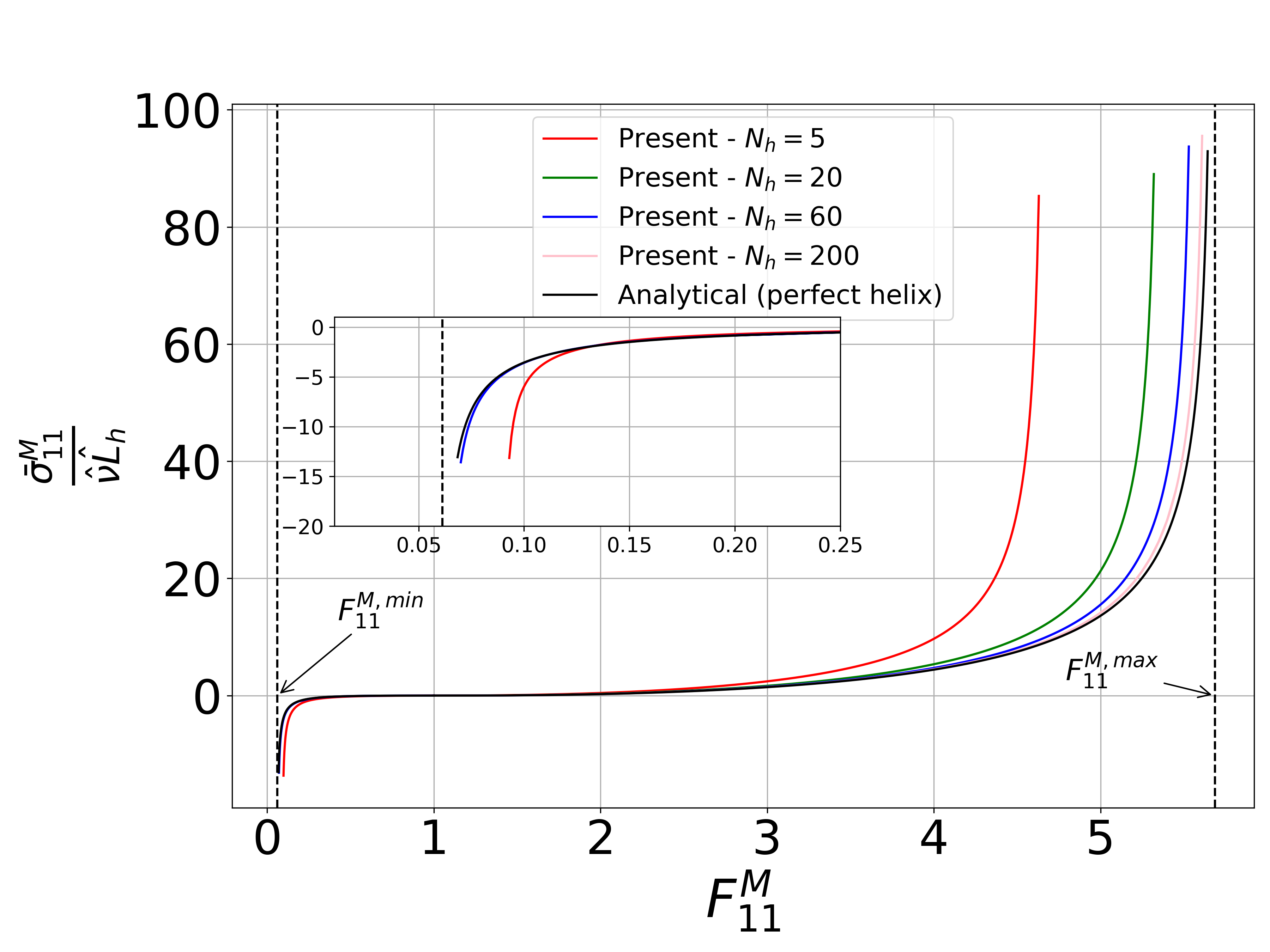}
    \caption{Stress-stretch response of helical 8-chain RVE. Here the non-dimensionalization length $\mathscr{L}=1.0$, thus $\bar{\sigma}^M_{11} = \frac{1}{EI}\sigma^M_{11}$.}
\end{subfigure}
    \caption{Uniaxial tension-compression response of 8-chain helical RVE and its variation with number of turns ($N_h$) in the helical fibers. The helical fibers are inextensible and unshearable, i.e, the rod model is of Kirchhoff type. The fiber parameters are $\hat{p}_h = 1.0$, $\hat{r}_h = 0.5$ and $\hat{p}_j = 1.0$. The undeformed end-to-end length of the helical fiber is $\hat{L}_h=N_h\hat{p}_h + 2\hat{p}_j$, and $\hat{\nu}$ is the number of fibers per unit undeformed volume of the RVE. For perfect helices, $\hat{p}_j=0$. The discretization in the helical portion of the imperfect helix depends on the number of turns: $N^{el}=50~(N_h=5)$, $N^{el}=500~(N_h=20)$, $N^{el}=2000~(N_h=60)$, $N^{el}=4000~(N_h=200)$. For the curves at helix ends, $N^{el}=50$.}
    \label{fig:8chain_helical_uniaxial_tension_compression}
\end{figure}

 Fibrin networks exhibit much larger stretch \citep{brown2009multiscale}. This has been attributed to the unfolding of fibrin monomers in the fibrin fibers during stretching. To capture this effect, an analytical model was proposed in \cite{brown2009multiscale}  by modifying the linear (in case of collagen) force-stretch relationship of individual fibers to a non-linear one. In addition to the molecular underpinnings, undulations or waviness of the fiber are important parameters governing the overall response of the fiber networks \citep{picu2011mechanics,lohr2025modeling}. Thus, uniaxial stretching in biopolymer networks can be regarded as a hierarchical multiscale phenomenon that involves nonlinearities at the fiber scale as well. In the architected materials domain, recently proposed woven metamaterials with entangled and/or interpenetrating helical fibers in a unit cell have shown that leveraging fiber-level non-linearity provides an enlarged design space of metamaterials for describing the response of complex materials such as biological tissues \citep{surjadi2025double,carton2026design}. In this context, we consider helical fibers instead of straight rods in our 8-chain RVE, introducing another scale in the homogenization problem. We model the helical fibers as Kirchhoff rods, i.e., as inextensible and unshearable using the constraint given in \eqref{eq:kirchhoff_constraint}. To create a cross-linked helical 8-chain RVE, we consider a modified helical fiber geometry \citep{chang2022mechanics}. This helical fiber is constructed such that it has a uniform/perfect helix portion in the middle, joined by curves at the two ends, such that the two ends of the helical fiber lie on the helix axis. Apart from the fiber cross-sectional radius, the helical fiber is characterized by four design parameters: number of turns $(N_h)$, radius $(\hat{r}_h)$, and pitch $(\hat{p}_h)$ of the perfect helix portion and $\hat{p}_j$ that denotes the length of the non-uniformly curved segment. The force-stretch relationship of such a fiber can be approximated by the force-stretch relationship of a perfect helix \citep{djurivckovic2013twist} (refer \ref{sec:uniaxial_tension_helix_appendix} for more details). With this topology of the helical fiber, the 8-chain RVE is constructed, and uniaxial tension (with incompressibility and traction-free lateral boundary condition) is performed. The analytical stress-strain response of an 8-chain RVE with helical fibers can be obtained by substituting the helical fiber force-stretch relationship in the 8-chain stress-strain relationship of \cite{brown2009multiscale}. We compare this analytical model with the present homogenization-based approach as shown in Figure \ref{fig:8chain_helical_uniaxial_tension_compression}. The RVE now exhibits much larger stretch before stiffening as compared to the collagen response in Figure \ref{fig:8chain_uniaxial_tension_validation} - approximately three times. This is an outcome of the additional non-linearity at the fiber scale. The fiber's ``effective" force-stretch response is now bending dominated resulting in a much softer response. As the number of turns $N_h$ increases, the numerical results start to match analytical result - this is because the helix in the numerical simulation is not a perfect helix. Figure \ref{fig:8chain_helical_uniaxial_tension_compression} also shows the uniaxial compression response ($F^M_{11}<1$). In this regime as well, the fibers themselves undergo bending-dominated deformation while the RVE is under overall compression along $\textbf{e}_1$ direction.
\subsection{Compression}
Compression experiments of blood clots have shown that fibrin networks exhibit foam-like behaviour \citep{kim2014structural, kim2016foam,liang2017phase}. Initially, the stress-strain relationship is linear followed by sudden flattening and eventually strain-stiffening due to densification upon contact. In this section, we study the compression response of the 8- and 14-chain RVEs. We neglect the third region (densification) because fiber-to-fiber contact has not been considered in this homogenization model, although such effects can be incorporated.

First, strain-driven compression of 8- and 14-chain RVEs is considered. The deformation gradient is of the form:
\begin{equation}
    \mathbf{F}^M = \begin{bmatrix}
    F^M_{11} & 0 & 0 \\
    0 & 1 & 0 \\
    0 & 0 & 1
    \end{bmatrix},\quad\quad 0<F^M_{11}<1.
\end{equation}
\begin{figure}[h!]
\centering
\begin{subfigure}{{0.49\textwidth}}
    \includegraphics[width=\textwidth]{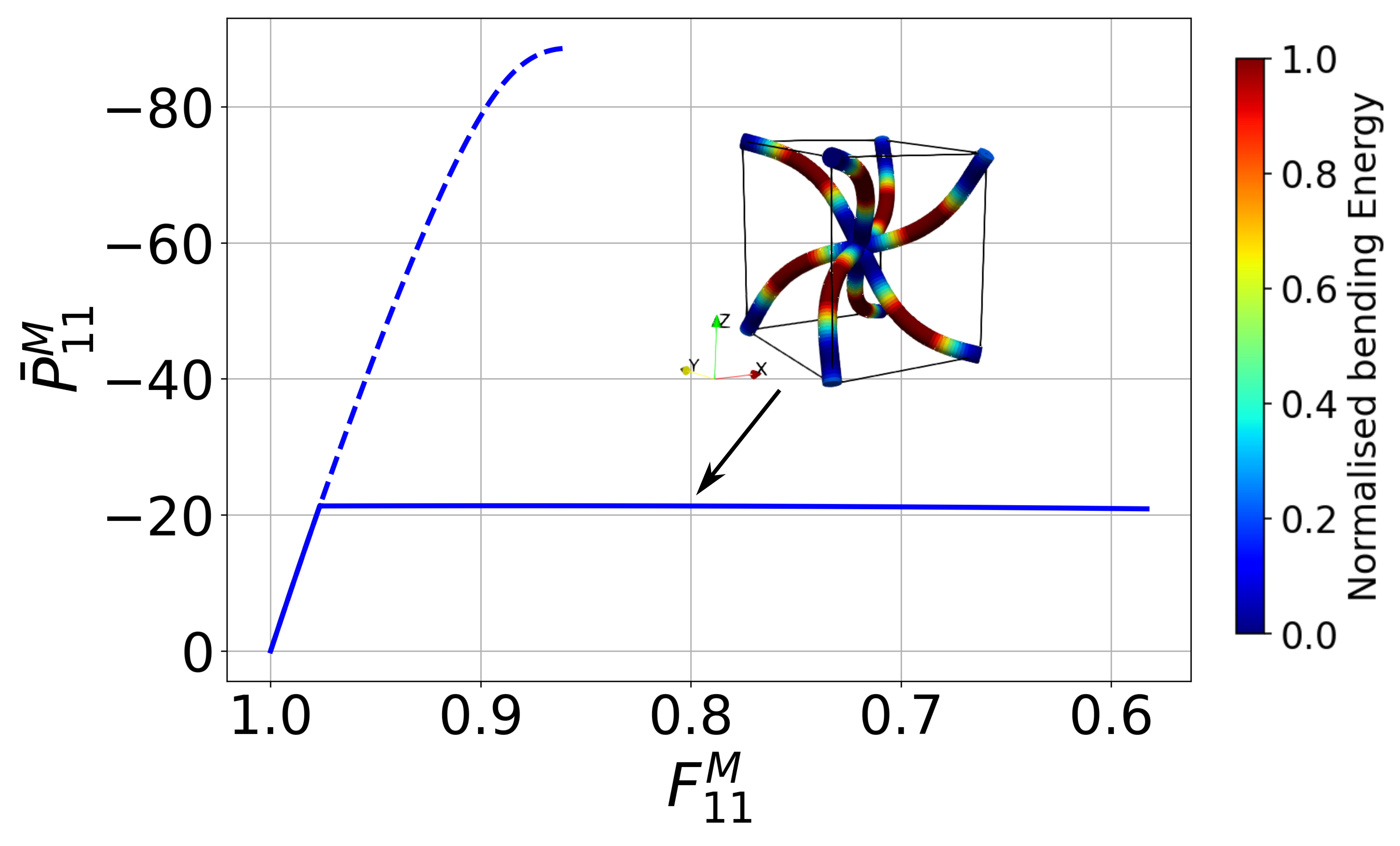}
    \caption{}
    \label{fig:8chain_strain_driven_compression}
\end{subfigure}
\begin{subfigure}{{0.49\textwidth}}
    \includegraphics[width=\textwidth]{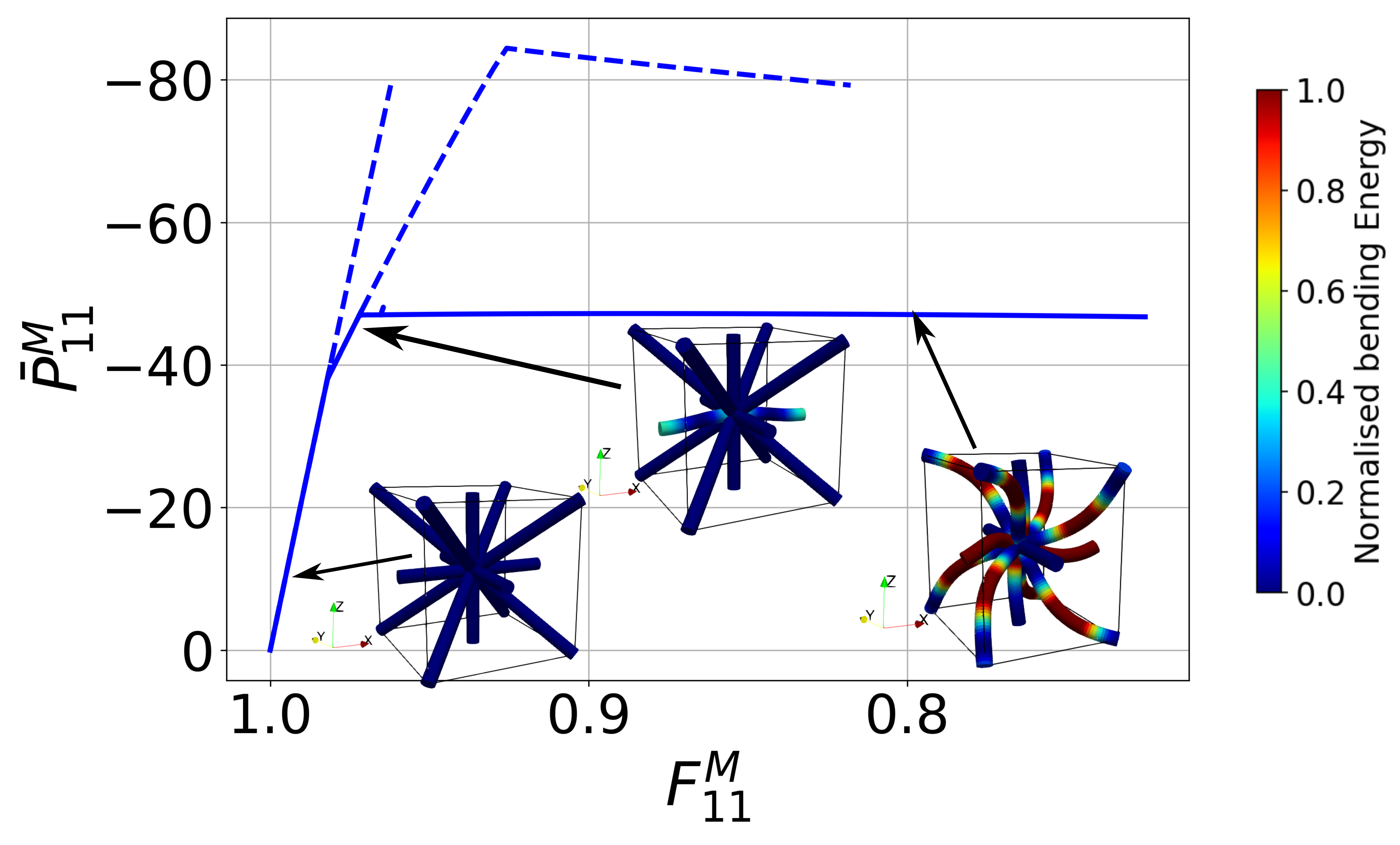}
    \caption{}\label{fig:14chain_strain_driven_compression}
\end{subfigure}
    \caption{Strain-driven compression of 8- and 14-chain RVEs. The dashed lines show the unstable paths whereas the solid lines show the stable paths. $N^{el} = 50$}
    \label{fig:strain_driven_compression}
\end{figure}
Figure \ref{fig:strain_driven_compression} shows the variation of the macroscopic stress component $\bar{P}^M_{11}$ with the applied macroscopic deformation gradient component $F^M_{11}$. The compression response initially exhibits a linear behavior, followed by a plateau region characterized by an approximately zero slope in the stress–strain curve for both RVEs, with a slight difference. For the 8-chain RVE, the flattening is obtained at a single bifurcation point when all the body diagonal rods buckle at the same time. At this point, three eigenvalues become zero simultaneously. However, the three mode shapes are similar with differences only in the plane of buckling of individual rods. They result in approximately the same stress-strain response. The buckled solution path is stable in both hard and soft loading scenarios for the 8-chain RVE. In the case of 14-chain RVE, we see that the flattening occurs in a stepwise manner, i.e., a trilinear $\bar{P}^M_{11}$ vs $F^M_{11}$ curve is obtained. At the first bifurcation point, two eigenvalues simultaneously become zero. The mode shapes corresponding to these are those of a buckled straight rod (along the compression direction). These modes lie in different planes because of the rod's isotropic cross-section. Perturbing the configuration using any of these mode shapes produces the same macroscopic response. We therefore select one of them to switch the branch along a secondary path. This results in a sudden decrease in the slope of $\bar{P}^M_{11}$ vs $F^M_{11}$ curve, yielding the second linear regime. The RVE response remains stretching (actually compression of inclined rods and no bending) dominated because of the unbuckled inclined rods. We note that the second null eigenvalue which had originated at first buckling point remains a null eigenvalue throughout. Upon further straining, a second bifurcation point is obtained. Here, two more eigenvalues simultaneously become zero in addition to the zero eigenvalue from the previous buckling point. The mode shapes of these two eigenvalues correspond to all the inclined rods undergoing buckling, similar to that in the 8-chain RVE. Again, both the mode shapes only differ in the plane of buckling of the inclined rods but result in the same stress-strain response. This solution path leads to the sudden flattening of the curve as all the rods have buckled (except the lateral straight rods which remain undeformed). The trilinear solution path of the 14-chain RVE is stable in both hard and soft loading criteria. Note that in the case of uniaxial compression that we have just discussed, the RVE response transitions from being stretching-dominated to bending-dominated (leads to strain softening), which is in contrast to the case of uniaxial tension (leads to strain stiffening).\\
\begin{figure}[h!]
    \centering
    \begin{subfigure}{{0.45\textwidth}}
    \includegraphics[width=\textwidth]{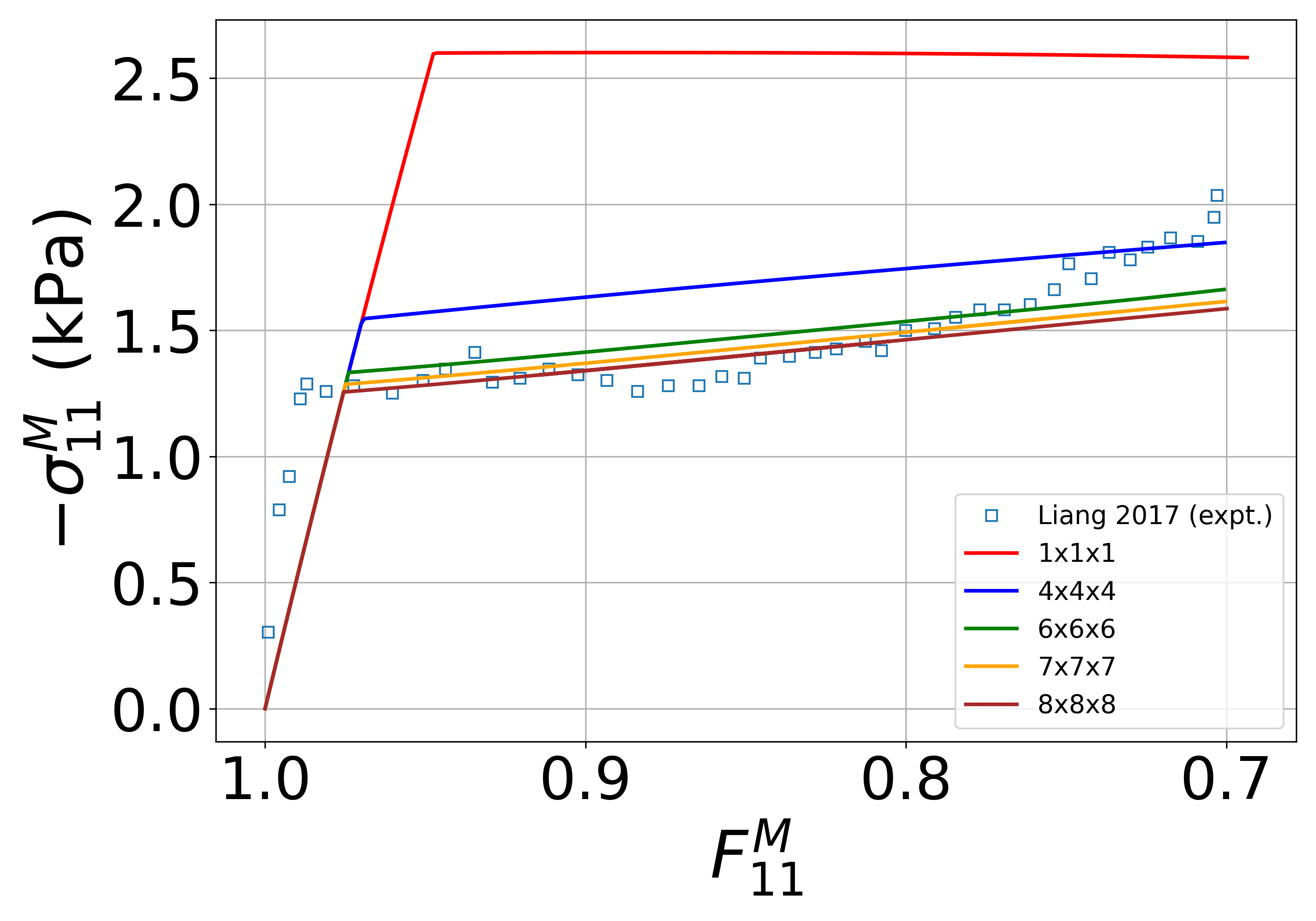}
    \caption{Effect of RVE size on the compressive response for $\frac{r}{l} = 0.075$.}
    \label{fig:8chain_uniaxial_compression_validation_rve_effect}
    \end{subfigure}
    \begin{subfigure}{{0.53\textwidth}}    
        \includegraphics[width=\textwidth]{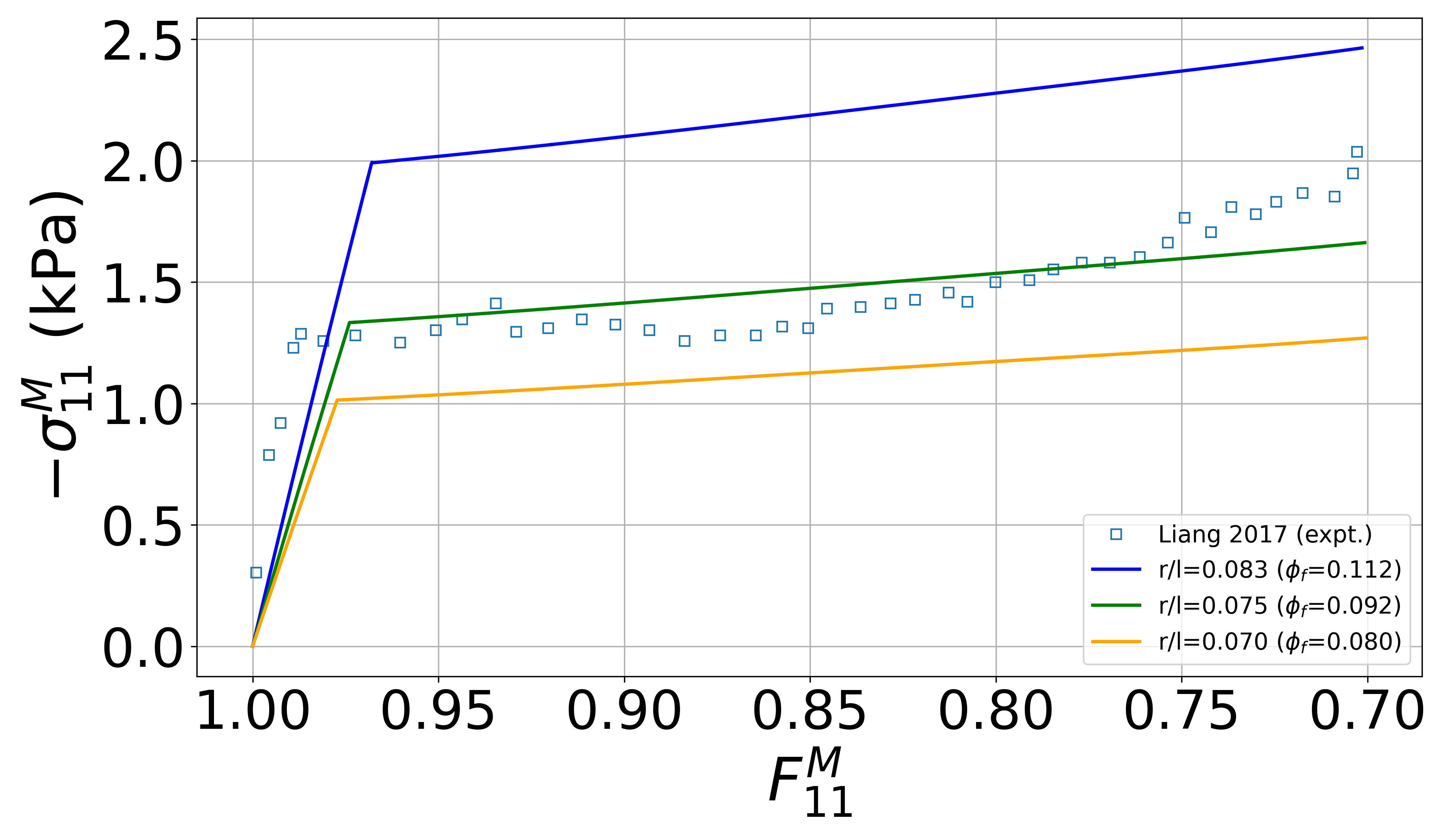}
        \caption{Effect of slenderness ratio $\frac{r}{l}$ on the compressive response of 6$\times$6$\times6$ 8-chain RVE.}
    \label{fig:8chain_uniaxial_compression_validation_r_by_l}
    \end{subfigure}
    \begin{subfigure}{\textwidth}
        \includegraphics[width=\textwidth]{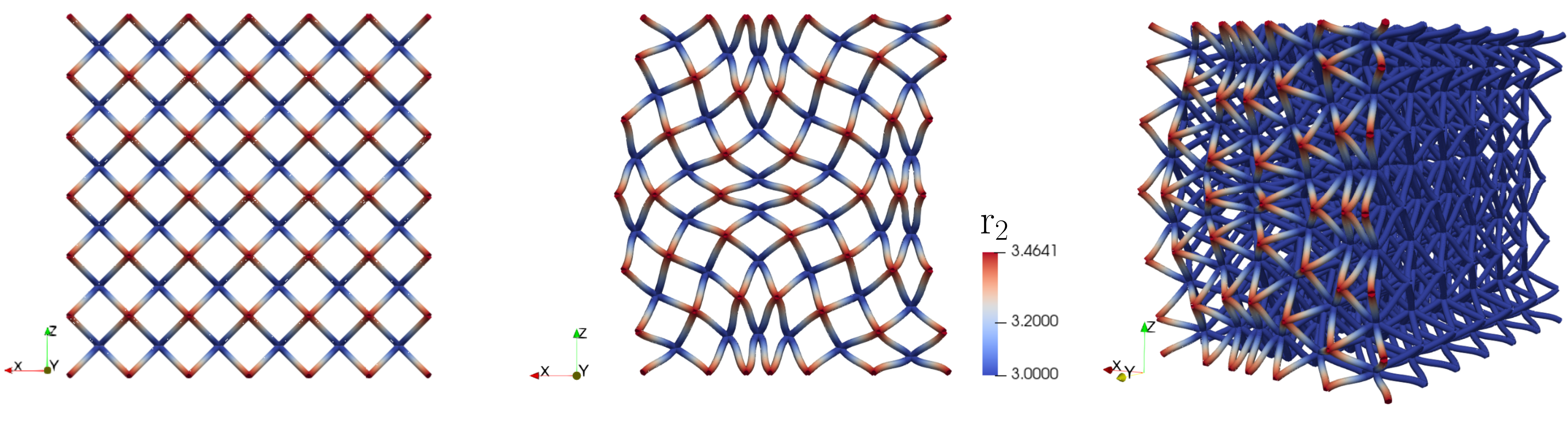}
        \caption{Undeformed (leftmost) and deformed configuration of 6$\times$6$\times6$ ($\frac{r}{l}=0.075$) RVE at $F^M_{11} = 0.85$. The color gradient shows y-component (or $\hat{\text{r}}_2$ component of rod's centerlines) of the top layer in the RVE.}
    \end{subfigure} 
    \begin{subfigure}{\textwidth}
            \centering
    \includegraphics[width=0.5\linewidth]{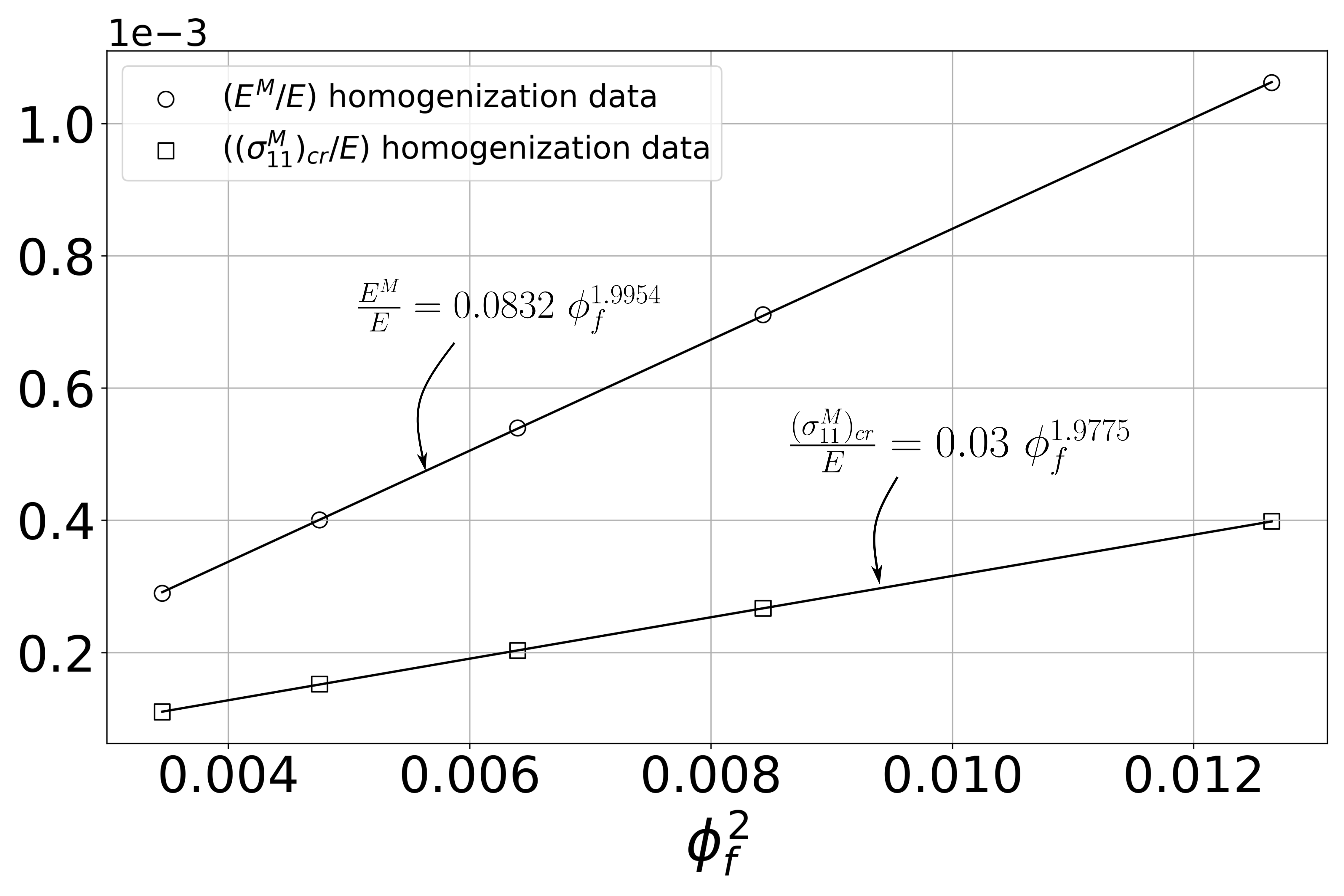}
    \caption{Foam-like behaviour of 8-chain RVE ($6\times6\times6$). Here $E^M$ is the Young's modulus at undeformed state and $(\sigma^M_{11})_{cr}$ is the stress at the buckling point. The curve fitting is done using the least square fitting technique. $N^{el} = 20$}
    \label{fig:8chain_6x6x6_foam_like}
    \end{subfigure}
    \label{fig:8chain_uniaxial_compression_validation}
    \caption{Comparison of uniaxial strain-driven compression of 8-chain RVE with experiment of \cite{liang2017phase} for blood clots. The parameters of fibrin fibers used here are: $E = 5~MPa$, $\text{Fibrin length }(l) = 1.32~\mu m$ and $\text{Fibrin diameter }(2r) = 0.220~\mu m$. The volume fraction for 8-chain RVE is given by: $\phi_f = \frac{8\pi r^2 l}{\hat{V}}$}
\end{figure}
Next, we investigate the macroscopic response of the present model by comparing it with the compression experiments on blood clots reported by \cite{liang2017phase}. Thus far, only a single unit cell has been considered in the RVE. However, \cite{herrnbock2022homogenization} showed that the post-buckling response depends on the size of the RVE. We also show it in Figure \ref{fig:8chain_uniaxial_compression_validation_rve_effect}. As the RVE size increases, although the unbuckled linear region remains the same, the buckling point is obtained earlier. Figure \ref{fig:8chain_uniaxial_compression_validation_r_by_l} considers a 6$\times$6$\times6$ RVE and shows that the compressive response depends also on the slenderness parameter $\frac{r}{l}$ or, in other words, on the volume fraction of the fibers ($\phi_f$) in the RVE. In Figure \ref{fig:8chain_6x6x6_foam_like}, for 8-chain RVE, we show the quadratic scaling behaviour of the Young's modulus in $[100]$ direction ($E^M = \frac{(\mathbb{C}^M_{1111}-\mathbb{C}^M_{1122})(\mathbb{C}^M_{1111}+2\mathbb{C}^M_{1122})}{\mathbb{C}^M_{1111}+\mathbb{C}^M_{1122}}$) and critical buckling Cauchy stress $(\sigma^M_{11})_{cr}$ with respect to $\phi_f$. This is very much like the open cell foam behaviour presented in \cite{gibson2003cellular} for foams and in \cite{liang2017phase} for fibrin networks. Finally, we consider stress-driven compression wherein the macroscopic first Piola-Kirchhoff stress is of the form:
\begin{equation}
    \mathbf{P}^M = \begin{bmatrix}
    P^M_{11} & 0 & 0 \\
    0 & 0 & 0 \\
    0 & 0 & 0
    \end{bmatrix},\quad \quad P^M_{11}<0
\end{equation}
\begin{figure}[h!]
    \centering
    \begin{subfigure}{0.6\textwidth}
    \centering
        \includegraphics[width=\textwidth]{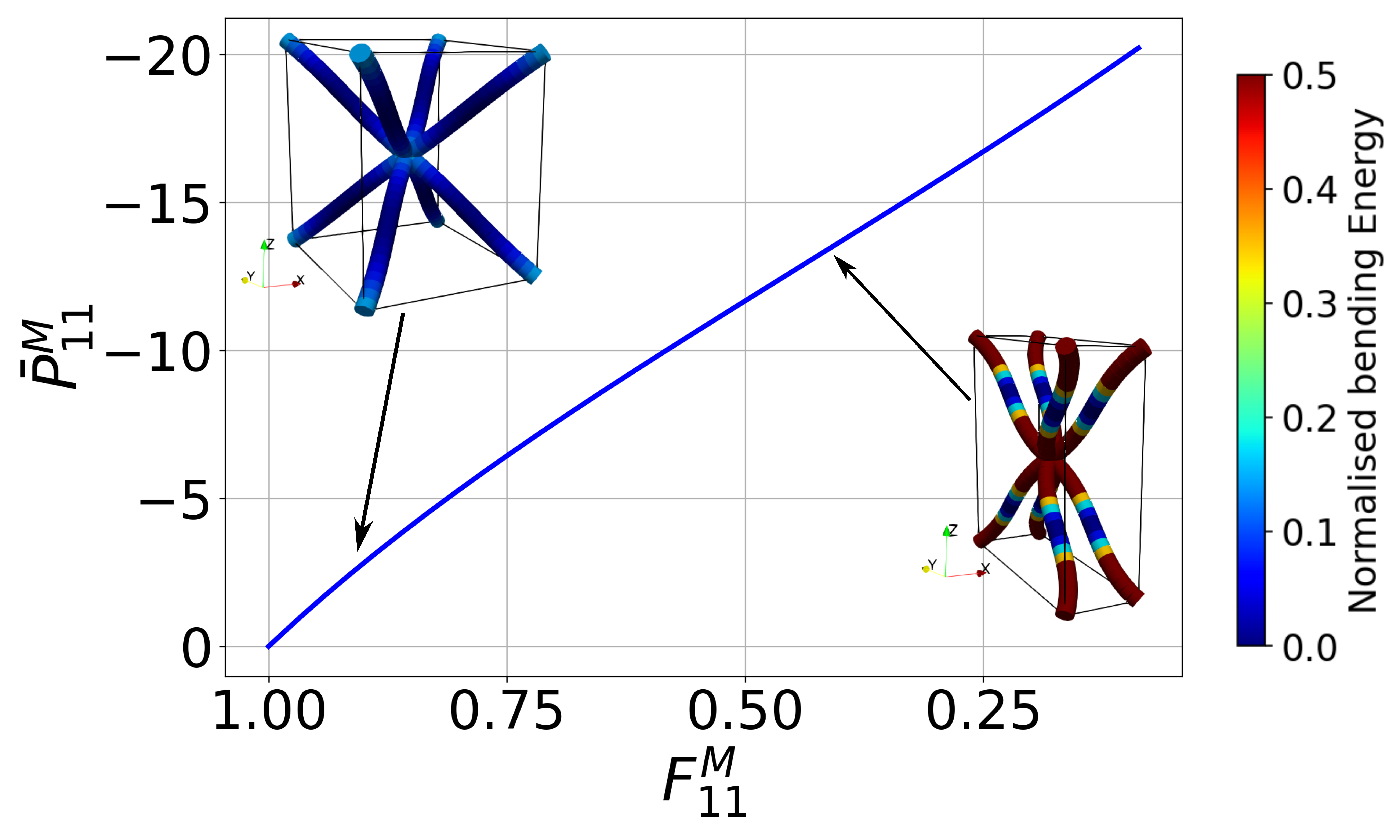}
        \caption{Stress-strain response of a 8-chain RVE under uniaxial compression}
        \label{fig:stress_driven_8chain_compression_P11_F11}
    \end{subfigure}
        \begin{subfigure}{0.49\textwidth}
        \includegraphics[width=\textwidth]{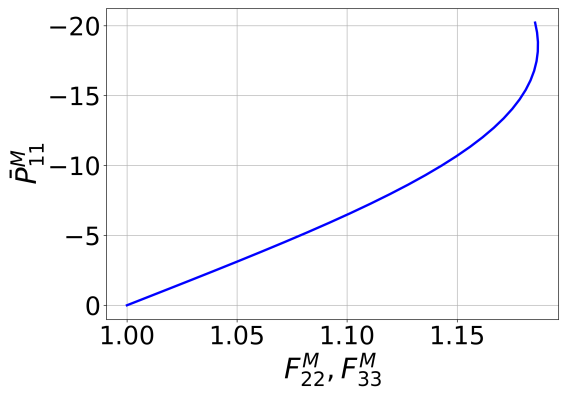}
        \caption{Stress-strain response of a 8-chain RVE under uniaxial compression}
        \label{fig:stress_driven_8chain_compression_P11_F22_F33}
    \end{subfigure}
    \begin{subfigure}{0.49\textwidth}
        \includegraphics[width=\textwidth]{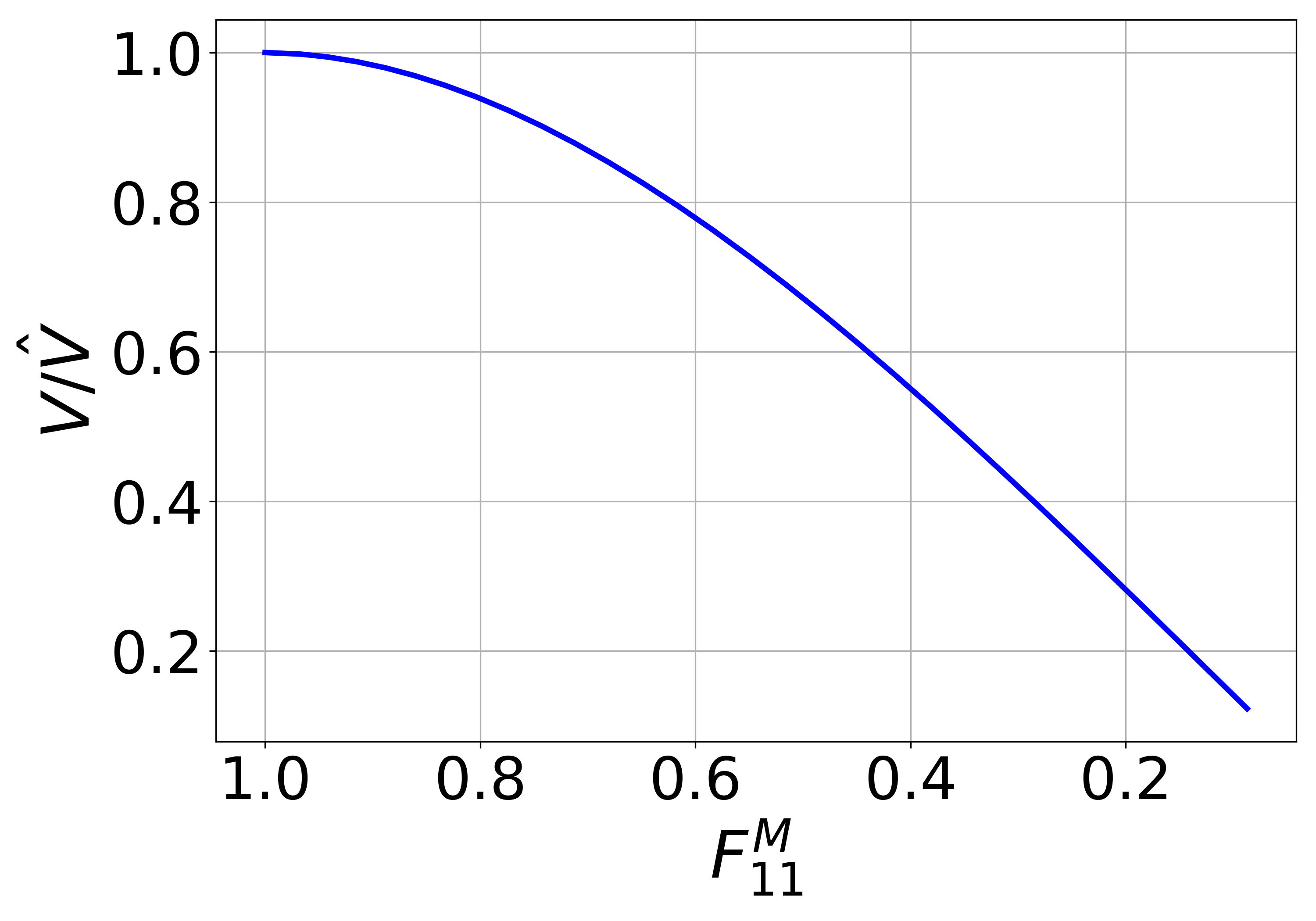}
        \caption{Volumetric stretch of a 8-chain RVE under stress driven uniaxial compression}
        \label{fig:stress_driven_8chain_compression_V_F}
    \end{subfigure}
    \caption{Stress driven compression of a 8-chain RVE. $N^{el} = 20$.}
    \label{fig:stress_driven_compression_8chain}
\end{figure}
\begin{figure}[h!]
      \centering
    \begin{subfigure}{\textwidth}
      \centering
        \includegraphics[width=0.6\textwidth]{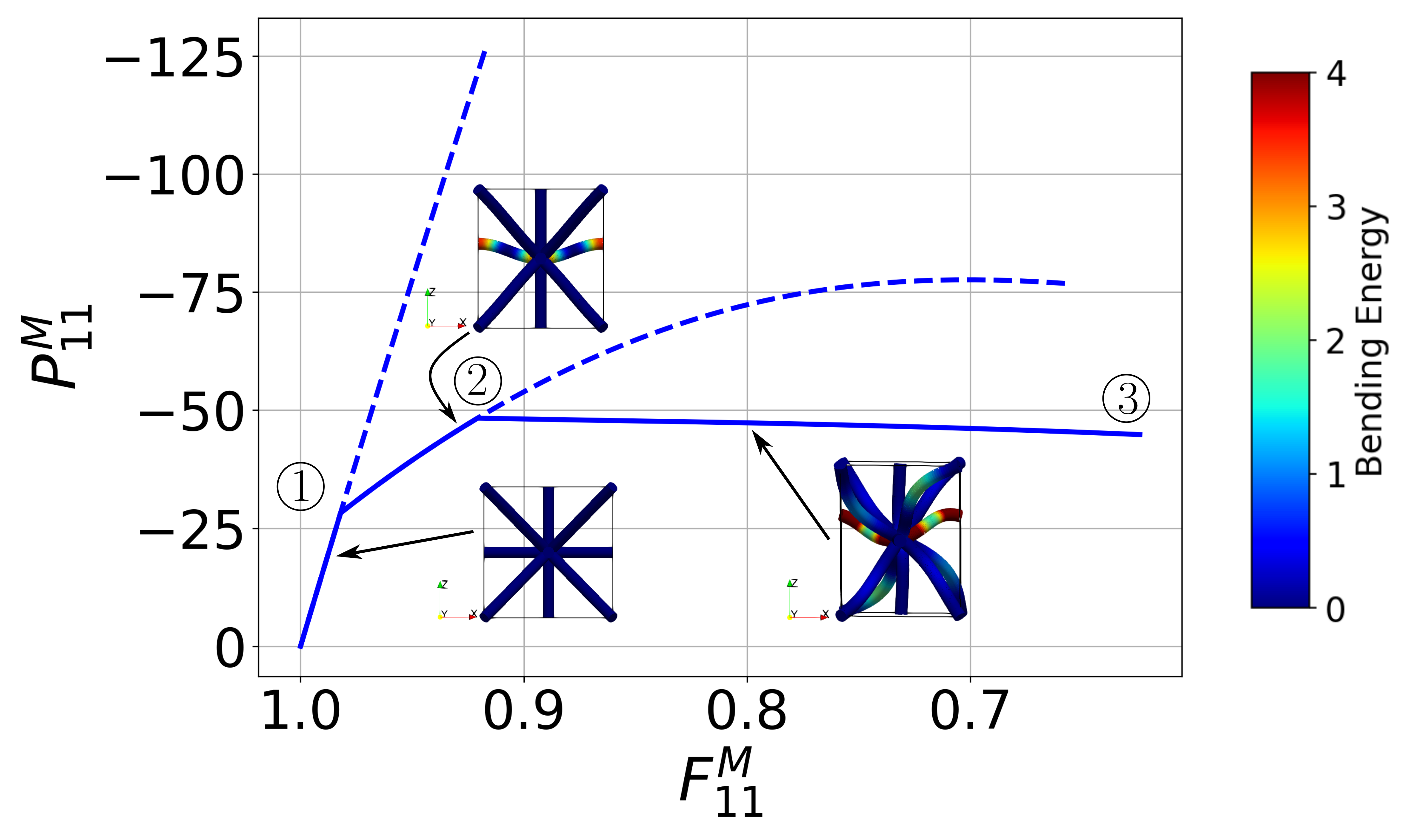}
        \caption{$P^M_{11}$ v/s $F^M_{11}$ response of 14-chain RVEs under stress-driven uniaxial compression. The branch 2-3 is stable under hard loading but unstable under soft loading stability criteria.}
        \label{fig:stress_driven_compression_14chain_P11_F11}
    \end{subfigure}
    \begin{subfigure}{0.49\textwidth}
        \includegraphics[width=\textwidth]{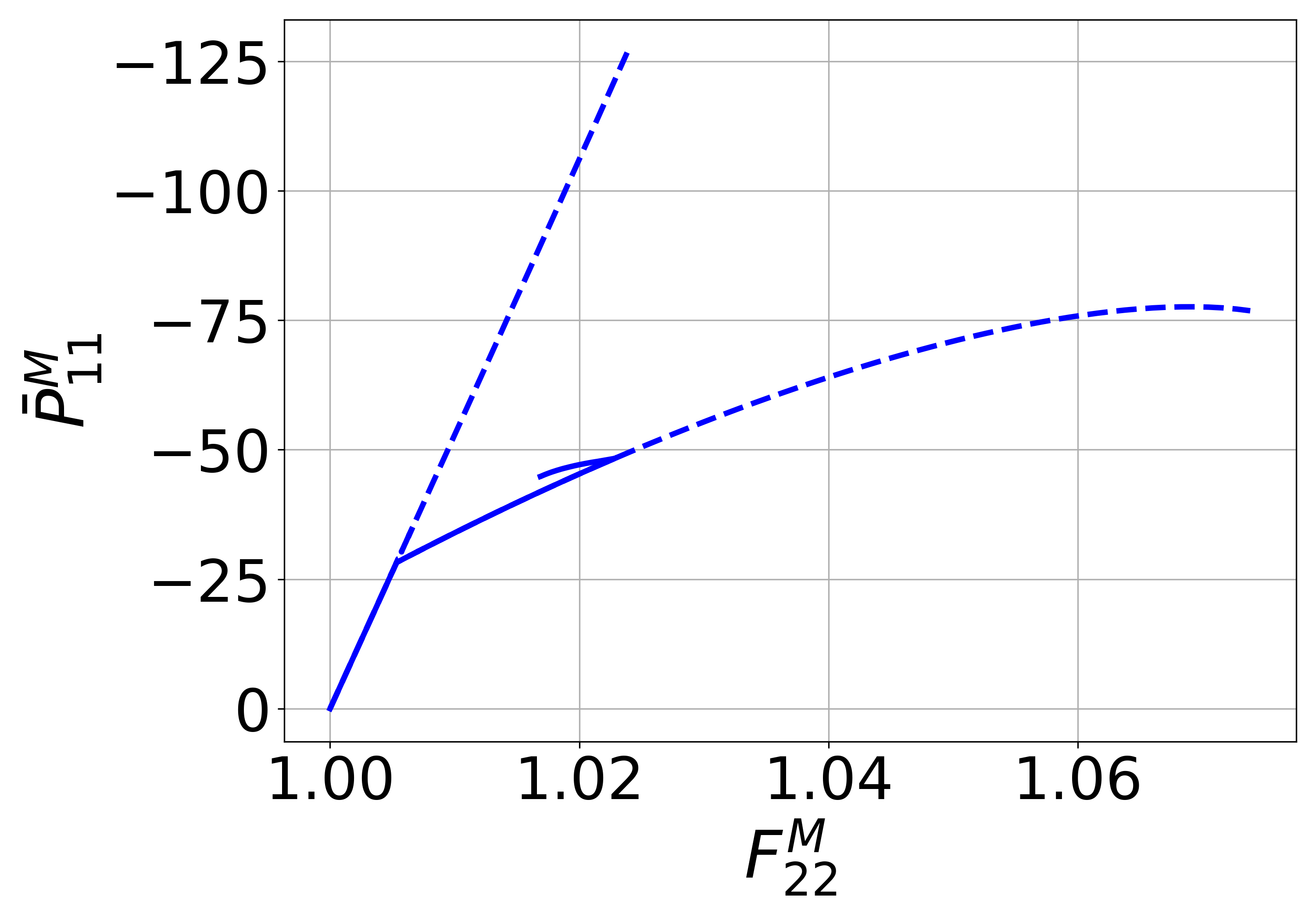}
        \caption{$P^M_{11}$ v/s $F^M_{22}$ response of 14-chain RVEs under uniaxial compression}
        \label{fig:stress_driven_compression_14chain_P11_F22}
    \end{subfigure}
    \begin{subfigure}{0.49\textwidth}
        \includegraphics[width=\textwidth]{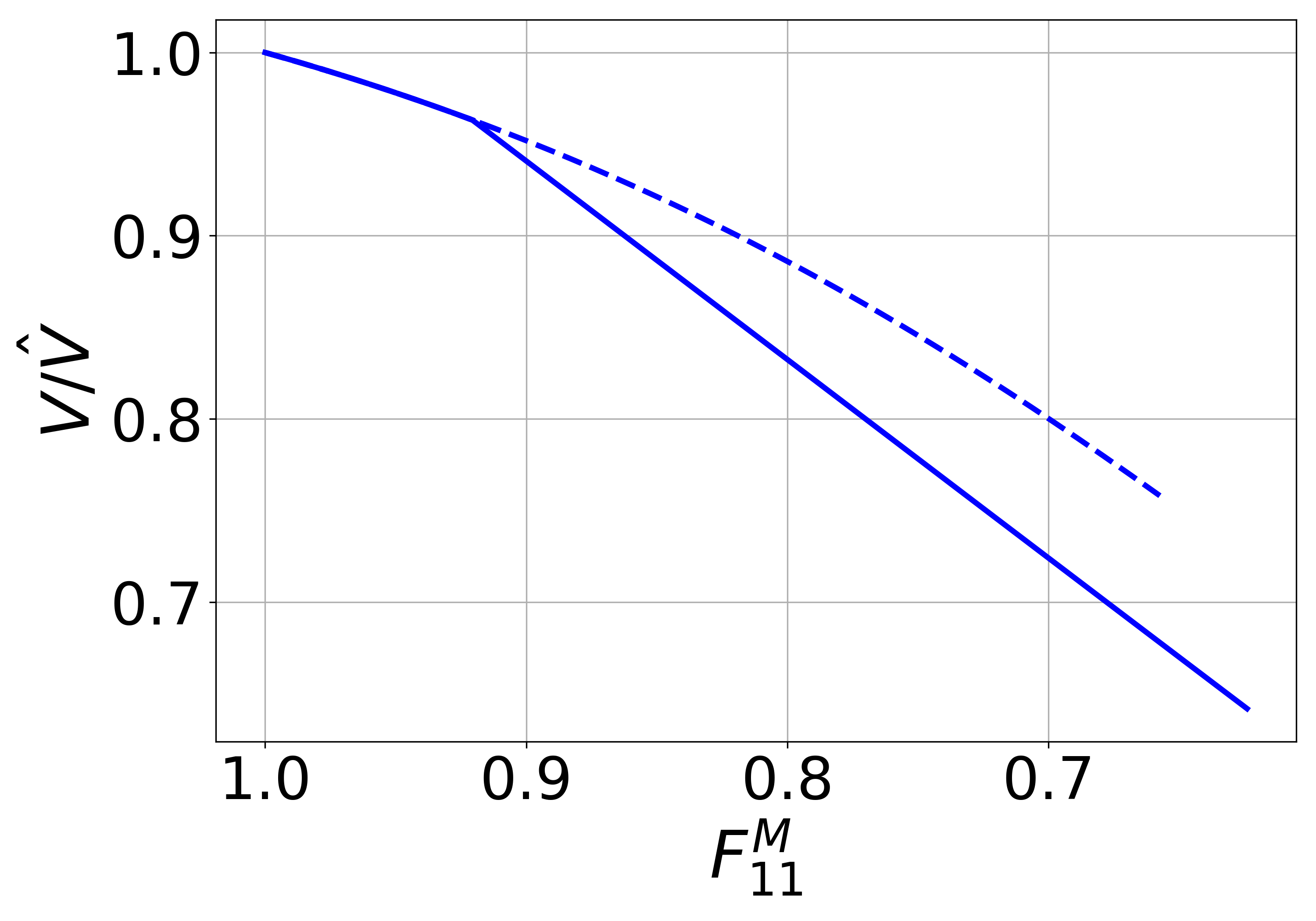}
        \caption{Volume change of 14-chain RVEs under stress-driven uniaxial compression}
        \label{fig:stress_driven_compression_14chain_VV}
    \end{subfigure}
    \begin{subfigure}{0.49\textwidth}
        \includegraphics[width=\textwidth]{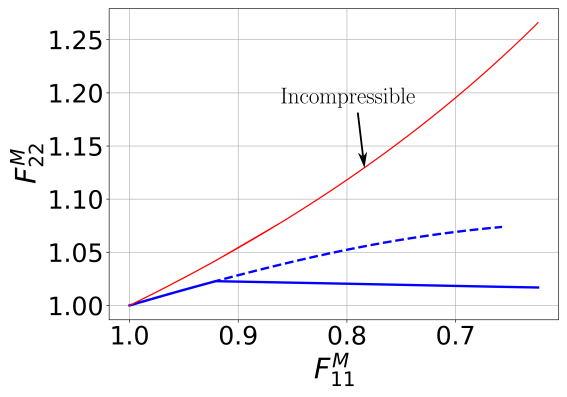}
        \caption{Lateral strain v/s longitudinal strain}
        \label{fig:stress_driven_compression_14chain_V_F}
    \end{subfigure}
    \caption{Stress-strain curve for 14-chain RVE under stress driven compression. $N^{el} = 50$}
    \label{fig:stress_driven_compression_14chain_P11_v_F11}
\end{figure}
 In Figure \ref{fig:stress_driven_compression_8chain}, the response of the 8-chain RVE under stress driven compression is shown. Similar to the stress driven tension case, the traction free boundary condition in the transverse direction ensures that the response is bending dominated. It remains so until fully compressed. This results in an approximately linear $\bar{P}^M_{11}$ vs $F^M_{11}$ curve (see Figure \ref{fig:stress_driven_8chain_compression_P11_F11}). The RVE expands in the two transverse directions but a large decrease in volume is also observed as shown in Figures \ref{fig:stress_driven_8chain_compression_P11_F22_F33} and \ref{fig:stress_driven_8chain_compression_V_F}, respectively. The stress-driven compression response of the 14-chain RVE is different due to the presence of straight rods. In this case, the response is initially stretching dominated until a buckling point is reached where two eigenvalues become zero simultaneously. Note that this buckling point is at a much lower stress than the strain-driven compression (see figure \ref{fig:14chain_strain_driven_compression}). Similar to the strain-driven case, the mode shapes correspond to buckling of only the rods along the compression direction and the two zero eigenvalues are due to the rod's isotropic cross-section. Switching the branch using one of the mode shapes, a second linear regime with a reduced slope is obtained. The other null eigenvalue remains zero along this path. The response in this region is still stretching-dominated because of the unbuckled inclined rods. Upon further compression, another buckling point is obtained on this solution path. At this point, two zero eigenvalues are obtained that correspond to mode shapes involving the buckling of inclined rods. The two mode shapes only differ in the plane of buckling of the inclined rods. We apply branch switching considering one of these mode shapes and obtain the third regime in which the $\bar{P}^M_{11}$ vs $F^M_{11}$ curve suddenly flattens (see Figure \ref{fig:stress_driven_compression_14chain_P11_F11}). We observe that the inclined rods have buckled here and the previously buckled straight rod undergoes non-planar deformation. Similar to the case of strain-driven case, stress-driven compression of 14-chain RVE leads to a solution path that is trilinear in $\bar{P}^M_{11}$ vs $F^M_{11}$. Here, the two linear regions (paths 0-1 and 1-2) remain stable under both hard and soft loading criteria. However, the flattened region (path 2-3) is unstable under soft loading criterion. The rate of volume shrinkage also increases after the second bifurcation point (see Figure \ref{fig:stress_driven_compression_14chain_VV}). It is mainly due to the fact that $F_{22}$ and $F_{33}$ which were both increasing earlier start to decrease after second bifurcation point - see Figures \ref{fig:stress_driven_compression_14chain_P11_F22} and \ref{fig:stress_driven_compression_14chain_V_F}.
\subsection{Simple shear}
Biopolymer networks undergo shear deformations during their function. For example, fibrin networks in blood clots rupture under hydrodynamic shear flow in a blood vessel leading to thrombotic embolization \citep{ramanujam2024mechanics}. Furthermore, biopolymer networks under shear show a unique effect known as the negative normal stress or reverse Poynting effect\footnote{The effect when the specimen under simple shear contracts and a tensile force is needed to be applied on it to maintain zero normal strain is known as the reverse or negative Poynting effect.} \citep{horgan2017poynting}. This has been shown to provide a stabilizing effect to the network \citep{janmey2007negative,destrade2015dominant,licup2015stress,horgan2017poynting,vahabi2018normal}. Therefore, we consider the shear deformation of 8- and 14-chain RVEs in this section. Considering strain-driven case (representative of simple shear), the macroscopic deformation gradient is taken to be of the form:
\begin{equation}
    \mathbf{F}^M = \begin{bmatrix}
    1 & F^M_{12} & 0 \\
    0 & 1 & 0 \\
    0 & 0 & 1
    \end{bmatrix},\quad \quad F^M_{12}>0.
\end{equation}
\begin{figure}[h!]
\centering
\begin{subfigure}{0.49\textwidth}
\centering
    \includegraphics[width=\textwidth]{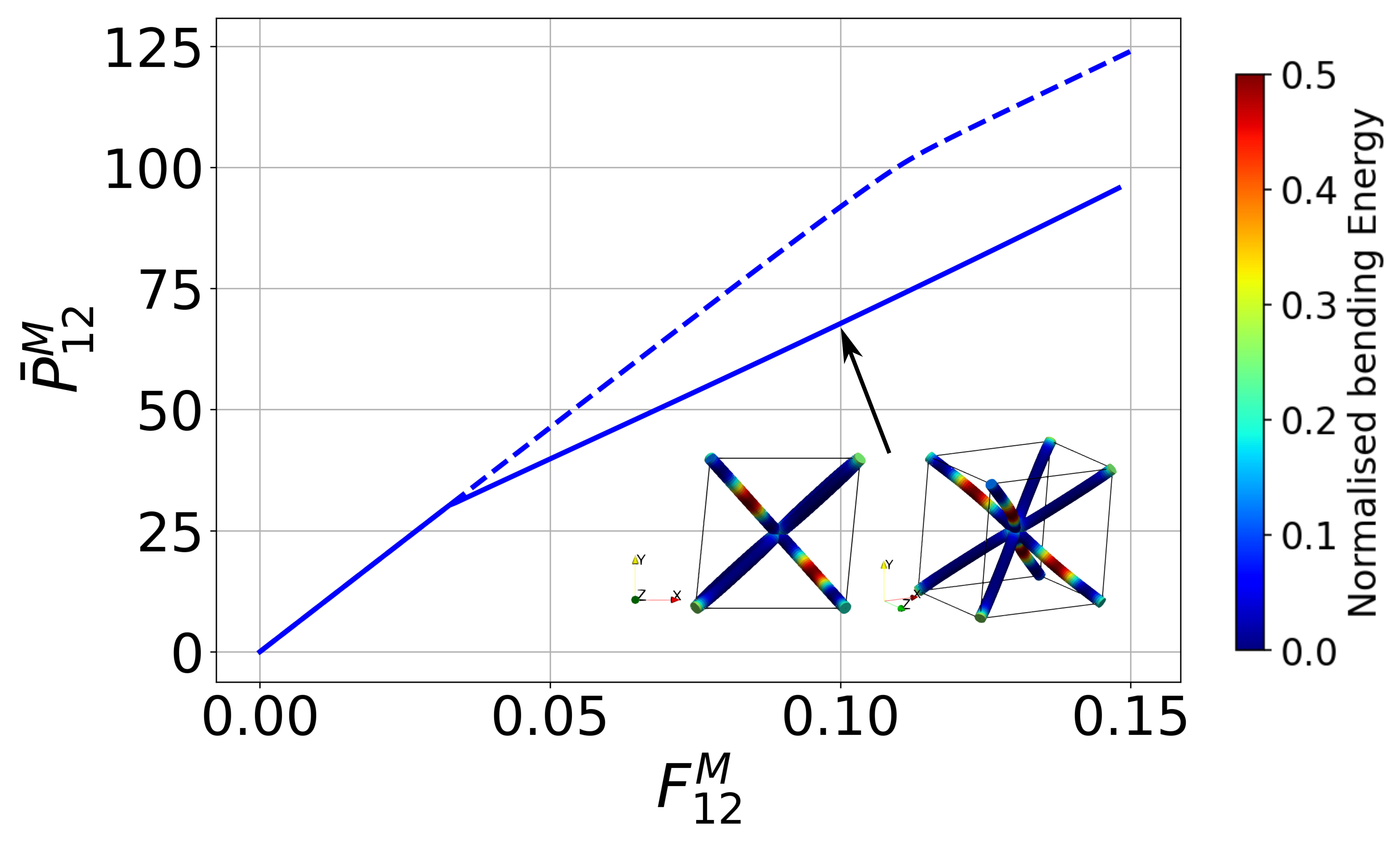}
    \caption{Shear stress v/s shear strain}
    \label{fig:simple_shear_8chain_P12_F12}
\end{subfigure}
\begin{subfigure}{0.40\textwidth}
\centering
    \includegraphics[width=\textwidth]{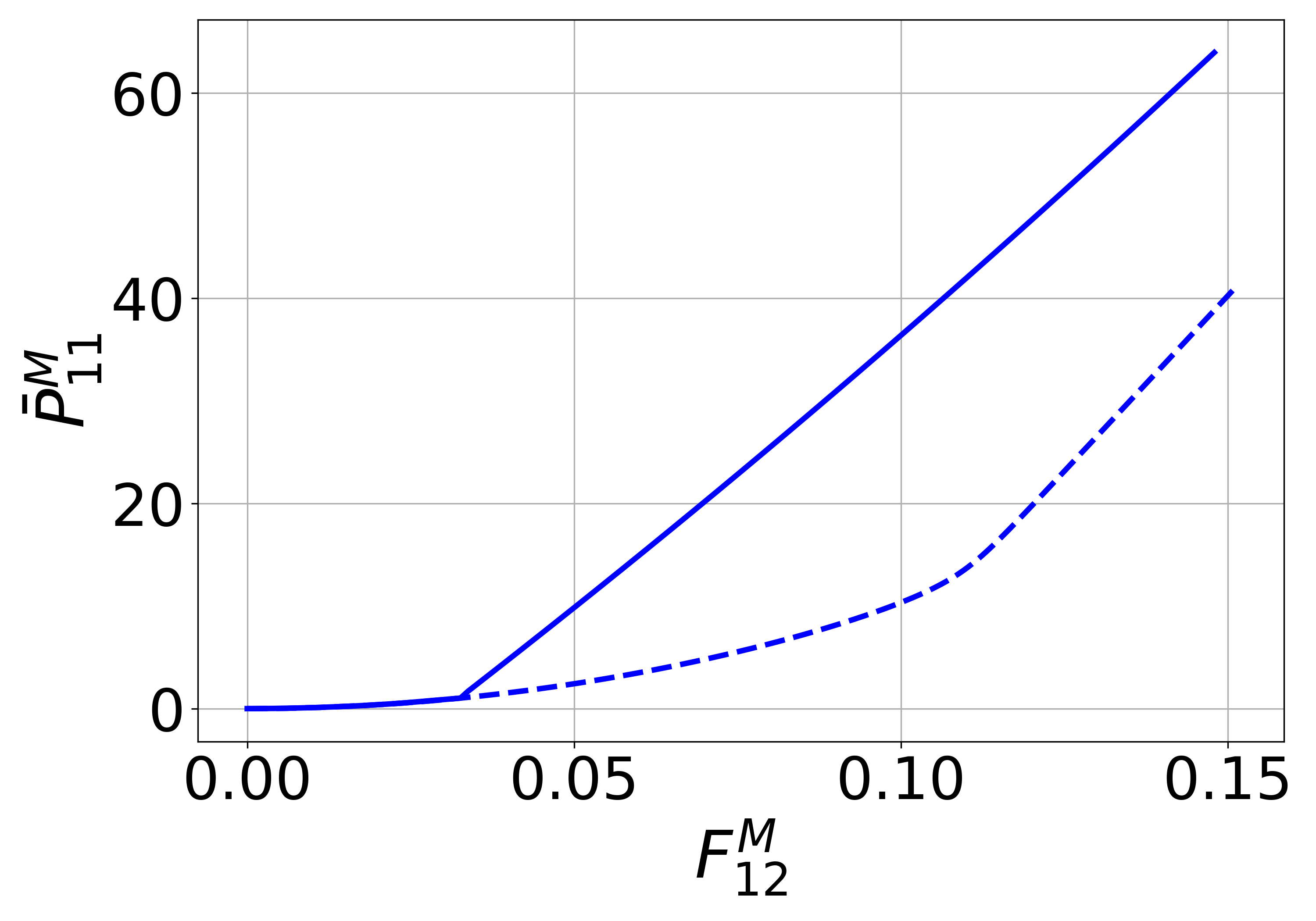}
    \caption{Transverse normal stress v/s shear strain.}
    \label{fig:simple_shear_8chain_P11_F12}
\end{subfigure}
    \caption{Simple shear response of 8-chain RVE. A deformed configuration is shown on the buckled branch. $N^{el}=20$}
    \label{fig:simple_shear_8chain}
\end{figure}
\begin{figure}[h!]
    \centering
    \begin{subfigure}{0.49\textwidth}
        \includegraphics[width=\textwidth]{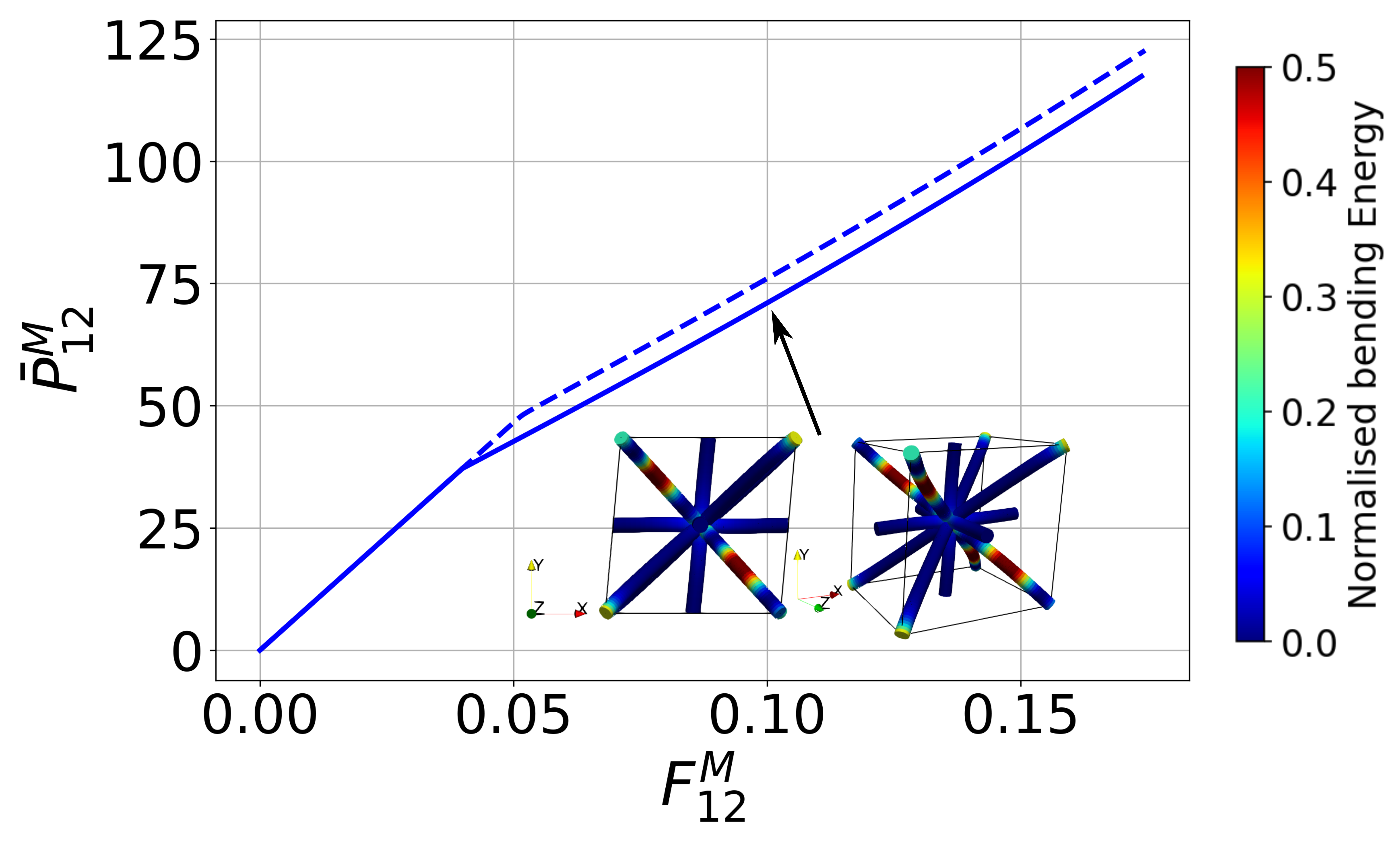}
        \caption{Shear stress v/s shear strain}
    \label{fig:simple_shear_14chain_P12_F12}
    \end{subfigure}
    \begin{subfigure}{0.40\textwidth}
        \includegraphics[width=\textwidth]{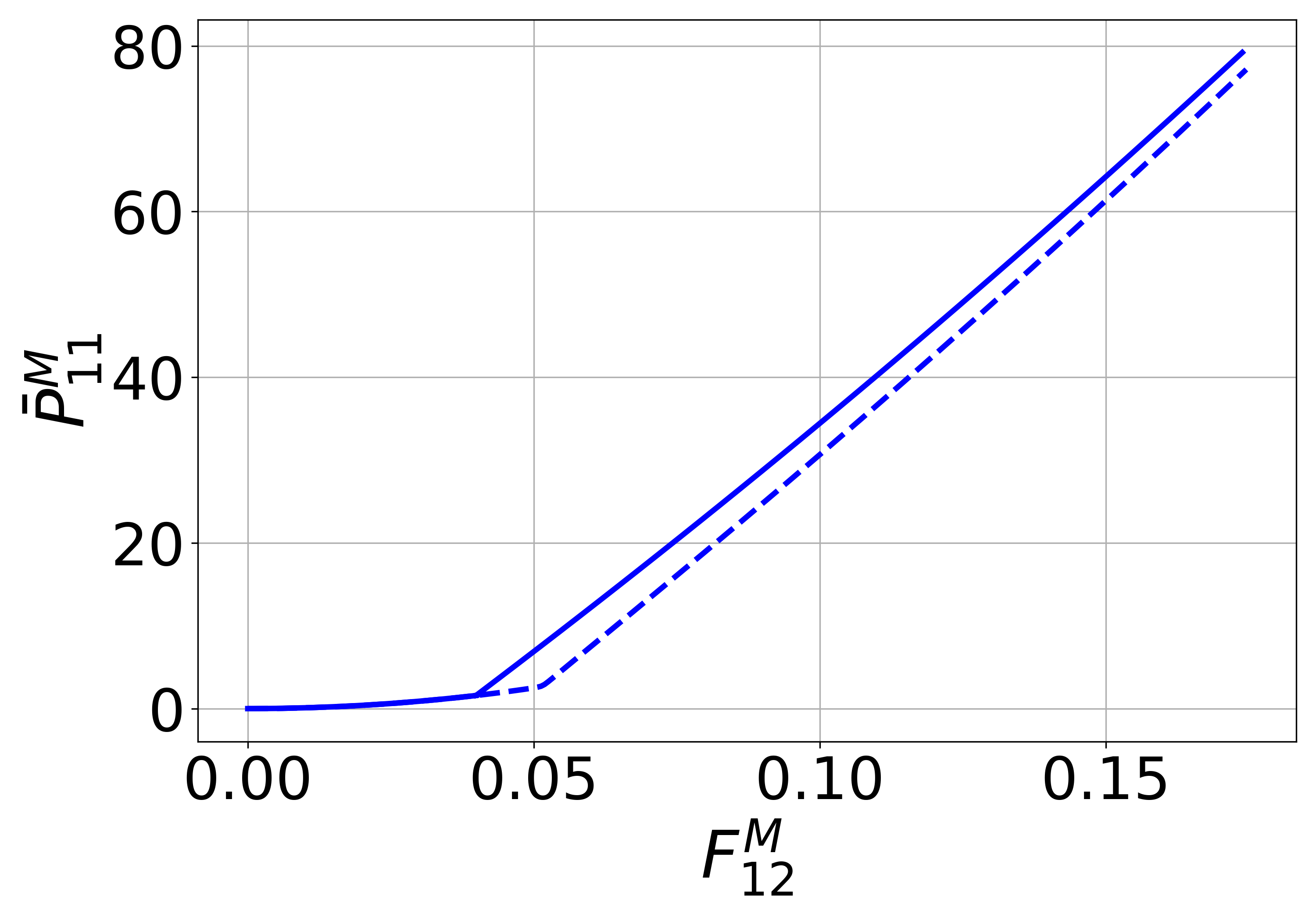}
        \caption{Transverse normal stress v/s shear strain}
    \label{fig:simple_shear_14chain_P11_F12}
    \end{subfigure}
    \caption{Simple shear response of 14-chain RVE. A deformed configuration is shown on the buckled branch. Note that, very close to the buckling point, the unstable branch (dashed line) bifurcates to another branch and remains unstable. $N^{el} = 20$}
    \label{fig:simple_shear_14chain}
\end{figure}
Figures \ref{fig:simple_shear_8chain} and \ref{fig:simple_shear_14chain} show the response of 8- and 14-chain RVEs under macroscopic simple shear. Their responses are similar. As shear strain is applied to the RVE, a bifurcation point is obtained. Here only one eigenvalue is zero for both RVEs and branch switching using the corresponding mode shape leads to softening of the stress-strain response (see Figures \ref{fig:simple_shear_8chain_P12_F12} and \ref{fig:simple_shear_14chain_P12_F12}). This is due to the buckling of the four rods along the body diagonals which undergo shortening. The other four rods undergo lengthening. \cite{zakharov2024clots}, through experiments and simulations on blood clots, also observed such a softening behaviour and attributed this to fiber buckling. The bilinear response here is stable under both hard and soft loading stability criteria. In addition to shear stress $P^M_{12}$, a normal tensile stress, $P^{M}_{11}$, is also generated as shown in Figures \ref{fig:simple_shear_8chain_P11_F12} and \ref{fig:simple_shear_14chain_P11_F12}. Transverse stresses $P^M_{22}$ and $P^M_{33}$ are also generated in the case of both RVEs. The tensile normal stress is very low initially but rises rapidly upon buckling of microscale rods. This is the well known reverse Poynting effect. Buckling of rods induces a contraction in the normal direction and since the periodic boundary condition needs to be maintained such that $F^M_{11} = 1.0$, a tensile normal stress is generated.   Moreover, the magnitude of normal stress is of similar order as shear stress. This has also been observed in shear experiments on polymer hydrogels \citep{vahabi2018normal} and also captured in the experiments and analytical model proposed in \citet{ramanujam2024mechanics} on blood clots. The model in \citet{ramanujam2024mechanics} accounted for fiber buckling in a simple phenomenological way but here the effects of fiber buckling are captured in a rigorous way.

\pagestyle{plain}

\noindent
\section{Conclusions} \label{sec:conclusions}
In this work, we have presented an analysis of the 8- and 14-chain periodic idealizations of fiber networks using a non-affine finite strain computational homogenization technique with stability analysis. An imperfection-free, non-linear path following approach was adopted to obtain post-buckled configurations of the RVEs. The stability of these solutions was analysed using hard and soft loading criteria. We employed both stress- and strain-driven deformation modes, and the results obtained reveal physical insights into how the idealized 8- and 14-chain models can capture the behaviours observed in complex biopolymer networks such as collagen or fibrin. The stress- and strain-driven approaches essentially manifest in the boundary conditions on the microscale problems and thus yield qualitatively different responses. We first highlight this in the case of extension of 8-chain RVEs, wherein the strain-driven extension is stretching-dominated, while stress-driven (uniaxial tension) results in a strain-stiffening behaviour. This is because, in case of stress-driven approach, the response is initially bending-dominated, which transitions to the stretching-dominated mode when the microscale rods align along the stretching direction. The strain-stiffening behaviour is characteristic of many polymer/biopolymer networks. We observe a non-linear Poisson effect that results in dramatic volume shrinkage in the case of stress-driven extension of both 8- and 14-chain RVEs. We show that the volume shrinkage in the case of 14-chain RVEs is a consequence of buckling and bending of fibers. We validate the homogenized uniaxial tension response of the 8-chain RVE with both analytical and experimental results on collagen. By considering helical fibers, we also highlight, both analytically and numerically, the role of fiber-level material model on the overall RVE response. Next, in case of compression, we show that buckling of fibers governs the flattening of the stress-strain curve as observed in experiments on biopolymer networks. By considering an appropriate size of the 8-chain RVE, we showed that the homogenization approach finds a good agreement with compression experiment results on blood clots. Furthermore, our non-linear branch following approach leads us to a trilinear compression response (both strain- and stress-driven) in case of a 14-chain RVE. Finally, in the case of simple shear, we again show that buckling plays an important role. It amplifies the reverse Poynting effect that is observed in shear experiments of several biopolymer networks, including fibrin gels.

Here, we focused on interesting non-linear, non-affine phenomena that are observed in biomaterials. Architected materials that aim to describe such behaviours can also be designed. We also studied purely elastic mechanical behaviour. However, biopolymer networks have also shown inelastic behaviours, such as plastic deformation due to permanent fiber elongation \citep{munster2013strain,ban2018mechanisms}. The homogenization model can be extended to incorporate such effects. Furthermore, the fibers in a fiber network undergo contact, especially under compression in the densification regime \citep{liang2017phase}. Therefore, including an efficient contact scheme is important to study the full stress-strain response. Likewise, the macroscopic homogenization model considered here is local and thus cannot capture the net bending moments that are generated in the microscale. Secondly, the stability analysis does not include perturbations of all wavelengths. For this, the phonon stability needs to be checked as well \citep{elliott2006stability}. We do consider the effect of RVE size in case of uniaxial compression to have a realistic comparison with the experimental data. Finally, the homogenization model can be incorporated in an $\text{FE}^2$ scheme or a machine learning model, utilizing it as a material model engine in a full-scale problem \citep{kalina2023fe}.

\noindent
\section*{Acknowledgements} PKP acknowledges partial support through a grant from the US National Science Foundation, NSF DMR 2212162.
\noindent

\biboptions{semicolon}
\bibliography{references_1d_3dh}

\appendix
\pagestyle{plain}
\section{Constraint sets of 8- and 14-chain RVEs}\label{appendix:14chain_rve_constraint_set}
\begin{figure}[h]
    \centering
    \includegraphics[width=0.5\linewidth]{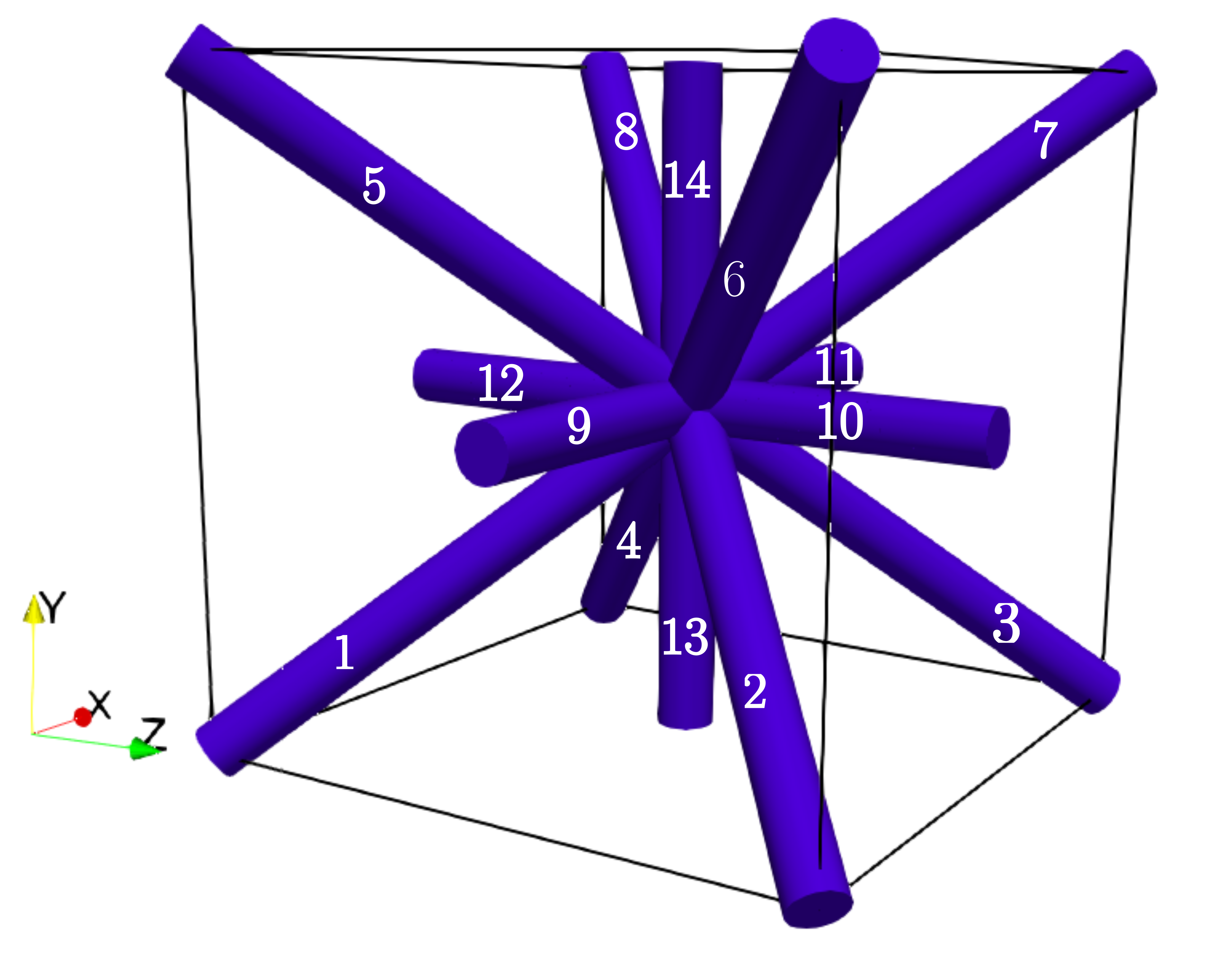}
    \caption{A 14-chain RVE with single unit cell with rod ids.}
    \label{fig:14chain_rve_with_rod_numbers}
\end{figure}
Consider the 14-chain RVE in Figure \ref{fig:14chain_rve_with_rod_numbers} with rod identification numbers annotated on each rod. The $+$ end for all rods is the one that lies on the boundary of the RVE box. The internal joint constraint for this RVE is:
\begin{align}
    \mathcal{J}_1 = &\{(1,-),(2,-),(3,-),(4,-),(5,-),(6,-),(7,-),(8,-),(9,-),(10,-),\nonumber\\
    &(11,-),(12,-),(13,-),(14,-)\}
\end{align}
Here the master node is $(1,-)$. The periodic sets of this RVE are:
\begin{align}
    \mathcal{P}_1 &= \{(1,+),(2,+),(3,+),(4,+),(5,+),(6,+),(7,+),(8,+)\}\nonumber\\
    \mathcal{P}_2 &= \{(9,+),(11,+)\}\nonumber\\
    \mathcal{P}_3 &= \{(10,+),(12,+)\}\nonumber\\
    \mathcal{P}_4 &= \{(13,+),(14,+)\}
\end{align}
And the global constraint set is
\begin{align}
    \mathcal{G} = &\{(1,+),(2,+),(3,+),(4,+),(5,+),(6,+),(7,+),(8,+)\}
\end{align}
In case of 8-chain RVEs, the rod numbers 9 to 14 will be removed from above sets.
\section{Volumetric stretch in 14-chain RVE}\label{sec:fractional_volume_change_14chain}
In section \ref{sec:numerical_examples_extension}, the increasing-decreasing trend of volumetric stretch of stress-driven tension of 14-chain RVE is discussed. Here, we derive the slope of volumetric stretch with respect to the stretch of the RVE and show that it will always show an increasing trend initially. Working in the principle stress space, the volumetric stretch of the 14-chain RVE is given by
\begin{align}
    \frac{V}{\hat{V}} = \lambda_1\lambda_2\lambda_3 .
\end{align}
Considering $\hat{V} = 1.0$ and $\lambda_2 = \lambda_3$ for the 14-chain RVE under uniaxial tensile stress $P_{11}^M$, we have
\begin{align}
    V = \lambda_1\lambda_2^2.
\end{align}
The volume change with respect to the induced stretch $\lambda_1$ (due to the applied tension $P^M_{11}$) is given by
\begin{align}\label{eq:dV_dlamba1}
    \frac{dV}{d\lambda_1}\bigg |_{\lambda_1=1} = \lambda_2^2 + 2\lambda_1\lambda_2\frac{d\lambda_2}{d\lambda_1}.
\end{align}
Now, since we only need the initial slope, we need the above quantity at zero stress i.e. $\lambda_1 = 1.0$. The initial
deformation of the 14-chain RVE is assumed to be mostly stretching dominated due to
the presence of straight rods. This means that the force in the body-diagonal rods as well as the straight transverse rods is along the centerline of the rod. Let us denote the force in the body--diagonal rods to be $F_1$ and $F_2$ is the force along the transverse rods. The force balance (in the small initial strain regime) of the RVE along the transverse direction (say $\textbf{e}_2$) is given by:
\begin{align}\label{eq:14chain_vol_fraction_force_bal}
    4F_1 \alpha + F_2 = 0 
\end{align}
where $\alpha$ is the direction cosine of the four inclined rods along the $\textbf{e}_2$ direction. We only need to consider one-half of the RVE because of symmetry. Assuming a linear force-stretch relationship of the fibers, we have
\begin{align}\label{eq:force}
    F_1 &= EA(\lambda_c -1),\nonumber\\
    F_2 &= EA(\lambda_2-1) .
\end{align}
Here, $\lambda_c$ is the stretch experienced by the inclined rods and it is given by
\begin{align}
    \lambda_c = \sqrt{\frac{\lambda_1^2+2\lambda_2^2}{3}}.
\end{align}
Its derivative with respect to longitudinal stretch $\lambda_1$ is as follows:
\begin{align}\label{eq:stretch_der}
    \frac{\partial\lambda_c}{\partial\lambda_1} &= \frac{1}{\lambda_c}\left(\lambda_1 + 2\lambda_2\frac{d\lambda_2}{d\lambda_1}\right).
\end{align}
Substituting \cref{eq:force} in \cref{eq:14chain_vol_fraction_force_bal}, we get
\begin{align}\label{eq:14chain_vol_fraction_force_bal_red}
    4(\lambda_c-1)\alpha + (\lambda_2-1) = 0.
\end{align}
Now, taking the derivative of \cref{eq:14chain_vol_fraction_force_bal_red}, we get
\begin{align}\label{eq:force_bal_derivative_1}
    4\frac{\partial\lambda_c}{\partial \lambda_1}\bigg |_{\lambda_1=1} \alpha + 4(\lambda_c-1)\frac{\partial\alpha}{\partial \lambda_1}\bigg |_{\lambda_1=1} + \frac{d\lambda_2}{d\lambda_1}\bigg |_{\lambda_1=1} = 0.
\end{align}
At zero stress or $\lambda_1 = 1.0$, we have
\begin{align}\label{eq:initial_state}
    \lambda_1 = \lambda_2 = \lambda_3 = \lambda_c = 1,\quad\text{and}\quad \alpha = \frac{1}{\sqrt{3}}.
\end{align}
From \cref{eq:stretch_der,eq:force_bal_derivative_1}, this results in 
\begin{align}
    \frac{d\lambda_2}{d\lambda_1}\bigg |_{\lambda_1=1} = \frac{-1}{2+\frac{3\sqrt{3}}{4}} \approx -0.303.
\end{align}
Substituting the above quantity in \cref{eq:dV_dlamba1} and using \cref{eq:initial_state}, we get
\begin{align}
    \frac{dV}{d\lambda_1}\bigg |_{\lambda_1=1} \approx 0.39 >0.
\end{align}
Therefore the volume will increase initially in case of 14-chain RVEs.
\section{Uniaxial tension with straight and helical fibers}\label{sec:uniaxial_tension_appendix}
In this section, we recall the 8-chain network model of \cite{brown2009multiscale} for uniaxial tension, i.e., traction free lateral boundary conditions. In their model, the macroscopic Cauchy stress tensor component
\begin{align}
\sigma^M_{11} =
\left(\frac{\lambda_1}{\lambda_*^{2}} - \frac{1}{\lambda_1}\right)
\frac{\nu L_f F(\lambda_c)}{3 \lambda_c},\quad \lambda_c = \sqrt{\frac{\lambda_1^2 + 2\lambda_*^2}{3}},\quad \lambda_{*} = \lambda_2 = \lambda_3 
\end{align}
where $\nu$ is the number of fibers in the unit cell per unit reference volume of the unit cell, $L_f$ is the fiber end-to-end length and $F(\lambda_c)$ is the force-stretch relation for a single fiber in the 8-chain. 
\subsection{Straight fiber 8-chain model}\label{sec:uniaxial_tension_straight_appendix}
If the fiber is assumed to be straight, then its force-stretch relationship is given by
\begin{align}\label{eq:straight_fiber_force_stretch_relation}
    F(\lambda_c) = EA(\lambda_c-1).
\end{align}
Substituting the above equation and assuming incompressibility, i.e., $\lambda_{\star}=\sqrt\frac{1}{\lambda_1}$ the following form of macroscopic true stress is obtained:
\begin{align}\label{eq:straight_fiber_8_chain}
    \sigma^M_{11} = \frac{\nu EAL_f}{3}\bigg(\frac{\lambda_1^3-1}{\lambda_1}\bigg)\bigg(1-\frac{1}{\lambda_c}\bigg)\quad\text{and} \quad \lambda_c = \sqrt{\frac{\lambda_1^3+2}{3\lambda_1}}.
\end{align}
 The engineering stress is given by
\begin{align}
    P^M_{11} =\frac{\sigma_{11}}{\lambda_1} &=  \frac{\nu EAL_f}{3}\bigg(\frac{\lambda_1^3-1}{\lambda_1^2}\bigg)\bigg(1-\frac{1}{\lambda_c}\bigg).
\end{align}
Next, we show how the constraints (namely traction free boundary condition and incompressibility) employed for the above formula can be incorporated in the present homogenization model. 
\subsection{Incompressible uniaxial tension using finite strain homogenization}
In order to obtain the response consistent with incompressible uniaxial tension of \cite{brown2009multiscale} using the present homogenization scheme, think of performing stress-driven homogenization with an additional incompressibility constraint $\det(\textbf{F}^M)=1$. Enforcing this in the stress-driven constrained energy functional through the Lagrange multiplier $p$, we get
\begin{align}
    \textbf{P}^M-\textbf{P}_0^M + p~\text{det}(\textbf{F}^M)\textbf{F}^{M^{-T}}~\textbf{I} = \textbf{0}
\end{align}
or in terms of Cauchy stress
\begin{align}\label{eq:incompressible_macro_stress}
    \boldsymbol{\sigma}^M - \boldsymbol{\sigma}_0^M + p\textbf{I} = \textbf{0}.
\end{align}
 Now consider the uniaxial tension case in which the prescribed macroscopic stress $\boldsymbol{\sigma}_0^M = diag(\sigma^M_0,0,0)$. From the second and third diagonal component of \cref{eq:incompressible_macro_stress}, we obtain the incompressibility Lagrange multiplier as follows:
\begin{align}\label{eq:incompressibility_lm}
    p = -\sigma^M_{22} = -\sigma^M_{33}.
\end{align}
Substituting the above relationship in the first diagonal component of  \cref{eq:incompressible_macro_stress}, we get
\begin{align}\label{eq:unixial_tension_cauchy_stress_incompressible}
    \sigma_{11}^M = \sigma_0^M - p =  \sigma_0^M + \sigma_{22}^M=  \sigma_0^M + \sigma_{33}^M.
\end{align}
Instead of performing stress-driven homogenization with the additional incompressibility constraint as described above, we perform strain-driven homogenization and account for the uniaxial tension response as a post processing step. We use the incompressible deformation gradient as the input to the strain-driven homogenization problem and then obtain the macroscopic Cauchy stress tensor using \cref{eq:macro_P_analytic} and $\boldsymbol{\sigma}^M = (1/\text{det}(\textbf{F}^M))\textbf{P}^M(\textbf{F}^M)^T$. This stress tensor does not correspond to the uniaxial tension response. It will have $\sigma^M_{22} = \sigma^M_{33} \neq 0$. However, using the interpretation of the Lagrange multiplier in \cref{eq:incompressibility_lm}, we can obtain the incompressible uniaxial tension stress tensor component $\sigma^M_{0}$ from the \cref{eq:unixial_tension_cauchy_stress_incompressible} even with a strain-driven approach. It is essentially given by
\begin{align}
    \sigma_0^M = \sigma^M_{11} - \sigma^M_{22} = \sigma^M_{11} - \sigma^M_{33}.
\end{align}
The results presented in \ref{sec:numerical_examples_extension} for the uniaxial tension validation against collagen experiments and for the helical 8-chain model under uniaxial tension were obtained using this approach.
\subsection{Helical 8-chain model}\label{sec:uniaxial_tension_helix_appendix}
\begin{figure}[h!]
    \centering
    \begin{subfigure}{\textwidth}
    \includegraphics[width=0.32\textwidth]{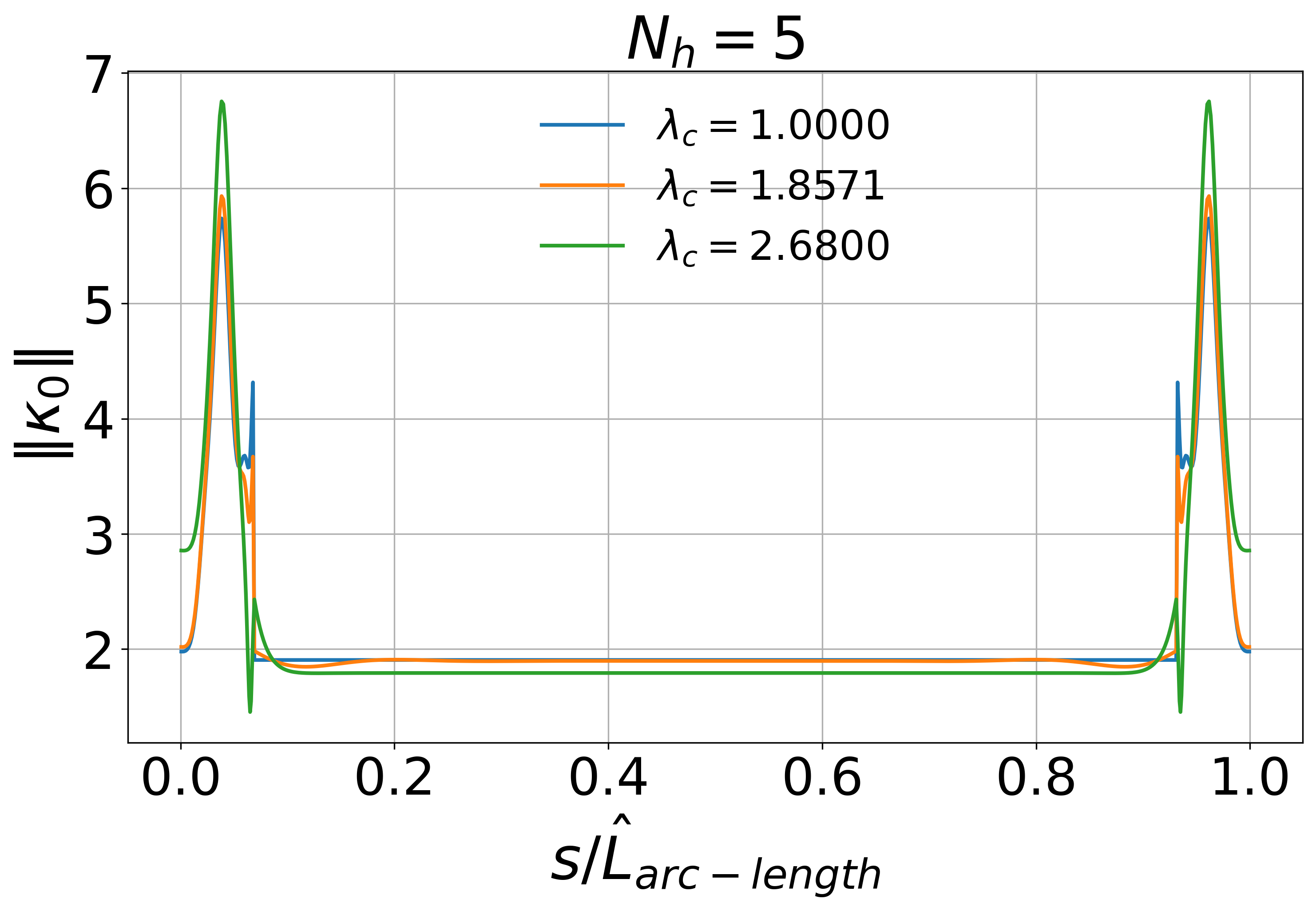}
    \includegraphics[width=0.32\textwidth]{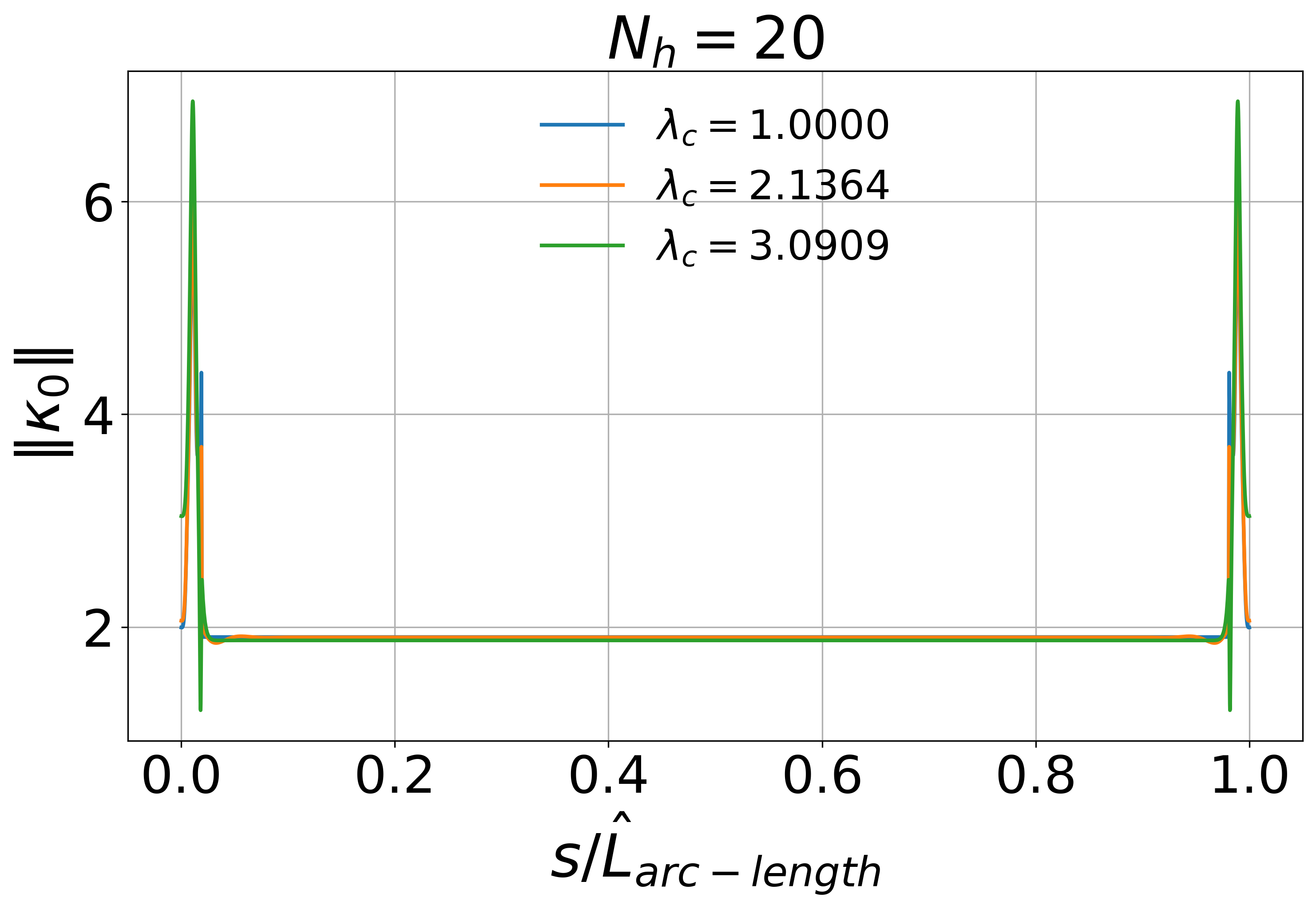}
    \includegraphics[width=0.32\textwidth]{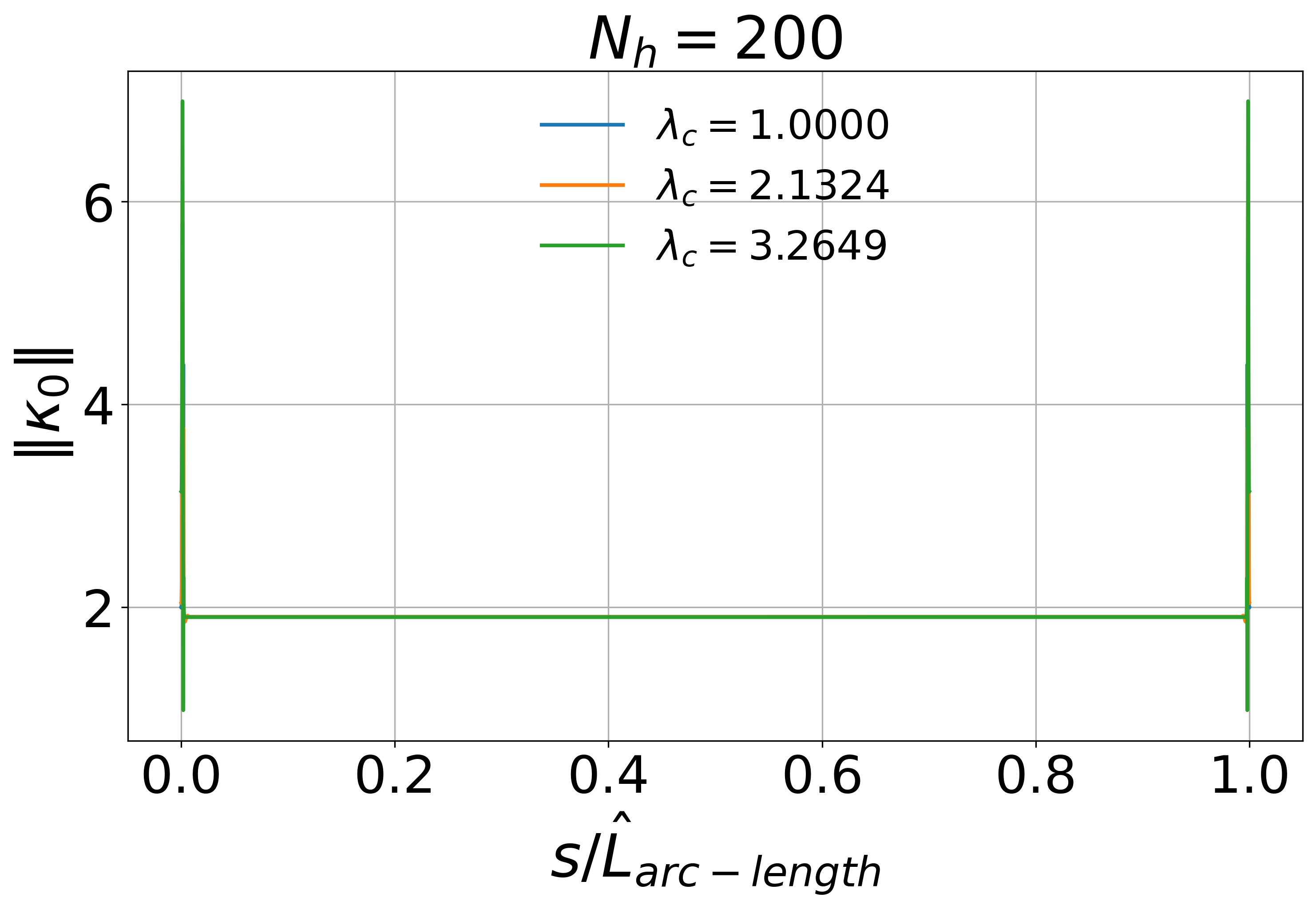}
    \caption{Effect of number of turns on the norm of curvature assumption for a single modified helix along its arc length. The zone of non-uniformity of curvature  decreases as the number of turns are increased.}
    \label{fig:single_helix_modified_curvature_norm}
\end{subfigure}
\begin{subfigure}{0.6\textwidth}
    \includegraphics[width=\textwidth]{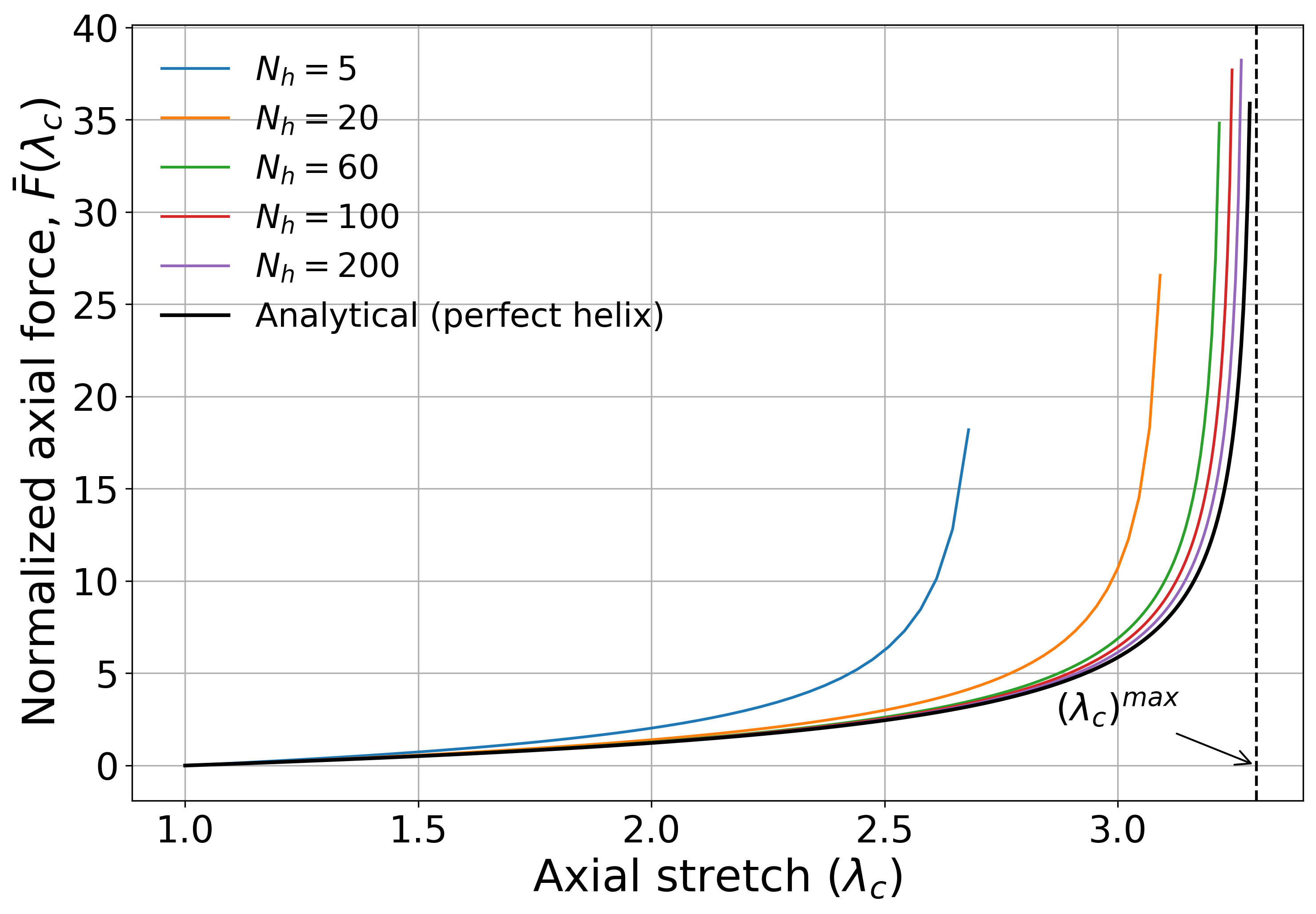}
    \caption{Effect of number of turns on the force-stretch response of a single modified helix}
    \label{fig:single_helix_modified_force_stretch}
\end{subfigure}
    \caption{Uniaxial stretching of single helical fiber - $\hat{p}_h = 1.0$, $\hat{r}_h = 0.5$ and $p_j = 1.0$. Here, the normalized axial force $\bar{F}(\lambda_c) = F(\lambda_c)/K_b$. In the perfect helix portion, $N^{el} = 500~(N_h=5)$, $N^{el} = 1000~(N_h=20)$, $N^{el} = 1000~(N_h=60)$, $N^{el} = 3000~(N_h=100)$, $N^{el} = 4000~(N_h=200)$. For the end curves, $N^{el}=50$.}
    \label{fig:single_helix_stretching}
\end{figure}

Instead of straight fibers in the 8-chain unit cell, let us assume each fiber to be a modified/imperfect helix of the form given in \cite{chang2022mechanics}. We call this the helical 8-chain model. We show below that the modified helix force-stretch relationship converges to that of a perfect helix in the limit of a large number of turns. Consider a perfect helix with undeformed radius $\hat{r}_h$, pitch $\hat{p}_h$, and number of turns $N_h$. The deformed arc length and end-to-end length of this helix is given by
\begin{align}
    \hat{L}_{\text{arc-length}} = N_h\sqrt{4\pi \hat{r}_h^2 + \hat{p}_h^2}\quad\quad \text{and} \quad     \hat{L}_h = N_h\hat{p}_h,
\end{align}
respectively. As the helix is stretched, assuming constant (during deformation) and uniform (along the helix arclength) norm of curvature, the force-stretch relationship of a perfect helix is given by \citep{djurivckovic2013twist}:
\begin{align}
    F(\lambda_c) = \eta^2 \frac{\hat{p}_h}{2\pi} \bigg[-\mathbb{C}^{rod}_{44}\lambda_c(1-\frac{\hat{r}_h}{r_h})+\mathbb{C}^{rod}_{66}(\lambda_c-1)\bigg]
\end{align}
where
\begin{align}\label{eq:helix_parameters_rel}
    \frac{\hat{r}_h}{r_h}  &= \frac{1}{\sqrt{1+(\frac{\hat{p}_h}{2\pi\hat{r}_h})^2(1-\lambda_c^2)}},\quad \text{and}\quad 
    \eta^2 =  \frac{1}{\hat{r}_h^2 + (\frac{\hat{p}_h}{2\pi})^2} =  \frac{1}{{r}^2 + (\frac{p_h}{2\pi})^2}.
\end{align}
The quantities $\mathbb{C}^{rod}_{44}$ and $\mathbb{C}^{rod}_{66}$ are bending and twisting stiffnesses of the rod, respectively. The above force-stretch relationship can now be directly substituted in \cref{eq:straight_fiber_force_stretch_relation} to obtain the uniaxial tension response of 8-chain helical (perfect) RVE. The assumption of constant curvature norm for the above formula is crucial and Figure \ref{fig:single_helix_modified_curvature_norm} shows it is only realized with a large number of turns in the helix. Figure \ref{fig:single_helix_modified_force_stretch} shows the comparison of the modified helix response with that of a perfect helix. As the number of turns in the modified helix is increased, the response approaches the ideal helix.
The following are some of the important points in the presented helical 8-chain model:
\begin{itemize}
    \item From \cref{eq:helix_parameters_rel}(a), the term inside the square root must be positive, thus the following must hold:
\begin{align}
    \lambda_c < \sqrt{1+\left(\frac{2\pi\hat{r}_h}{\hat{p}_h}\right)^2}
\end{align}
The above inequality gives a limit on the maximum stretch ($(\lambda_c)^{max})$) that a uniform helical fiber can exhibit. For $\hat{p}_h = 1.0$ and $\hat{r}_h=0.5$
\begin{align}\label{eq:perfect_helix_max_stretch}
    (\lambda_c)^{max} \approx 3.2969
\end{align}
\item  From \cref{eq:straight_fiber_8_chain}(b), this also means
\begin{align}\label{eq:helical_8chain_stretch_constraint}
    \lambda_1^2 + \frac{2}{\lambda_1} < 3(\lambda_c)^{max}
\end{align}. Using \cref{eq:perfect_helix_max_stretch}, the two positive roots of \cref{eq:helical_8chain_stretch_constraint} give
\begin{align}
    \lambda_1^{max} = 5.68595,\quad \quad \lambda_1^{min} = 0.06134
\end{align}
These are the maximum stretch and minimum compression an 8-chain RVE (with perfect helix) can exhibit (as shown in Figure \ref{fig:8chain_helical_uniaxial_tension_compression}).
\item The inclined rods undergo tension even when $0<\lambda_1<1$, i.e., during uniaxial compression of 8-chain RVE. This is clear from the fact that the expression in \cref{eq:straight_fiber_8_chain}(b) has a minima at $\lambda_1 = 1.0$.
\end{itemize}
\end{document}